\documentclass[twocolumn]{aastex63}
\usepackage{rotating}
\usepackage{color, colortbl}
\usepackage{tikz}
\usepackage{afterpage}
\usepackage{pifont}
\usepackage[para]{threeparttable}
\usepackage{tabularx}
\usepackage{graphicx}

\shorttitle{SN progenitor stars}
\shortauthors{Strotjohann et al.}
\graphicspath{{./}{figures/}}

\begin{document}

\title{Search for supernova progenitor stars with ZTF and LSST}

\author[0000-0002-0786-7307]{Nora L. Strotjohann}
\affiliation{Department of Particle Physics and Astrophysics, Weizmann Institute of Science, 76100 Rehovot, Israel}

\author{Eran O. Ofek}
\affiliation{Department of Particle Physics and Astrophysics, Weizmann Institute of Science, 76100 Rehovot, Israel}

\author{Avishay Gal-Yam}
\affiliation{Department of Particle Physics and Astrophysics, Weizmann Institute of Science, 76100 Rehovot, Israel}

\author[0000-0003-1546-6615]{Jesper Sollerman}
\affiliation{Department of Astronomy, The Oskar Klein Center, Stockholm University, AlbaNova, 10691 Stockholm, Sweden}

\author{Ping Chen}
\affiliation{Department of Particle Physics and Astrophysics, Weizmann Institute of Science, 76100 Rehovot, Israel}

\author{Ofer Yaron}
\affiliation{Department of Particle Physics and Astrophysics, Weizmann Institute of Science, 76100 Rehovot, Israel}

\author{Barak Zackay}
\affiliation{Department of Particle Physics and Astrophysics, Weizmann Institute of Science, 76100 Rehovot, Israel}

\author[0000-0002-5683-2389]{Nabeel Rehemtulla}
\affiliation{Department of Physics and Astronomy, Northwestern University, 2145 Sheridan Road, Evanston, IL 60208, USA}
\affiliation{Center for Interdisciplinary Exploration and Research in Astrophysics (CIERA), 1800 Sherman Ave., Evanston, IL 60201, USA}

\author{Phillipe Gris}
\affiliation{Laboratoire de Physique de Clermont, IN2P3/CNRS,F-63000 Clermont-Ferrand, France}

\author[0000-0002-8532-9395]{Frank J. Masci}
\affiliation{IPAC, California Institute of Technology, 1200 E. California
             Blvd, Pasadena, CA 91125, USA}

\author[0000-0001-7648-4142]{Ben Rusholme}
\affiliation{IPAC, California Institute of Technology, 1200 E. California
             Blvd, Pasadena, CA 91125, USA}

\author[0000-0003-1227-3738]{Josiah Purdum}
\affiliation{Caltech Optical Observatories, California Institute of Technology, Pasadena, CA 91125, USA}

\begin{abstract}
The direct detection of core-collapse supernova (SN) progenitor stars is a powerful way of probing the last stages of stellar evolution. However, detections in archival Hubble Space Telescope images are limited to about one detection per year. Here, we explore whether we can increase the detection rate by using data from ground-based wide-field surveys. Due to crowding and atmospheric blurring, progenitor stars can typically not be identified in pre-explosion images alone. Instead, we combine many pre-SN and late-time images to search for the disappearance of the progenitor star.
As a proof of concept, we implement our search for ZTF data. For a few hundred images, we achieve limiting magnitudes of $\sim23\,\text{mag}$ in the $g$ and $r$ band. However, no progenitor stars or long-lived outbursts are detected for 29 SNe within $z\leqslant0.01$, and the ZTF limits are typically several magnitudes less constraining than detected progenitors in the literature.
Next, we estimate progenitor detection rates for the Legacy Survey of Space and Time (LSST) with the Vera C. Rubin telescope by simulating a population of nearby SNe. The background from bright host galaxies reduces the nominal LSST sensitivity by, on average, $0.4\,\text{mag}$. Over the ten-year survey, we expect the detection of $\sim50$ red supergiant progenitors and several yellow and blue supergiants. The progenitors of SNe Ib and Ic are detectable if they are brighter than $-4.7\,\text{mag}$ or $-4.0\,\text{mag}$ in the LSST $i$ band, respectively. In addition, we expect the detection of hundreds of pre-SN outbursts depending on their brightness and duration.
\end{abstract}


\section{Introduction} \label{sec:intro}

While thousands of core-collapse supernovae (SNe) are discovered and classified every year\footnote{see, e.g., \url{https://www.wis-tns.org/stats-maps}}, detecting their faint progenitor stars is much more challenging. Therefore, we cannot be certain that they are similar to the well-studied stars in the Milky Way or Magellanic clouds. 

Direct progenitor star detections so far (see, e.g., \citealt{smartt2015} or \citealt{vandyk2017} for reviews) have established that the progenitors of SNe II are red supergiants (RSG) and SNe IIb have been observed to arise from yellow supergiants (YSG). Slowly rising, SN1987A-like SNe II are the explosions of more compact blue supergiants (BSG), and at least some interacting SNe IIn are believed to originate from luminous blue variables (see, e.g., \citealt{gal-yam2007_2005gl_lbv_progenitor} or \citealt{smith2017_lbv_fate}). Less is known about the progenitor stars of other, rarer SN types, like SNe Ibc or Ibn (see, e.g., \citealt{eldridge2016_disappearance_Ib_progenitor_iPTF13bvn, xiang2019_sn2017ein_progenitor, kilpatrick2021} for potential detections). SN observations indicate that their progenitors are partially or completely stripped, massive stars. The stripping presumably requires either strong winds or a binary partner. 

The Hubble Space Telescope (HST) has been very successful at detecting progenitor stars. However, detections are only attained at a rate of about one progenitor per year (see e.g. \citealt{davies2018}), such that increasing the sample size substantially would require decades of observations. Archival HST observations are available for about 25\% of the closest SNe \citep{smartt2015}, and identifying a progenitor star securely requires both precise astrometry and additional late-time HST observations to verify that the progenitor candidate has indeed vanished (see, e.g., \citealt{crockett2011, maund2014, maund2015, vandyk2023_disapperance_of_progenitor_stars}). This confirmation is crucial as some of the brightest progenitor candidates have turned out to be stellar clusters rather than single stars (e.g., \citealt{maund2009_progenitor_disappearance}, \citealt{maund2015}, \citealt{vandyk2023_disapperance_of_progenitor_stars}).

\citet{smartt2015} (and earlier \citealt{smartt2009}) compile a sample of detected progenitor stars and compare their inferred masses to predictions by stellar models. \citet{smartt2015} find that all 26 RSG mass estimates and upper limits are fainter than $\log_{10}(L/L_\odot)\leqslant5.2$, corresponding to a bolometric magnitude of $-8.2\,\text{mag}$. Based on stellar evolution models, they conclude that all progenitors had initial masses of $<18\,\text{M}_\odot$, while they would expect that $30\%$ of the progenitors are more massive. This discrepancy, pointed out earlier by \citet{li2006_rsgproblem} and \citet{kochanek2008_failedsne}, was coined the RSG problem, and \citet{smartt2015} suggests that the most massive stars collapse into black holes without producing a bright SN. Such failed SNe are also predicted by studies that simulate stellar cores (see, e.g., \citealt{patton2020} for a recent result), and a few candidates have been reported \citep{reynolds2015_HST_disappeared_star, basinger2021_lbt_disappeared_star, neustadt2021_LBT_missingBSG}.

However, the lack of bright SN progenitors was diagnosed based on sparse observations. SN progenitor stars are usually only detected in one or few HST observations, often in a single band. Thus, the star's surface temperature and bolometric luminosity cannot be estimated reliably (see, e.g., \citealt{smartt2015, davies2018}). Other uncertainties are induced by host extinction, circumstellar dust (e.g., \citealt{kochanek2012}), uncertain SN distances, and the small number of progenitor detections \citep{davies2018}. Consequently, it is under debate whether the RSG problem is significant (e.g., \citealt{davies2018, davies2020, kochanek2020, rodriguez2022}).

An additional complication is that the impending core collapse might trigger mass-loss events that change the progenitor's temperature and luminosity: Violent stellar eruptions are common prior to SNe IIn \citep{ofek2014, strotjohann2021}, and similar, but fainter, outbursts were also detected prior to SNe II \citep{jacobson-galan2022}, SNe Ibn \citep{pastorello2007, strotjohann2021}, broad-lined SNe Ic \citep{ho2019}, and potentially SNe IIb \citep{strotjohann2015_precursors_sneIIb}. In addition, \citet{margutti2017} and \citet{sollerman2020} observed late-time interaction for three SNe Ib, which indicates a major mass-loss event shortly before the SN explosion. The spectra of young SNe indicate that a large fraction of them explode within a confined shell of circumstellar medium \citep{khazov2016, bruch2021, bruch2022}, which points to increased mass loss in the last years before the core collapse. Outbursts that inflate stellar envelopes have also been proposed to explain the fast rise times and hot temperatures of young SNe \citep{morazova2020, foerster2018_early_snlcs_csm}. Mass-loss events can boost the progenitor luminosity, e.g., due to interaction of the ejected material. However, absorption or inflated envelopes can also redden the stellar spectrum and reduce the progenitor luminosity \citep{davies2022}. Pre-SN outbursts can hence cause dimming or brightening in a single band.

Most of the described challenges could be mitigated by a larger sample of SN progenitors detected in several bands and epochs. Therefore, we explore here whether ground-based, large field-of-view surveys are sensitive enough to detect the closest SN progenitor stars.
We consider two surveys: the Zwicky Transient Facility (ZTF; \citealt{bellm2019, graham2019, dekany2020_ztf_observing_system}) that has been running since 2018 and the planned, more sensitive Legacy Survey of Space and Time (LSST; \citealt{ivezic2019}).
Due to blurring by the atmosphere, stars in nearby galaxies are blended with each other, such that pre-SN data alone is usually not sufficient to pinpoint the progenitor. The angular resolution of ZTF observations is typically limited by the median seeing of $2\,\text{arcsec}$, while a median seeing of $0.65\,\text{arcsec}$ is expected for the LSST site \citep{ivezic2019}. In contrast, the HST wide field camera 3 has a much smaller PSF width of $\sim0.1\,\text{arcsec}$ in the infrared \citep{dressel2023_wfc3handbook}. To detect the progenitor despite the lower resolution, we combine many images before the SN and after it has faded and search for a flux difference between these two time windows. The search does not require dedicated observations and can be done for any position monitored by a survey.

Ground-based, wide-field surveys have so far detected a few progenitor stars: The bright YSG progenitor of SN\,2011dh was detected by the Palomar Transient Factory (PTF; \citealt{strotjohann2015_precursors_sneIIb}), the Large Binary Telescope \citep{szczgiel2012_variability_of_sn2011dh_progenitor}, and the Nordic Optical Telescope \citep{ergon2015_sn2011dh}. Combining hundreds of PTF images before and after the SN confirmed that the progenitor of SN\,2011dh had indeed disappeared. The same search was sensitive enough to disfavor a progenitor candidate for another SN IIb, SN\,2012P, as it was still present after the SN had faded. In a more recent search, we detected the progenitor star of the IIn SN\,2019cmy at a seemingly constant luminosity of $-14\,\text{mag}$ in the ZTF $g$ and $r$ bands in the last year before the SN explosion \citep{strotjohann2021}. However, the star is fainter in earlier PTF observations (Soumagnac et al. in prep.); we hence likely observe a long-lasting outburst rather than a quiescent progenitor star. 
Progenitors in their quiescent state are many magnitudes fainter and the low temperatures of RSG make their detection in visible bands even more challenging.

In Sect.~\ref{sec:methods}, we conduct a progenitor search for ZTF data as a proof of concept and compare the ZTF results to earlier progenitor detections. In Sect.~\ref{sec:lsst}, we quantify how sensitive LSST will be to progenitor stars and pre-SN outbursts and we conclude in Sect.~\ref{sec:conclusion}.

\section{Search for Progenitor stars in ZTF data} \label{sec:methods}

As a test, we implement our progenitor search for the closest SNe in ZTF data. Section~\ref{sec:sample} describes the sample selection, and Sect.~\ref{sec:search} explains the details of the search. Results are presented in Sect.~\ref{sec:results}, and we compare to progenitor detections in literature in Sect.~\ref{sec:compare_to_hst}. Finally, in Sect.~\ref{sec:ztf_sensitivity} we quantify whether the search is as sensitive as expected. \ref{sec:ztf_details} provides more details on the photometric pipeline and error sources, and we verify the sensitivity of the search by injecting faint, artificial sources into the images and quantifying their recoverability.

\subsection{Sample selection} \label{sec:sample}

\begin{figure*}[tp]
\centering
\includegraphics[width=\textwidth]{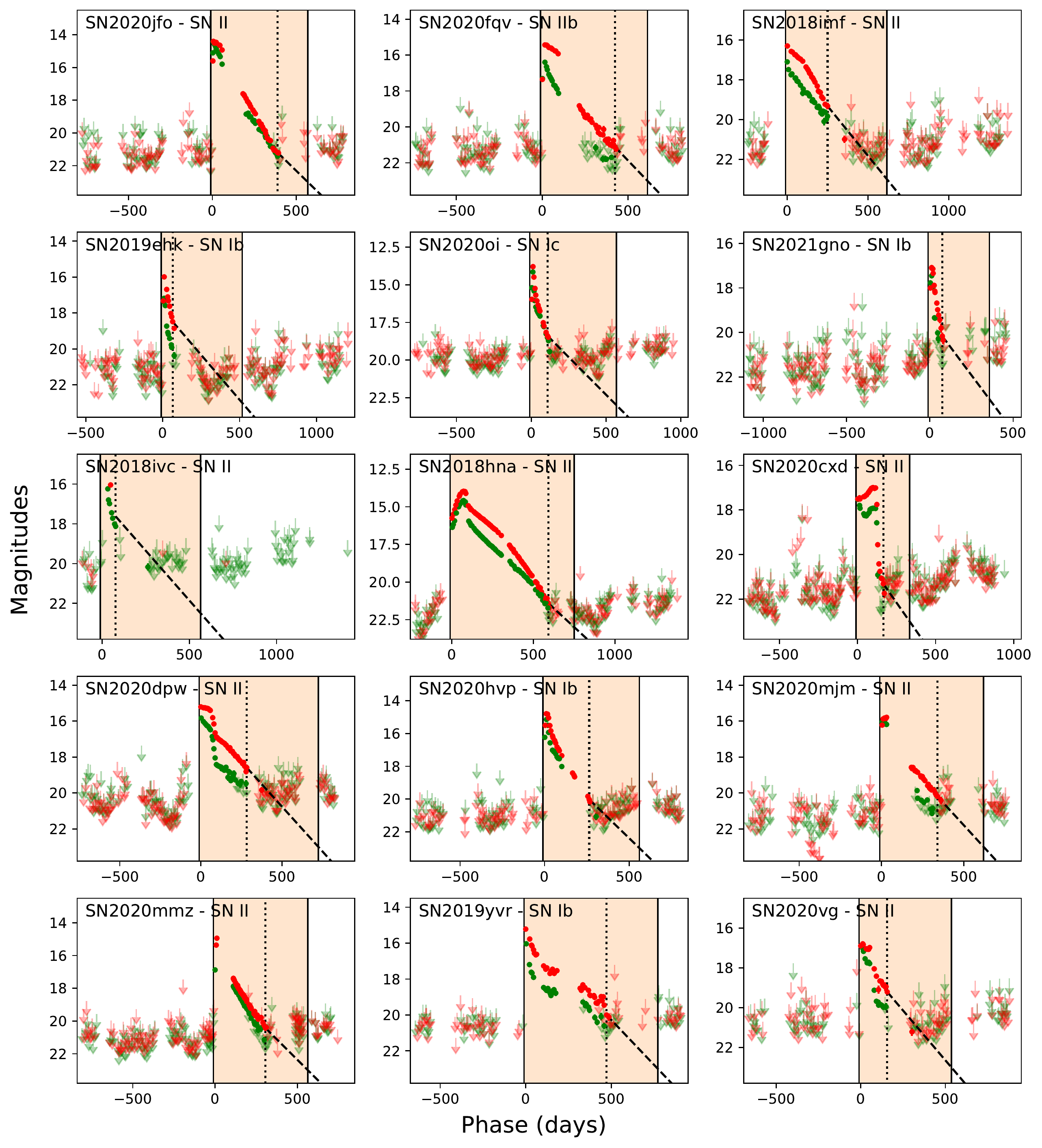}
\\
\caption{Light curves in week-long bins for 29 nearby SNe with enough observations before and after the SN (continued in Fig.~\ref{fig:sn_lcs2}). To determine whether the SN is still present at late times, we identify the second-last $5\,\sigma$ detection in the $r$ band and extrapolate the light curve with a slope of 1\,mag per 100\,days (shown as a dashed line). SN\,2018ivc and SN\,2018ebt do not have enough $r$-band observations, so we extrapolate the $g$-band light curve, but shift the curve by $0.5$ mag, as most SNe are brighter in the $r$ band at this time. The orange shaded area marks the time when the SN is likely brighter than $23\,\text{mag}$ in the ZTF $r$ band and we combine observations before and after the SN, respectively, to measure both the early- and late-time flux at the SN position.}
\label{fig:sn_lcs}
\end{figure*}

\setcounter{figure}{0}
\begin{figure*}[tp]
\centering
\includegraphics[width=\textwidth]{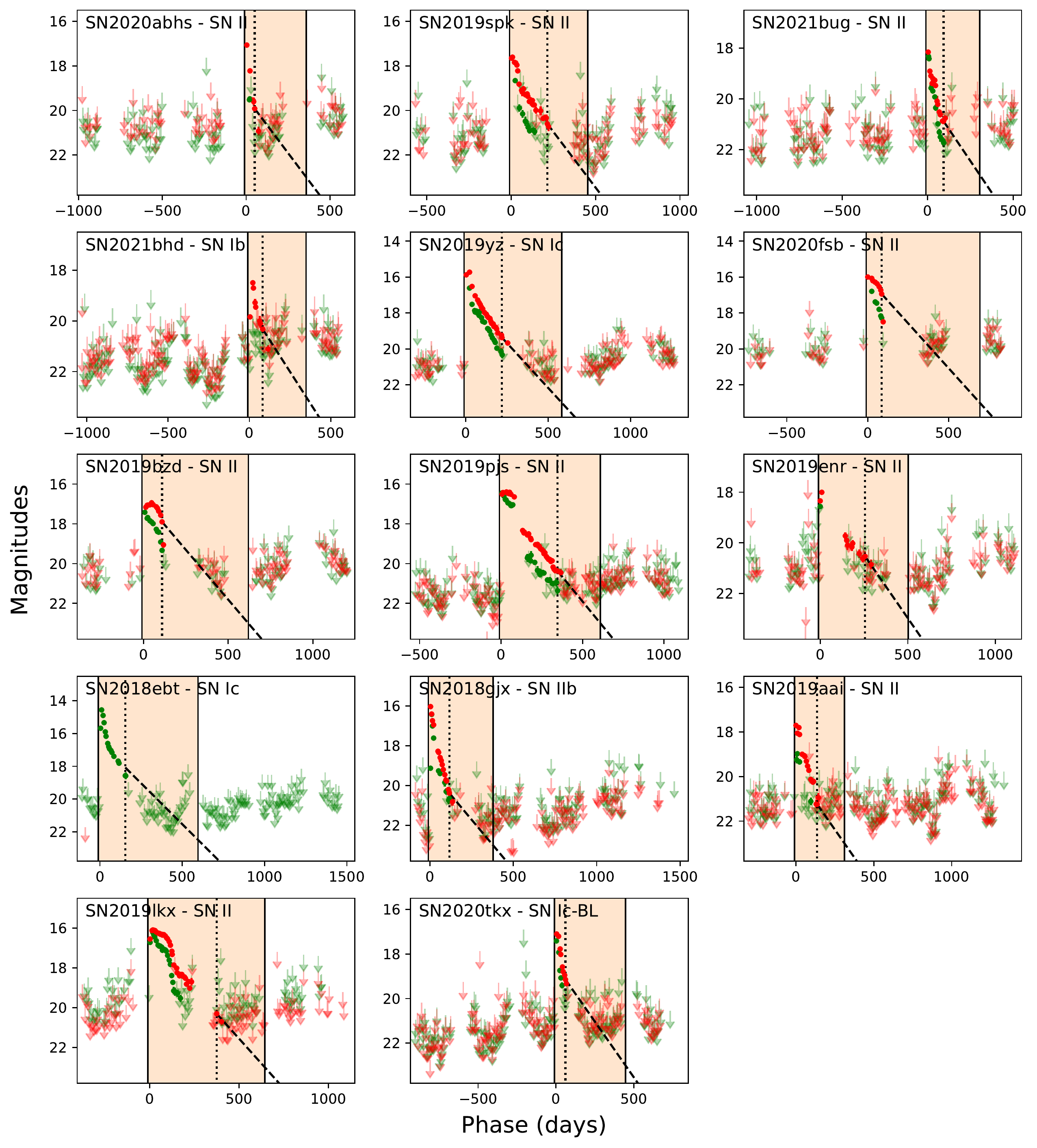}
\caption{Continuation of Fig.~\ref{fig:sn_lcs}.}
\label{fig:sn_lcs2}
\end{figure*}

Our ZTF search is based on SNe detected by the Bright Transient Survey (BTS; \citealt{fremling2020, perley2020}). The ZTF pipeline \citep{masci2019} uses the ZOGY image subtraction algorithm \citep{zackay2016} and yields $>10^6$ potential detections per night \citep{patterson2019}, but quality cuts reduce the number of SN candidates to $\sim50$ per night \citep{perley2020}. An on-duty astronomer searches the remaining candidates for genuine, SN-like transients that surpass the brightness threshold of $18.5\,\text{mag}$ in the $g$ or $r$ band. The selected objects are classified with the SEDM spectrograph on the P60 telescope \citep{blagorodnova2018, rigault2019, kim2022}, and discoveries are reported to the Transient Name Server\footnote{\url{https://www.wis-tns.org/}} and the BTS sample explorer\footnote{\url{https://sites.astro.caltech.edu/ztf/bts/explorer.php}}. For our search, we select 60 SNe that exploded between 2018 and 2021 within $z\leqslant0.01$ or $45\,\text{Mpc}$.

The luminosity distances of nearby SNe are rather uncertain, and, if available, we use the more precise host galaxy redshift instead of the SN redshift. 
We here adopt the preferred redshift from the NASA/IPAC Extragalactic Database\footnote{\url{https://ned.ipac.caltech.edu/}} and convert it to the infall-corrected distance. For SN\,2018ebt, we measure a host redshift of $z=0.0095$ from galaxy lines in a late-time spectrum. After refining the redshifts, we are left with 55 SNe within $45\,\text{Mpc}$.

Next, we download IPAC difference images \citep{irsa} \footnote{Including data until August 2022.} and calculate forced-photometry light curves for the remaining SNe as described in Appendix~\ref{sec:pipeline}. We inspect the light curves visually and, if necessary, correct the approximate explosion date, $t_0$, to ensure that the selected pre-SN observations do not contain any SN light. Conservatively, we only combine observations obtained more than ten days before $t_0$ to calculate the pre-explosion flux. IPAC produces a single ZTF reference image for each combination of ZTF field, filter, CCD number, and CCD quadrant \citet{masci2019}. In the following, we call each unique combination of these parameters a field and only compare difference images in the same field, i.e., that were produced with the same reference image. We require at least 15 pre-SN observations in the same field and discard three out of 55 SNe because they do not have enough pre-SN observations. We also use the selected pre-SN observations to do a baseline correction to ensure that the pre-SN light curve corresponds to zero flux and we scale up the error bars if they are too small to account for the observed scatter (details given in \ref{sec:pipeline}).

\movetabledown=40mm
\begin{rotatetable}
\begin{deluxetable*}{llcclcclcc}
\tablecaption{\label{tab:sample}Nearby BTS SNe}
\tablehead{\colhead{IAU name} & \colhead{ZTFID} & \colhead{R.A.} & \colhead{Dec.} & \colhead{Type} & \colhead{$t_0$} & \colhead{$t_{23}$} & \colhead{host} & \colhead{distance} & \colhead{progenitor det.}\\ 
\colhead{} & \colhead{} & \colhead{(deg)} & \colhead{(deg)} & \colhead{} & \colhead{(JD)} & \colhead{(days)} & \colhead{} & \colhead{(Mpc)} & \colhead{}} 
\startdata
SN\,2020jfo	&	ZTF\,20aaynrrh	&	185.460355	&	  4.481697	&   SN\,II      &   2458974.7   & 569 & M\,61             & 14.7	        & $25.47\pm0.07$ (F814W)$^a$ \\
SN\,2020fqv	&	ZTF\,20aatzhhl	&	189.138575	&	 11.231653	&	SN\,II     &	2458937.8	& 612 & NGC\,4568       & 14.9          &	$>24.8$ (F606W)$^b$\\
SN\,2018imf	&	ZTF\,18acqsqrg	&	190.672493	&	 13.265236	&	SN\,II     &	2458438.0	& 617 & NGC\,4639       & 15.0          & \\
SN\,2019ehk	&	ZTF\,19aatesgp	&	185.733956	&	 15.826127	&	SN\,Ib (Ca rich)	    &	2458602.7	& 517 & NGC\,4321       & 15.1          & $>28.7$ (F555W)$^c$\\ 
SN\,2020oi	&	ZTF\,20aaelulu	&	185.728855	&	 15.823611	&	SN\,Ic	    &	2458855.9	& 570 & NGC\,4321       & 15.1          & bright cluster$^d$\\
SN\,2021gno	&	ZTF\,21aaqhhfu	&	183.042900	&	 13.249178	&	SN\,Ib (Ca rich)$^e$      &   2459293.7	& 358 & NGC\,4165	    & 15.1          &	\\
SN\,2018ivc	&	ZTF\,18acrcogn	&	 40.671990	&	$-0.008900$	&	SN\,II	    &	2458447.9	& 565 & NGC\,1068       & 16.4          & $>25.4$ (F606W)$^f$\\
SN\,2018hna	&	ZTF\,18acbwaxk	&	186.550338 &	 58.314126	&	SN\,II (BSG)$^g$ &	2458420.9	& 751 & UGC\,07534      & 16.5          & \\ 
SN\,2020cxd	&	ZTF\,20aapchqy	&	261.621952	&	 71.094062	&	SN\,II      &	2458898.0	& 333 & NGC\,6395       & 23.6          & \\
SN\,2020dpw	&	ZTF\,20aauhbvu	&	309.293939	&    66.102971	&	SN\,II	    &	2458936.0	& 724 & NGC\,6951       & 26.1          &	\\
SN\,2020hvp	&	ZTF\,20aavzffg	&	245.439146	&	$-2.289270$ &	SN\,Ib 	    &	2458962.9	& 561 & NGC\,6118       & 25.2          & \\
SN\,2020mjm	&	ZTF\,20abeohfn	&	217.372229	&	$-0.021626$ &	SN\,II	    &	2459012.2	& 620 & UGC\,09299      & 28.3	        &   \\
SN\,2020mmz	&	ZTF\,20abevbxv	&	140.298238	&	 64.253968	&	SN\,II	    &	2459012.7	& 563 & NGC\,2814       & 28.9          &	\\
SN\,2019yvr	&	ZTF\,20aabqkxs	&	191.283891	&	$-0.459114$ &	SN\,Ib	    &	2458846.0	& 772 & NGC\,4666       & 29.4          & $24.9 \pm 0.02$ (F635W)$^h$\\
SN\,2020vg	&	ZTF\,20aafclxb	&	177.226830	&   $-4.681626$ &	SN\,II 	    &	2458860.0	& 536 & NGC\,3915       & 29.4          & \\
SN\,2020abhs	&ZTF\,20actodrq	&	193.739146	&  $-13.541537$ &	SN\,II	    &	2459174.0	& 359 & --              & 31.2$^i$      &	\\
SN\,2019spk	&	ZTF\,19acecluy	&	150.726656	&	$-6.207211$ &	SN\,II 	    &	2458756.0	& 454 & PGC\,166103     & 32.6          & \\
SN\,2021bug	&	ZTF\,21aagydmn	&	188.594858	&     2.317296	&	SN\,II  	&	2459249.9	& 305 &	NGC\,4533       & 32.6          & \\
SN\,2021bhd	&	ZTF\,21aagbpvf	&	207.061475	&	 68.089699	&	SN\,Ib  	&	2459230.9	& 349 &	UGC\,08737      & 33.2          & \\
SN\,2019yz	&	ZTF\,19aadttht	&	235.488760	&	  0.710949	&	SN\,Ic	    &	2458501.0	& 583 & UGC\,09977      & 34.6          & \\
SN\,2020fsb	&	ZTF\,20aaunfpj	&	234.765872	&  $-25.974543$ &	SN\,II	    &	2458936.9	& 692 & ESO\,515$-$G004 & 34.7          &	\\
SN\,2019bzd	&	ZTF\,19aamwhat	&	221.883498	&  $-19.766047$	&	SN\,II	    &	2458560.8	& 619 & ESO\,580$-$029  & 39.9          & \\
SN\,2019pjs	&	ZTF\,19abwztsb	&	271.168089	&    21.634511	&	SN\,II	    &	2458730.7	& 610 & UGC\,11105      & 40.1          & \\
SN\,2019enr	&	ZTF\,19aatwvft	&	143.698711	&	 10.286955	&	SN\,II      &	2458608.2	& 503 & NGC\,2919       & 40.6          & \\
SN\,2018ebt	&	ZTF\,18abjrbza	&	310.479145	&	 64.214621	&	SN\,Ic	    &	2458315.9	& 596 & --              & 42.4          & \\
SN\,2018gjx	&	ZTF\,18abwkrbl	&	 34.064926	&	 28.591298	&	SN\,IIb	    &	2458374.9	& 378 & NGC\,0865       & 43.2          & \\ %
SN\,2019aai	&	ZTF\,19aadtqcd	&	248.086436	&	 19.840120	&	SN\,II	    &	2458502.0	& 311 & NGC\,6181       & 43.2          & \\
SN\,2019lkx	&	ZTF\,19abgiwkt	&	 55.262133	&	$34.652071$	&	SN\,II	    &	2458677.9	& 645 & --              & 44.6\,$^{i}$  & \\ 
SN\,2020tkx   &   ZTF\,20abzoeiw&   280.037525  &	 34.116527  &   SN\,Ic-BL   &   2459108.8   & 446 & --              & 44.6\,$^{i}$  &\\
\enddata
\tablecomments{All SNe with $\geqslant20$ data points before and after the SN in the same field. Data obtained more than 10 days before $t_0$ are used to measure the pre-explosion flux. At $t_{23}$ the SN has likely faded to $23$rd magnitude in the $r$ band and we combine later observations to measure the late-time flux.
$^a$ \citet{sollerman2021}; $^b$ \citet{tinyanont2022_sn2020fqv}; $^c$ \citet{jacobson_galan2020}; $^d$ \citet{gagliano2022}; $^e$ \citet{jacobson-galan2022_sn2021gno}; $^f$ \citet{bostroem2020};
$^g$ \citet{singh2019}; $^h$ \citet{kilpatrick2021}, likely a binary system \citet{sun2022_sn2019yvr_binaryprogenitor}; $^i$ No host galaxy redshift available. We use the less precise SN redshift.}
\end{deluxetable*}
\end{rotatetable}
\clearpage

The expected sensitivity of our search is $\sim23\,\text{mag}$, and we extrapolate the late-time SN light curves to determine when they have faded below this threshold. To gain sensitivity to the faint late-time fluxes, we combine the $g$- and $r$-band fluxes in 7-day-long bins as described in Appendix~\ref{sec:pipeline}. To avoid marginal detections, we select the second-last $r$-band detection and extrapolate it with a slope of $1\,\text{mag}$ per 100 days, the decay rate of $^{56}\text{Co}$. Two SNe, SN\,2018ivc and SN\,2018ebt, have no, or few, $r$-band observations that pass our quality cuts (described in Appendix~\ref{sec:pipeline}) and we extrapolate the $g$-band light curve instead but increase the flux by $0.5\,\text{mag}$ as SNe are typically brighter in the $r$ band at this time. We acknowledge that some SNe might fade more slowly at late times, e.g., due to the late-time interaction \citep{sollerman2020, weil2020_sn2017eaw_latetime} or light echos \citep{maund2019} which might yield false non-detections. The best way to mitigate this is inspecting the late-time light curves carefully, and using several different time windows to calculate the late-time flux.

Of the 52 nearby SNe with pre-SN observations 29 have already faded below magnitude 23 and Table~\ref{tab:sample} lists their properties. Unbinned forced-photometry light curves are provided in Table~\ref{tab:fp_fluxes}. Figs.~\ref{fig:sn_lcs} and~\ref{fig:sn_lcs2} show the forced-photometry light curves in 7-day-long bins. The orange shaded regions in the figures indicate the time when the SN is likely brighter than 23rd magnitude and we use the observations before and after to calculate the pre-SN and late-time fluxes, respectively. Our sample includes 18 SNe II or IIP. SN\,2018hna \citep{singh2019} has an SN\,1987A-like light curve and its progenitor is presumably a more compact BSG. The remaining 11 SNe originate from, at least partially, stripped progenitor stars: two are classified as SNe IIb, five are SNe Ib, three are SNe Ic, and one is a broad-lined SN Ic. The two closest SNe Ib in our sample belong to the subclass of low-energy, Calcium-rich SNe.

\begin{deluxetable}{l c c c c}
\tablecaption{\label{tab:fp_fluxes}Unbinned ZTF forced-photometry light curves for 29 nearby SNe}
\tablewidth{\columnwidth}
\tablehead{
\colhead{SN name} & \colhead{JD} & \colhead{fcqfid} & \colhead{flux} & \colhead{flux error} \\
\colhead{} & \colhead{($2450000 + $)} & \colhead{} & \colhead{($10^{-9}\,\text{mgy}$)} & \colhead{($10^{-9}\,\text{mgy}$)}
}
\startdata
SN2020jfo & 8199.80343 & 4730622 & $-5.3$ & $3.6$ \\
SN2020jfo & 8204.80090 & 4730621 & $1.2$ & $2.7$ \\
SN2020jfo & 8204.81700 & 4730621 & $-1.2$ & $2.5$ \\
SN2020jfo & 8210.80147 & 4730621 & $-8.1$ & $4.7$ \\
SN2020jfo & 8214.73870 & 4730622 & $4.5$ & $3.4$ \\
...
\enddata
\tablecomments{We removed all data points that do not pass our quality cuts, applied a baseline correction, and rescaled flux uncertainties based on the observed scatter of pre-SN observations (see \ref{sec:ztf_details}). The third column encodes the ZTF field (first three to five digits), CCD ID (fourth and third last digits), quadrant ID (second last digit), and filter (last digit) of the ZTF observations. Fluxes are given in maggies. The full table contains 29\,305 observations and is published online.}
\end{deluxetable}

\subsection{The progenitor search}
\label{sec:search}

The low spatial resolution of ground-based surveys like ZTF does not allow us to reliably identify the progenitor star in pre-SN images alone as each pixel contains the light from many stars. But the vanishing of the progenitor star reduces the flux at the SN position and we search for this flux residual by comparing pre-SN images to observations taken after the SN has faded.

Table~\ref{tab:fields} lists all fields with at least 20 observations before and after the SN and their light curves are shown in Figs.~\ref{fig:sn_lcs} and~\ref{fig:sn_lcs2}. For each field, we calculate the mean flux, weighted according to the flux error bars, before and after the SN. We only compare observations that have the same reference image.
We estimate the flux error with the bootstrap method \citep{efron1982} by resampling both the pre-SN and late-time fluxes while allowing for repetitions. For each randomized light curve we calculate the weighted mean flux before and after the SN and the difference between the two. The standard deviation of the resulting distribution of residuals is used as the error on the flux residual. The advantage of the bootstrap method is that the results are valid even if the flux errors are inaccurate or not Gaussian. By using the weighted mean, we assume, however, that the relative size of the error bars is correct.

Combining hundreds of observations reduces statistical fluctuations, therefore, our search might be limited by additional errors and systematics, such as residuals due to bright host galaxies, shallow reference images, a bias mismatch, or errors on the PSF. In Appendix~\ref{sec:ztf_syserr}, we quantify the impact of these error sources empirically: We select 20 positions in the same images that have host fluxes similar to the SN position and perform a progenitor search at each position. For appropriately large error bars the standard deviation of these flux residuals should be close to one. If it is larger, the errors are likely underestimated, and we scale them up, i.e., we scale down the significance of the progenitor detection (see Appendix~\ref{sec:ztf_syserr} for details). The resulting scaling factors are given in the seventh column of Table~\ref{tab:fields}. A few fields have scaling factors as large as two, sometimes due to a bright host, while in other cases, we cannot pinpoint what limits the sensitivity. However, most scaling factors are close to one, indicating that the sensitivity is determined by well-understood statistical fluctuations and that additional ZTF observations might improve the depth of the search.

\startlongtable
\begin{deluxetable*}{lcccccccccc}
\tablecaption{\label{tab:fields}Measured flux residuals}
\tablehead{\colhead{IAU name} & \colhead{band} & \colhead{FCQFID} & \colhead{$n_{\text{early}}$} & \colhead{$n_{\text{late}}$} & \colhead{host flux} & \colhead{scaling} & \colhead{flux} & \colhead{sig.} & \colhead{limmag} & \colhead{abs. limmag} \\ 
\colhead{} & \colhead{} & \colhead{} & \colhead{} & \colhead{} & \colhead{($10^{-10}$\,mgy\,arcsec$^{-2}$)} & \colhead{} & \colhead{($10^{-10}$\,mgy)} & \colhead{} & \colhead{(mag)} & \colhead{(mag)} }
\startdata
SN\,2020jfo & $ g $ & 4730621 & 149 & 65 & $ 22.2 \pm 0.6 $ & $ 1.8 $ & $ -0.4 \pm 2.7 $ & $ -0.2 $ & 22.2 & $ -8.6 $ \\
 & $ r $ & 4730622 & 191 & 76 & $ 33.6 \pm 0.8 $ & $ 1.5 $ & $ -0.4 \pm 2.2 $ & $ -0.2 $ & 22.4 & $ -8.4 $ \\
SN\,2020fqv & $ g $ & 5250521 & 168 & 109 & $ 39.6 \pm 0.9 $ & $ 1.0 $ & $ -0.1 \pm 1.1 $ & $ -0.1 $ & 23.1 & $ -7.8 $ \\
 & $ r $ & 5250522 & 203 & 120 & $ 80.4 \pm 2.0 $ & $ 1.6 $ & $ 3.6 \pm 2.5 $ & $ 1.4 $ & 22.3 & $ -8.6 $ \\
SN\,2018imf & $ g $ & 5261221 & 21 & 94 & $ 4.7 \pm 0.5 $ & $ 1.8 $ & $ -3.2 \pm 3.2 $ & $ -1.0 $ & 22.0 & $ -8.9 $ \\
 & $ r $ & 5261222 & 29 & 166 & $ 6.1 \pm 0.6 $ & $ 1.0 $ & $ -1.0 \pm 1.8 $ & $ -0.6 $ & 22.6 & $ -8.3 $ \\
SN\,2021gno & $ g $ & 5240911 & 200 & 22 & $ 3.0 \pm 0.5 $ & $ 1.2 $ & $ -0.7 \pm 2.5 $ & $ -0.3 $ & 22.3 & $ -8.6 $ \\
 & $ r $ & 5240912 & 242 & 23 & $ 5.5 \pm 0.7 $ & $ 1.1 $ & $ -1.1 \pm 2.5 $ & $ -0.4 $ & 22.3 & $ -8.6 $ \\
SN\,2019ehk & $ g $ & 5760331 & 108 & 145 & $ 30.0 \pm 0.9 $ & $ 1.4 $ & $ -2.1 \pm 1.5 $ & $ -1.4 $ & 22.8 & $ -8.1 $ \\
 & $ r $ & 5760332 & 148 & 218 & $ 63.1 \pm 2.0 $ & $ 1.2 $ & $ -0.7 \pm 1.7 $ & $ -0.4 $ & 22.7 & $ -8.2 $ \\
SN\,2020oi & $ r $ & 5760332 & 189 & 97 & $ 560.4 \pm 13.9 $ & $ 1.0 $ & $ 3.0 \pm 1.6 $ & $ 1.8 $ & 22.7 & $ -8.2 $ \\
SN\,2018ivc & $ g $ & 4021631 & 46 & 143 & $ 298.9 \pm 19.3 $ & $ 1.9 $ & $ 1.5 \pm 3.0 $ & $ 0.5 $ & 22.1 & $ -9.0 $ \\
SN\,2018hna & $ g $ & 7891311 & 116 & 146 & $ 1.4 \pm 0.5 $ & $ 1.0 $ & $ -3.0 \pm 1.3 $ & $ -2.2 $ & 22.9 & $ -8.1 $ \\
 & $ g $ & 7901611 & 126 & 101 & $ 1.2 \pm 0.5 $ & $ 1.2 $ & $ -0.1 \pm 1.9 $ & $ -0.1 $ & 22.6 & $ -8.5 $ \\
 & $ g $ & 8190141 & 101 & 157 & $ 1.4 \pm 0.5 $ & $ 1.1 $ & $ -0.3 \pm 1.3 $ & $ -0.2 $ & 23.0 & $ -8.1 $ \\
 & $ g $ & 8200431 & 128 & 113 & $ 1.4 \pm 0.4 $ & $ 1.6 $ & $ 0.2 \pm 2.2 $ & $ 0.1 $ & 22.4 & $ -8.7 $ \\
 & $ r $ & 7891312 & 123 & 176 & $ 1.8 \pm 0.6 $ & $ 1.7 $ & $ -11.1 \pm 3.1 $ & $ -3.6 $ & 22.0 & $ -9.0 $ \\
 & $ r $ & 7901612 & 121 & 156 & $ 1.5 \pm 0.4 $ & $ 1.2 $ & $ -2.4 \pm 1.4 $ & $ -1.8 $ & 22.9 & $ -8.2 $ \\
 & $ r $ & 8190142 & 88 & 167 & $ 1.5 \pm 0.6 $ & $ 1.1 $ & $ 0.3 \pm 1.5 $ & $ 0.2 $ & 22.8 & $ -8.3 $ \\
 & $ r $ & 8200432 & 115 & 142 & $ 1.8 \pm 0.5 $ & $ 1.2 $ & $ -1.2 \pm 1.7 $ & $ -0.7 $ & 22.6 & $ -8.4 $ \\
SN\,2020cxd & $ g $ & 8481121 & 564 & 232 & $ 18.0 \pm 1.6 $ & $ 1.1 $ & $ 8.1 \pm 1.5 $ & $ 5.3 $ & 22.8 & $ -9.1 $ \\
 & $ r $ & 8481122 & 478 & 234 & $ 27.1 \pm 1.8 $ & $ 1.0 $ & $ 0.2 \pm 0.7 $ & $ 0.3 $ & 23.6 & $ -8.3 $ \\
SN\,2020dpw & $ g $ & 8510431 & 262 & 24 & $ 14.4 \pm 0.5 $ & $ 1.0 $ & $ -0.4 \pm 4.7 $ & $ -0.1 $ & 21.6 & $ -10.5 $ \\
 & $ r $ & 8510432 & 266 & 33 & $ 42.6 \pm 1.0 $ & $ 1.0 $ & $ -2.6 \pm 4.3 $ & $ -0.6 $ & 21.7 & $ -10.4 $ \\
SN\,2020mjm & $ g $ & 4271631 & 130 & 33 & $ 0.2 \pm 0.3 $ & $ 1.2 $ & $ -0.7 \pm 1.8 $ & $ -0.4 $ & 22.6 & $ -9.7 $ \\
 & $ r $ & 4271632 & 341 & 36 & $ 0.2 \pm 0.4 $ & $ 1.0 $ & $ -2.7 \pm 2.0 $ & $ -1.3 $ & 22.5 & $ -9.8 $ \\
SN\,2020mmz & $ g $ & 8161431 & 290 & 39 & $ 48.4 \pm 2.1 $ & $ 1.0 $ & $ -2.3 \pm 2.1 $ & $ -1.1 $ & 22.4 & $ -9.9 $ \\
 & $ r $ & 8161432 & 316 & 58 & $ 79.9 \pm 1.5 $ & $ 1.2 $ & $ -0.6 \pm 1.9 $ & $ -0.3 $ & 22.6 & $ -9.7 $ \\
SN\,2020hvp & $ g $ & 4300941 & 66 & 30 & $ 10.1 \pm 0.7 $ & $ 1.0 $ & $ -7.5 \pm 3.9 $ & $ -1.9 $ & 21.8 & $ -10.5 $ \\
 & $ g $ & 4311231 & 67 & 31 & $ 9.4 \pm 0.5 $ & $ 1.2 $ & $ -1.9 \pm 4.1 $ & $ -0.5 $ & 21.7 & $ -10.6 $ \\
 & $ r $ & 4300942 & 107 & 33 & $ 19.9 \pm 0.6 $ & $ 1.4 $ & $ -1.4 \pm 4.1 $ & $ -0.3 $ & 21.7 & $ -10.6 $ \\
 & $ r $ & 4311232 & 119 & 35 & $ 18.9 \pm 0.9 $ & $ 1.5 $ & $ 0.1 \pm 4.0 $ & $ 0.0 $ & 21.7 & $ -10.6 $ \\
SN\,2019yvr & $ g $ & 4231531 & 43 & 29 & $ 83.8 \pm 1.6 $ & $ 1.3 $ & $ -3.7 \pm 3.2 $ & $ -1.2 $ & 22.0 & $ -10.3 $ \\
 & $ r $ & 4231532 & 79 & 22 & $ 160.4 \pm 3.4 $ & $ 1.0 $ & $ -0.4 \pm 2.7 $ & $ -0.1 $ & 22.2 & $ -10.2 $ \\
SN\,2020abhs & $ g $ & 3211411 & 49 & 19 & $ 9.2 \pm 0.7 $ & $ 1.1 $ & $ -1.5 \pm 3.3 $ & $ -0.4 $ & 21.9 & $ -10.5 $ \\
 & $ g $ & 3720441 & 66 & 21 & $ 8.5 \pm 0.9 $ & $ 1.3 $ & $ -6.7 \pm 4.3 $ & $ -1.6 $ & 21.7 & $ -10.8 $ \\
 & $ r $ & 3211412 & 50 & 31 & $ 10.5 \pm 0.9 $ & $ 1.1 $ & $ 0.9 \pm 4.5 $ & $ 0.2 $ & 21.6 & $ -10.8 $ \\
SN\,2019spk & $ g $ & 3661611 & 45 & 58 & $ 0.6 \pm 0.3 $ & $ 1.2 $ & $ 2.4 \pm 3.1 $ & $ 0.8 $ & 22.0 & $ -10.5 $ \\
 & $ g $ & 4170241 & 48 & 98 & $ 0.6 \pm 0.2 $ & $ 1.4 $ & $ -0.7 \pm 2.1 $ & $ -0.3 $ & 22.4 & $ -10.1 $ \\
 & $ r $ & 3661612 & 45 & 54 & $ 0.6 \pm 0.6 $ & $ 1.3 $ & $ 4.0 \pm 3.3 $ & $ 1.2 $ & 21.9 & $ -10.6 $ \\
 & $ r $ & 4170242 & 55 & 126 & $ 0.9 \pm 0.5 $ & $ 1.3 $ & $ -1.3 \pm 2.8 $ & $ -0.5 $ & 22.1 & $ -10.4 $ \\
SN\,2021bug & $ g $ & 4730111 & 171 & 69 & $ 8.4 \pm 0.5 $ & $ 1.2 $ & $ 2.3 \pm 1.7 $ & $ 1.4 $ & 22.7 & $ -9.9 $ \\
 & $ r $ & 4730112 & 228 & 75 & $ 13.3 \pm 0.6 $ & $ 1.1 $ & $ -2.6 \pm 1.9 $ & $ -1.4 $ & 22.6 & $ -10.0 $ \\
SN\,2021bhd & $ g $ & 8450841 & 555 & 42 & $ 20.9 \pm 0.7 $ & $ 1.1 $ & $ -1.3 \pm 1.4 $ & $ -1.0 $ & 22.9 & $ -9.7 $ \\
 & $ r $ & 8450842 & 506 & 50 & $ 48.7 \pm 1.4 $ & $ 1.1 $ & $ 2.8 \pm 1.5 $ & $ 1.8 $ & 22.8 & $ -9.8 $ \\
SN\,2019yz & $ g $ & 4291421 & 25 & 82 & $ 7.7 \pm 0.5 $ & $ 1.0 $ & $ -2.3 \pm 2.5 $ & $ -0.9 $ & 22.3 & $ -10.4 $ \\
 & $ r $ & 4291422 & 40 & 116 & $ 14.1 \pm 0.5 $ & $ 1.0 $ & $ -2.6 \pm 2.8 $ & $ -0.9 $ & 22.1 & $ -10.6 $ \\
SN\,2020fsb & $ g $ & 2770731 & 27 & 29 & $ 46.8 \pm 1.7 $ & $ 1.0 $ & $ 1.0 \pm 6.8 $ & $ 0.2 $ & 21.2 & $ -11.5 $ \\
 & $ r $ & 2770732 & 30 & 19 & $ 78.5 \pm 2.2 $ & $ 1.1 $ & $ 6.7 \pm 6.1 $ & $ 1.1 $ & 21.3 & $ -11.4 $ \\
SN\,2019bzd & $ g $ & 3250211 & 20 & 64 & $ 30.2 \pm 0.9 $ & $ 1.6 $ & $ 5.8 \pm 5.3 $ & $ 1.1 $ & 21.4 & $ -11.6 $ \\
 & $ r $ & 3250212 & 37 & 82 & $ 60.0 \pm 1.3 $ & $ 1.0 $ & $ 2.2 \pm 2.8 $ & $ 0.8 $ & 22.1 & $ -10.9 $ \\
SN\,2019pjs & $ g $ & 5881631 & 159 & 99 & $ 0.8 \pm 0.3 $ & $ 1.0 $ & $ -0.1 \pm 1.2 $ & $ -0.1 $ & 23.0 & $ -10.0 $ \\
 & $ r $ & 5881632 & 487 & 107 & $ 1.5 \pm 0.3 $ & $ 1.0 $ & $ 2.4 \pm 1.1 $ & $ 2.1 $ & 23.1 & $ -9.9 $ \\
SN\,2019enr & $ g $ & 5190631 & 39 & 114 & $ 28.3 \pm 0.8 $ & $ 1.1 $ & $ -0.3 \pm 1.8 $ & $ -0.1 $ & 22.6 & $ -10.5 $ \\
 & $ r $ & 5190632 & 219 & 164 & $ 52.1 \pm 1.2 $ & $ 1.7 $ & $ 0.3 \pm 1.4 $ & $ 0.2 $ & 22.9 & $ -10.1 $ \\
SN\,2018ebt & $ g $ & 8291531 & 18 & 233 & $ 2.6 \pm 0.8 $ & $ 1.1 $ & $ 0.5 \pm 6.5 $ & $ 0.1 $ & 21.2 & $ -11.9 $ \\
SN\,2020vg & $ g $ & 4210311 & 39 & 20 & $ 34.5 \pm 2.0 $ & $ 1.4 $ & $ 2.8 \pm 3.5 $ & $ 0.8 $ & 21.9 & $ -11.3 $ \\
 & $ r $ & 4210312 & 60 & 61 & $ 50.6 \pm 3.2 $ & $ 1.4 $ & $ 1.8 \pm 3.0 $ & $ 0.6 $ & 22.1 & $ -11.1 $ \\
SN\,2019aai & $ g $ & 5840931 & 35 & 361 & $ 13.1 \pm 1.3 $ & $ 2.1 $ & $ -1.1 \pm 4.0 $ & $ -0.3 $ & 21.7 & $ -11.4 $ \\
 & $ r $ & 5840932 & 63 & 457 & $ 21.4 \pm 1.3 $ & $ 1.5 $ & $ -3.1 \pm 2.2 $ & $ -1.4 $ & 22.4 & $ -10.8 $ \\
SN\,2018gjx & $ g $ & 6041331 & 29 & 291 & $ 5.7 \pm 0.7 $ & $ 1.2 $ & $ -2.4 \pm 1.7 $ & $ -1.4 $ & 22.7 & $ -10.5 $ \\
 & $ r $ & 6041332 & 50 & 493 & $ 10.0 \pm 0.9 $ & $ 1.8 $ & $ -1.9 \pm 2.2 $ & $ -0.9 $ & 22.4 & $ -10.8 $ \\
SN\,2019lkx & $ g $ & 6551021 & 59 & 43 & $ 1.6 \pm 0.5 $ & $ 1.1 $ & $ 7.2 \pm 8.0 $ & $ 0.9 $ & 21.0 & $ -12.3 $ \\
 & $ r $ & 6551022 & 73 & 156 & $ 3.2 \pm 0.4 $ & $ 1.1 $ & $ 0.6 \pm 4.1 $ & $ 0.2 $ & 21.7 & $ -11.5 $ \\
SN\,2020tkx & $ g $ & 6841241 & 737 & 37 & $ 0.9 \pm 0.4 $ & $ 1.4 $ & $ 0.2 \pm 2.9 $ & $ 0.1 $ & 22.1 & $ -11.1 $ \\
& $ r $ & 6841242 & 968 & 47 & $ 1.4 \pm 0.4 $ & $ 1.3 $ & $ -4.9 \pm 1.6 $ & $ -3.0 $ & 22.7 & $ -10.5 $ \\
\enddata
\tablecomments{Fields with early and late data. The FCQFID encodes the ZTF field ID (first three digits), the CCDID (following two digits), the quadrant ID (second-last digit), and the filter ID (last digit), where 1 stands for the $g$ band and 2 for the $r$ band. Difference images with the same FCQFID have the same reference image. $n_{\text{early}}$ is the number of images obtained more than ten days before the SN discovery at $t_{0}$ and $n_{\text{late}}$ is the number of images available after the SN has faded to at least 23rd magnitude. The sixth column lists the host magnitude in maggies$\,\text{arcsec}^{-2}$. The scaling factor quantifies how much the residuals scatter at different positions in the image and the errors on the flux residuals are scaled up by this factor. The following column lists the flux residuals $f$ when subtracting late from early observations and its scaled-up error in units of maggies. They can be converted to AB magnitudes via $m=-2.5\times \log10(f)$. The third-last column shows the significance of the progenitor detection and the last two columns display the limiting magnitudes that the search reaches.}
 \end{deluxetable*}

\subsection{Results}
\label{sec:results}

\begin{figure}[t]
\centering
\includegraphics[width=\columnwidth]{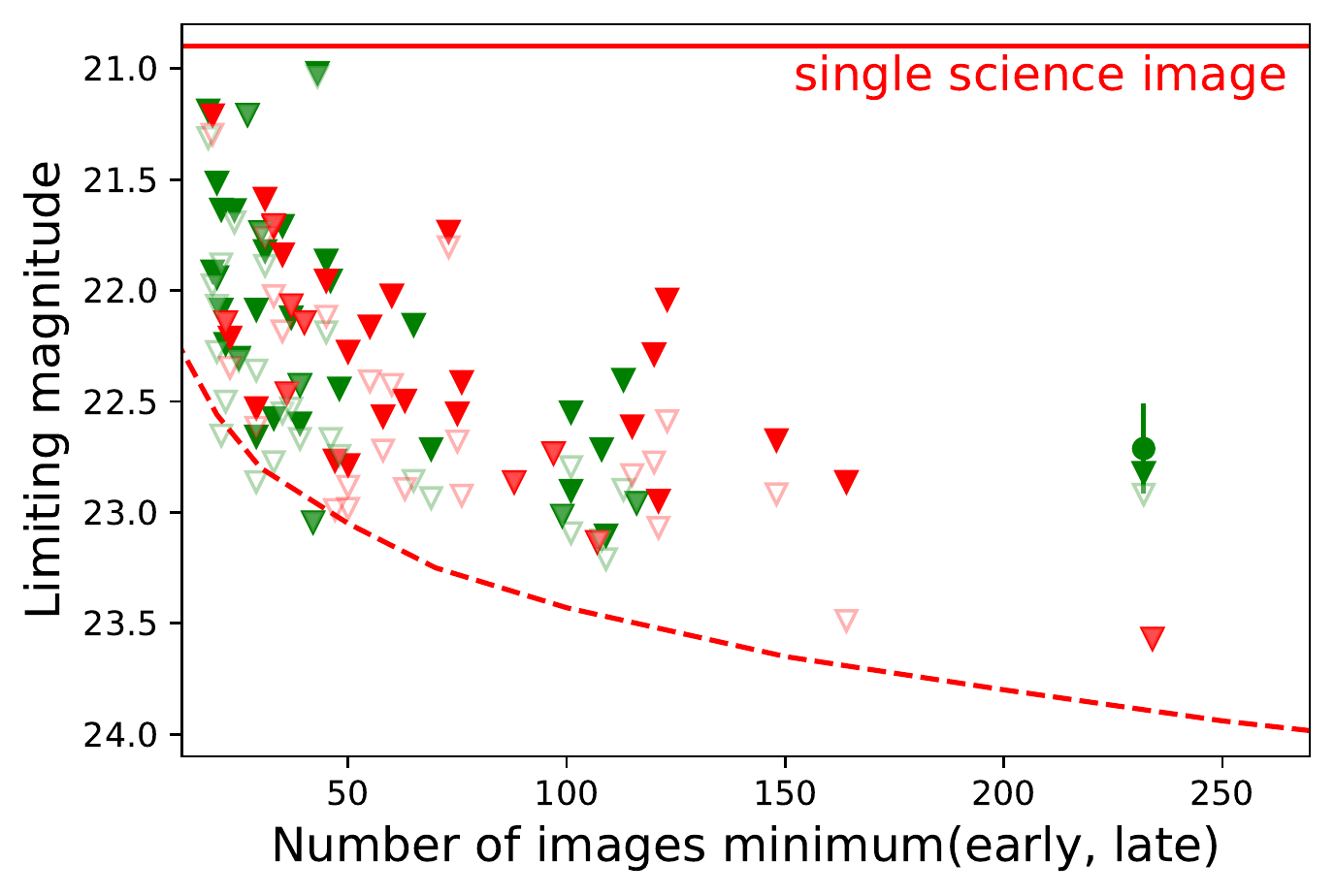} \\
\caption{Apparent limiting magnitudes of the ZTF progenitor search in the $g$ (green) and $r$ (red) bands versus the number of images in the time window, before or after the SN, that contains fewer ZTF images and, hence, limits the sensitivity of the search. Open triangles show the limits that are obtained with bootstrap error propagation. We then repeat our search 20 positions throughout the image and increase the error bars if the flux residuals scatter more than expected (see Appendix~\ref{sec:ztf_syserr}). The resulting scaled limits are shown as solid triangles. The red dashed line indicates the ideal $r$-band sensitivity based on typical ZTF zeropoints, sky background, and seeing values (described in Sect.~\ref{sec:ztf_sensitivity}). The limits that we obtain are less constraining due to host galaxies, non-optimal image processing, or additional systematic errors. For comparison, the horizontal red line indicates the median $r$-band limiting magnitude of single difference images.}
\label{fig:ztf_scaling_factors}
\end{figure}

All flux residuals, their errors, and their significances are listed in Table~\ref{tab:fields}. Our search yields a single $5\,\sigma$ detection: When combining the $g$-band observations of SN\,2020cxd \citep{yang2021, kozyreva2022, valerin2022} we measure a flux residual of $22.7 \pm 0.2\,\text{mag}$ (shown as a green data point in Fig.~\ref{fig:ztf_scaling_factors}). We inspect the 30-day-bin light curve of SN\,2020cxd, and Fig.~\ref{fig:lc_sn2020cxd} shows that both the $g$- and $r$-band flux are fading over the four years during which ZTF monitored the position. The trend seems to continue after the SN explosion, implying that it is either due to an unrelated background source or due to long-term changes in the calibration. We  did not observe similar variability for any of the 20 background positions in the same galaxy, so it is likely a local effect. A single ZTF pixel covers $1.012\,\text{arcsec}$ in the sky, corresponding to $110\,\text{kpc}$ in the host galaxy NGC\,6395. Therefore, it is possible that the fading originates from an unrelated source. 
We thus conclude that unrelated variable sources and progenitor outbursts are a potential contamination for our search. We do not find any genuine detections and the ZTF limiting magnitudes of our progenitor search are shown in Fig.~\ref{fig:ztf_scaling_factors}.

\subsection{SN progenitor detections in the literature}
\label{sec:compare_to_hst}

\begin{figure*}[t]
\centering
\includegraphics[width=\textwidth]{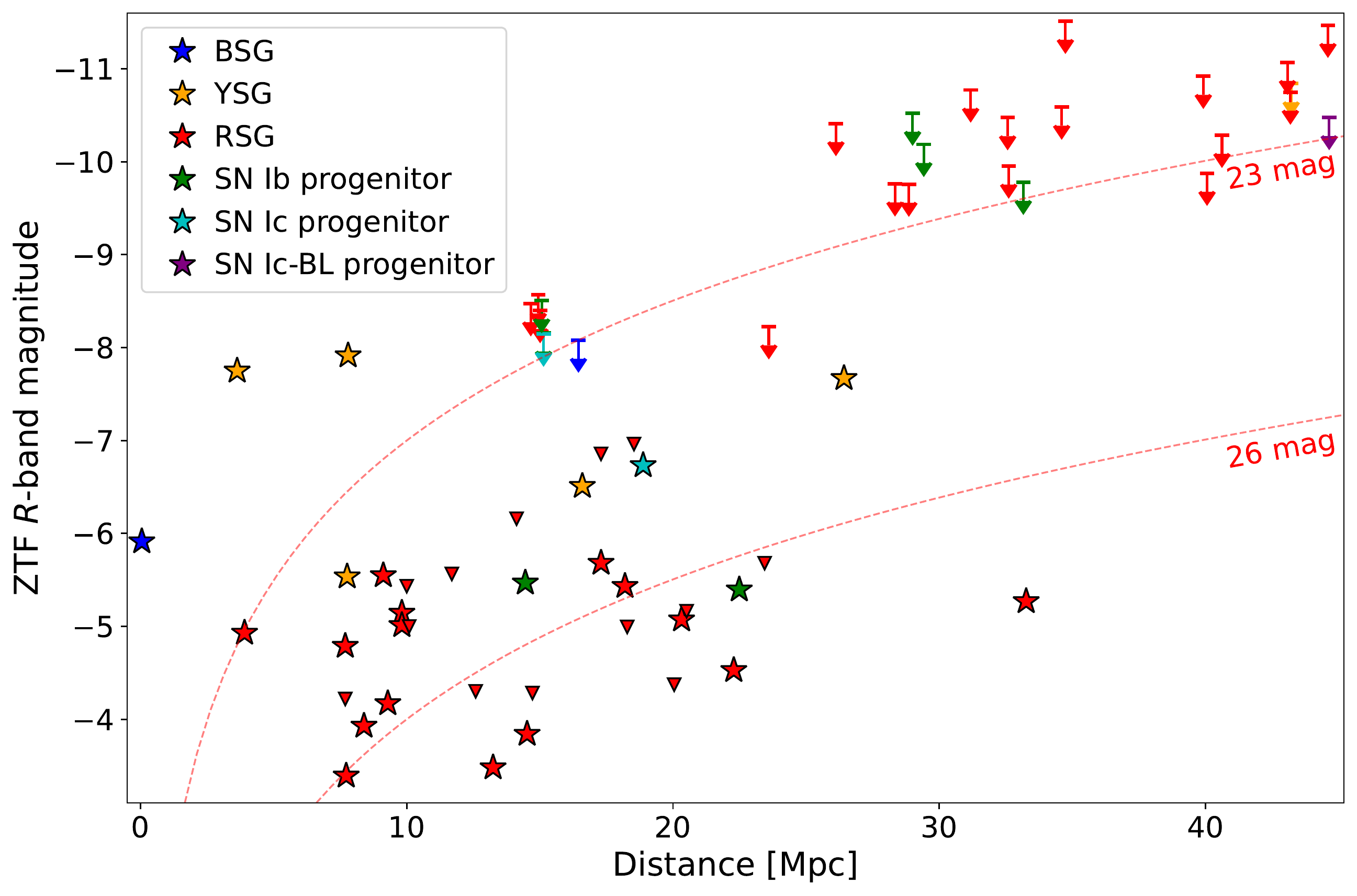}
\\
\caption{Arrows mark $5\,\sigma$ $r$-band upper limits on the progenitor magnitudes from the ZTF search (also given in Table~\ref{tab:fields}). Colors indicate the expected progenitor type based on the SN type. For comparison, we show detections (stars) and non-detections (triangles; $3\,\sigma$ upper limits) of SN progenitors from the literature (listed in Tables~\ref{tab:rsg_progenitors} and \ref{tab:other_progenitors}). All magnitudes are converted to ZTF $r$-band AB magnitudes and we do not correct for extinction because we want to compare to the sensitivity of surveys. Stars above the red, dashed lines are detectable in searches with limiting magnitudes of $23\,\text{mag}$ or $26\,\text{mag}$, respectively.
}
\label{fig:mag_limits}
\end{figure*}

So far, progenitor searches in HST images have been published for six of the SNe in the ZTF sample (see Table~\ref{tab:sample}). The closest object, SN\,2020jfo, has a rather faint progenitor star with $25.47\pm0.07\,\text{mag}$ or $-5.35\,\text{mag}$ in the F814W band \citep{sollerman2021}. Moreover, the detection of an SN Ib progenitor was reported for SN\,2019yvr \citep{kilpatrick2021} and the colors imply an F-type spectrum. Late-time observations will reveal, whether it is a single star, binary system, or dense stellar cluster \citep{sun2022_sn2019yvr_binaryprogenitor}. The progenitors of SN\,2018ivc, SN\,2019ehk, and SN\,2020fqv remain undetected (see \citealt{bostroem2020}, \citealt{jacobson_galan2020}, and \citealt{tinyanont2022_sn2020fqv}), while a bright stellar cluster was detected at the position of SN\,2020oi \citep{gagliano2022}.

Samples of progenitor searches have, for example, been presented by \citet{smartt2015} and \citet{vandyk2017} (for all SN types) and \citet{davies2018} (for RSG progenitors). We update the \citet{davies2018} and \citet{davies2020} sample by adding more recent detections and limits and we discard three progenitor candidates that are not detected securely (SN\,2009hd, SN\,2009kr, and SN\,2009md; \citealt{elias-rosa2011, maund2015}). We also compile a sample of detected, non-RSG progenitors in Table~\ref{tab:other_progenitors}. Host extinction estimates are taken from the literature and, like the SN distance estimates, they are often uncertain (see, e.g., \citealt{maund2017_progenitor_stellar_environment}). We do not consider dust which might reduce the impact of extinction in some bands \citep{kochanek2012}. 

We aim to compare the ZTF limits to direct observations and not to derived quantities. Therefore, we look up the original progenitor observations in individual bands and convert them to ZTF $r$-band AB magnitudes as described in \ref{sec:bolo_corr}. If the progenitor is detected in several bands, we pick the band that is closest to the ZTF $r$ band. We use filter profiles from the pyphot library\footnote{\url{https://mfouesneau.github.io/pyphot/}} if available or from the database of the Spanish Virtual Observatory\footnote{\url{https://svo.cab.inta-csic.es}} and adopt the spectral shapes of giant or supergiant stars in the X-shooter Spectral Library \citep{verro2022}.

Figure~\ref{fig:mag_limits} shows the ZTF upper limits compared to detections and upper limits reported in the literature (mostly from HST). The ZTF limits are several magnitudes less sensitive than most detected progenitor stars. The upper dashed red line indicates an apparent magnitude of $23\,\text{mag}$, the limiting magnitude that we obtained for ZTF fields with many observations (see Sect.~\ref{sec:results} and Table~\ref{tab:fields}). Only three detected progenitors have brighter apparent magnitudes in the $r$ band. The ZTF search would hence have a very rough detection rate of one progenitor per decade. It is sensitive to bright YSG progenitors out to $\sim20\,\text{Mpc}$, BSG progenitors within $\sim5\,\text{Mpc}$, and even closer RSGs.

One reason for the ZTF non-detections are the low RSG surface temperatures. If the adopted M4 spectral shape is typical, RSG progenitors are $\sim1.5\,\text{mag}$ fainter in the $r$ band compared to the F814W band. ZTF also has an $i$ band, but the camera is less sensitive at these longer wavelengths (see Fig.~12 by \citealt{masci2019}) and the $i$ band suffers from fringing. Furthermore, it is not used as often, and the small number of observations reduces the sensitivity even further. In conclusion, the ZTF $g$ and $r$ bands are not red enough to detect RSG progenitors, therefore the detection of hotter, less extended progenitors such as YSG or BSG stars is more likely even though they are rarer.

\subsection{Expected ZTF sensitivity}
\label{sec:ztf_sensitivity}

Next, we check whether the progenitor search is as good as expected for the P48 telescope or whether we lose sensitivity due to unaccounted-for errors, such as registration errors or biases in the background estimation (see Appendix~\ref{sec:ztf_syserr}). 

We start by quantifying the sensitivity of single ZTF science images. For each science image, we look up the zeropoint, $ZP$, and seeing, $S$, and measure the standard deviation, $\sigma_{\text{bg}}$, of the sky background, $B$, from the image. For a source with magnitude $m$, the expected total number of signal counts is $n_{\text{signal}}=10^{0.4 (ZP - m)}$. The standard deviation of the Gaussian PSF is $\sigma=S/2.35$. Based on Eq.~1 by \citet{ofek2020_seeing_limited_surveys}, the expected signal-to-noise ratio in a single image can hence be approximated as: 
\begin{equation} \label{eq:snr}
\frac{S}{N} = \sqrt{\sum \left( \frac{n_{\text{sig, pix}}}{n_{\text{{bg, pix}}}} \right)^2} = \frac{10^{0.4 (ZP - m)}}{\sqrt{4 \pi} (S/2.35) \sigma_{\text{bg}}}
\end{equation}
where $n_{\text{sig, pix}}$ and $n_{\text{bg, pix}}$ are the number of signal and background counts per pixel. We obtain the last expression by integrating over all pixels assuming a Gaussian PSF. We verify Eq.~\ref{eq:snr} by injecting simulated sources and find that the signal-to-noise ratio is 22\% lower because the PSF is not perfectly Gaussian. Except for this correction, Eq.~\ref{eq:snr} is a good measure for the sensitivity.
To estimate the improvement when combining many images, we draw ZTF images randomly and add their signal-to-noise ratios in quadrature. The resulting ideal sensitivity as a function of the number of images is shown as a dashed red line in Fig.~\ref{fig:ztf_scaling_factors}.

The triangles in Fig.~\ref{fig:ztf_scaling_factors} represent the limiting magnitudes of the progenitor search (also given in Table~\ref{tab:fields}) and they show that the sensitivity of our search is sometimes close to the ideal sensitivity, but can be up to $1.5\,\text{mag}$ worse for other fields. The ideal sensitivity does not consider the host background which limits our search for SNe with bright hosts, such as SN\,2020oi or SN\,2018ivc (compare Appendix~\ref{sec:ztf_syserr}). Moreover, we assume that the sensitivity is purely determined by the time window that contains the fewest observations. But if both time windows contain the same number of observations, the sensitivity would decrease by a factor of $\sqrt{2}$ or by $0.36\,\text{mag}$. We neglect all fields where either of these effects is relevant, but find that a discrepancy remains: On average, the ZTF search could be improved by $1\,\text{mag}$. 

Reaching the ideal sensitivity would likely require custom image coaddition and subtraction as well as a more careful calibration that reduces the scatter between epochs and between different positions within the same image. If we reach the ideal sensitivity, we would expect limiting magnitudes of $24\,\text{mag}$ in the $r$ band for more than $250$ images in both time windows (see Fig.~\ref{fig:ztf_scaling_factors}). 
However, this is not sufficient to detect progenitors on a regular basis, as shown in Sect.~\ref{sec:compare_to_hst}. Thus, detecting a sample of SN progenitors requires either a larger telescope, that detects more signal photons or a better site with a smaller seeing or darker sky (compare Eq.~\ref{eq:snr}).

\section{Prospects for direct progenitor detections with LSST}
\label{sec:lsst}

We explore here how many progenitor detections we expect with LSST. For this purpose, we simulate a population of nearby SNe in Sect.~\ref{sec:lsst_sim} and estimate the impact of bright host galaxies in Sect.~\ref{sec:lsst_sensitivity}. We calculate the expected number of progenitor detections in Sect.~\ref{sec:lsst_results} and discuss pre-SN outbursts in Sect.~\ref{sec:lsst_precursors}. \ref{sec:bolo_corr} shows the bolometric corrections that we use to convert between different spectral bands. Table~\ref{tab:survey_prop} summarizes the LSST properties taken from \citet{ivezic2019} and we compare them to the ZTF data analyzed in Sect.~\ref{sec:methods}.

\begin{deluxetable*}{lcc}
\tablewidth{\columnwidth}
\tablecaption{\label{tab:survey_prop} Properties of the ZTF BTS data (analyzed in Sect.~\ref{sec:methods}) and assumed LSST properties (Sect.~\ref{sec:lsst})}
\tablehead{\colhead{} & \colhead{ZTF BTS} & \colhead{LSST}\\
\colhead{} & \colhead{($r$ band)} & \colhead{}
}
\startdata
survey duration & $4.6\,\text{yrs}$ & $10\,\text{yrs}$\\ 
surveyed area & $13\,000\,\text{sqdeg}$ & $18\,000\,\text{sqdeg}$ \\
number of visits & $110\,\text{yr}^{-1}$ & 8/18/18$\,\text{yr}^{-1}$ ($g$/$r$/$i$) \\
limiting mag. (single img.) & $20.9\,\text{mag}$ & $24.9$/$24.7$/$24\,\text{mag}$ ($g$/$r$/$i$)\\
limiting mag. (1 yr) & $22-23\,\text{mag} ^a$ & $26.2/26.4/25.7\,\text{mag}$ ($g$/$r$/$i$)\\
sky brightness & $20.2\,\text{mag}\,\text{arcsec}^{-2}$ & $21.2\,\text{mag}\,\text{arcsec}^{-2}$ ($r$)\\
median seeing & $2\,\text{arcsec}$ & $0.65\,\text{arcsec}$ ($r$) \\
\enddata
\tablecomments{The expected LSST properties are taken from \citet{ivezic2019}. \\$^a 23.3\,\text{mag}$ expected (see Sect.~\ref{sec:ztf_sensitivity})}
\end{deluxetable*}

\subsection{Simulating a population of nearby supernovae}
\label{sec:lsst_sim}

To simulate a population of nearby SNe, we randomly draw a distance, explosion time, peak magnitude, fading time, progenitor magnitude, and the host background at the SN position.

The distances of very nearby SNe are determined by individual galaxies, as exemplified by SN\,2019ehk and SN\,2020oi, which exploded in the same host (see Table~\ref{tab:sample}). We adopt the local galaxy distribution from the GLADE+ catalog \citep{dalya2022} and roughly select the part of the extragalactic sky that LSST can observe.
We consider galaxies within $70\,\text{Mpc}$ and the catalog is approximately complete out to this distance \citep{dalya2022}. The $B$-band flux of galaxies mostly originates from young, hot stars and is closely correlated with the star-formation and SN rate. We, therefore, draw the distances of the simulated SNe according to the $B$-band luminosity of galaxies in the GLADE+ catalog.

Conservatively, we only consider bright SNe with peak magnitudes of at least $18.5\,\text{mag}$, i.e., SNe that a survey similar to BTS would find and classify. For this purpose we draw absolute peak magnitudes from the $r$-band SN luminosity function presented by \citet{perley2020}, convert them to apparent magnitudes, and discard 22\% of the simulated SNe that are fainter than $18.5\,\text{mag}$.

The SN fading time is crucial because the progenitor search requires late-time observations as we search for a vanishing source rather than trying to identify the progenitor in pre-SN images alone. In Sect.~\ref{sec:sample}, we extrapolate the $r$-band light curves of 50 ZTF SNe. They fade to $23\,\text{mag}$ within 0.8 to 2.5 years with a median of 1.6 years. But when extrapolating to fainter magnitudes, slower radioactive decays might start to contribute. SN\,1987A is one of few SNe with very-late-time detections \citep{seitenzahl2014} and we here assume that all light curves develop similarly after they have reached $23\,\text{mag}$. Compared to a pure $^{56}$Co decay, the slower $^{57}$Co decay only increases the median fading time to $27.5\,\text{mag}$ by 70 days. $^{44}$Ti starts to dominate the light curve at even later times and is not relevant to our search. Slower radioactive decays hence only have a very minor impact on the progenitor search, if the ratios between the isotopes are similar to the ones for SN\,1987A. SNe fade to $27.5\,\text{mag}$ within 1.9 to 6 years with a median duration of 3 years.

We caution that some SNe might fade away more slowly especially in the optical bands: Most SNe in the sample by \citet{smith2023_latetime_nearbysne} did not fade below their progenitor magnitude for a decade. However, in infrared bands, where RSG progenitors are brighter the disappearance of progenitor stars has been confirmed for many SNe (see e.g.~\citealt{vandyk2023_disapperance_of_progenitor_stars}). Slower fading rates can for example be cause by late-time interaction with a circumstellar medium (see, e.g., \citealt{sollerman2020} for extreme examples, or \citealt{weil2020_sn2017eaw_latetime}), light echos \citep{maund2019}, or prolonged magnetar emission. The LSST candence will allow to monitor the SN fading rates at least as long as they are detectable in single images.
Of all SNe that explode during the ten-year survey 80\% fade to at least $24.7\,\text{mag}$ before the end of the survey, i.e., they are no longer detectable in single LSST images. Within the survey footprint, LSST is scheduled to collect 184 $r$- and $i$-band observations \citep{ivezic2019}, while $80$ $g$-band visits are planned. SNe that are bright in the middle of the survey have 60 to 80 LSST $r$- and $i$-band images both before and after the SN. However, SNe that happen earlier or later have fewer observations in one of the time windows resulting in a lower sensitivity (compare Fig.~\ref{fig:ztf_scaling_factors}). As a consequence, the average SN has 35 $r$- and $i$-band and $16$ $g$-band observations in the shorter time window.

Finally, we draw absorbed $i$-band magnitudes for RSG progenitors from the luminosity function by \citet{davies2020} and convert the to the LSST $g$, $r$, and $i$-band magnitudes using the bolometric corrections given in \ref{sec:bolo_corr}.

\subsection{The impact of bright hosts on the LSST sensitivity}
\label{sec:lsst_sensitivity}

The design limiting magnitude of a single LSST visit is $24.7\,\text{mag}$ in the $r$ band (Table~1 of \citealt{ivezic2019}) while the expected limiting magnitude after 10 years and 184 observations is $27.5\,\text{mag}$. Figure~2 by \citet{ivezic2019} shows how the sensitivity improves when coadding several $r$-band observations. The improvement is similar for $g$- and $i$-band images, but $g$-band observations are $0.2\,\text{mag}$ deeper while $i$-band observations are $0.7\,\text{mag}$ less constraining \citep{ivezic2019}. The LSST $z$ and $y$ bands can likely also be used for progenitor searches and RSGs are bright in these bands (see \ref{sec:bolo_corr}). However, the SN fading times might be longer, e.g., due to dust formation, and we, therefore, do not consider these bands here.

However, the limiting magnitudes quoted by \citet{ivezic2019} are only valid for isolated sources. In such locations, the background is dominated by the sky brightness of on average $21.2\,\text{mag}\,\text{arcsec}^{-2}$ in the LSST $r$ band (Table 2 by \citealt{ivezic2019}) or $3.3\times10^{-9}\,\text{maggies}\,\text{arcsec}^{-2}$. But many SNe explode on top of bright host galaxies. As shown in Eq.~\ref{eq:snr}, the signal-to-noise ratio is proportional to $\sigma_{\text{bg}}^{-1}$ and the standard deviation of the sky background is $\sigma_{\text{bg}} = B^{-0.5}$ where $B$ is the sum of all relevant backgrounds at the SN position. The host background, therefore, degrades the limiting magnitude by:
\begin{equation} \label{eq:host_bg}
\Delta l = 1.25 \log10{\left(1+\frac{B_{\text{host}}}{B_{\text{sky}}}\right)}
\end{equation}
where $B_{\text{host}}$ and $B_{\text{sky}}$ are the host and sky background in counts per area.

\begin{figure}[t]
\centering
\includegraphics[width=0.45\textwidth]{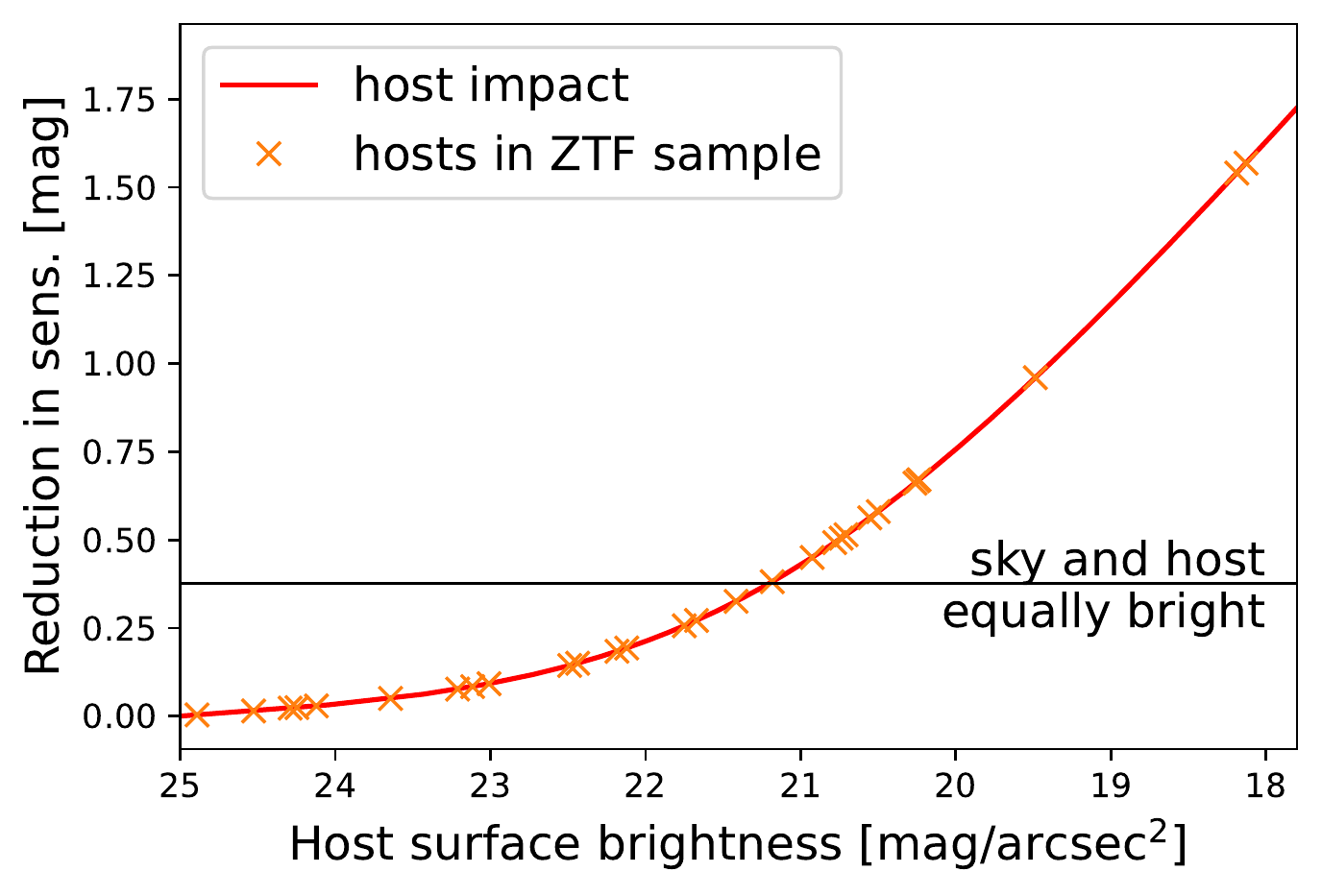}
\\
\caption{Reduction in limiting magnitude due to the host background for the expected LSST sky background of $21.2\,\text{mag}\,\text{arcsec}^{-2}$. The orange crosses indicate the $r$-band host backgrounds at the locations of 29 nearby ZTF SNe (see Sect.~\ref{sec:methods}).}
\label{fig:host_impact}
\end{figure}

In Sect.~\ref{sec:search}, we measured the $r$-band host surface brightnesses at the positions of 29 ZTF SNe using late-time ZTF data (see Table~\ref{tab:fields}; details on the calculation described in Appendix~\ref{sec:ztf_syserr}). We assume that LSST SNe will explode in as bright galaxies and randomly draw host backgrounds from this distribution. To first order, the host surface brightness is independent of the distance: Nearby galaxies are brighter, but they are also more resolved, such that the number of photons per pixel is roughly preserved. We use Eq.~\ref{eq:host_bg} to calculate by how much these hosts would reduce the sensitivity of our search in LSST data. Figure~\ref{fig:host_impact} shows that 46\% of the galaxies are brighter than the average LSST sky background, i.e., the host background reduces the limiting magnitude by more than $0.38\,\text{mag}$ and for 7\% of the hosts the sensitivity is reduced by more than $1\,\text{mag}$. 

Compared to LSST, ZTF is less affected by host backgrounds because the sky background is on average $1\,\text{mag}\,\text{arcsec}^{-2}$ brighter than at the LSST site. Nevertheless, the ZTF search is host-dominated for $14\%$ or four of the 29 SNe in our sample: the host backgrounds of SN\,2019yvr, SN\,2020oi, and SN\,2020fqv are larger than the average sky background and the bright host of SN\,2018ivc produces large residuals during image subtraction, such that quality cuts reject most $r$-band observations (compare Sect.~\ref{sec:ztf_syserr}). 

Even the brightest host galaxies are most likely not saturated in LSST images. For a seeing of $0.7\,\text{arcsec}$ and an exposure time of $15\,\text{s}$, a point source is saturated if it is brighter than $15.8\,\text{mag}$ in the $r$ band \citep{lsst2009_sciencebook}. For this seeing and the LSST pixel size \citep{ivezic2019} about $10\%$ of the photons fall on the brightest pixel. An extended source, like a galaxy, produces as many photons per pixel if it has a surface brightness of $14.8\,\text{mag}\,\text{arcsec}^{-2}$. In the ZTF sample, SN\,2020oi has the brightest host background with $18.1\,\text{mag}\,\text{arcsec}^{-2}$, and is hence a factor of 20 fainter than the LSST saturation limit for extended sources.

\subsection{Expected LSST progenitor detections}
\label{sec:lsst_results}

LSST is planning to monitor $18\,000\,\text{deg}^2$ of extragalactic sky over ten years in the $ugrizy$ filters \citep{ivezic2019}. For an SN rate of $10^{-4}\,\text{Mpc}^{-3}\,\text{yr}^{-1}$ \citep{perley2020} and the distribution of nearby galaxies \citep{dalya2022}, we expect in total $\sim600$ SNe within $70\,\text{Mpc}$. Out of these, 490 have peak magnitudes brighter than $18.5\,\text{mag}$ in the $g$ or $r$ band and fade within the survey duration to at least $24.7\,\text{mag}$ in the $r$ band. 56\% or about 270 of them are SNe II with RSG progenitors \citep{li2011_loss_sn_fractions}.

\begin{figure}[t]
\centering
\includegraphics[width=0.45\textwidth]{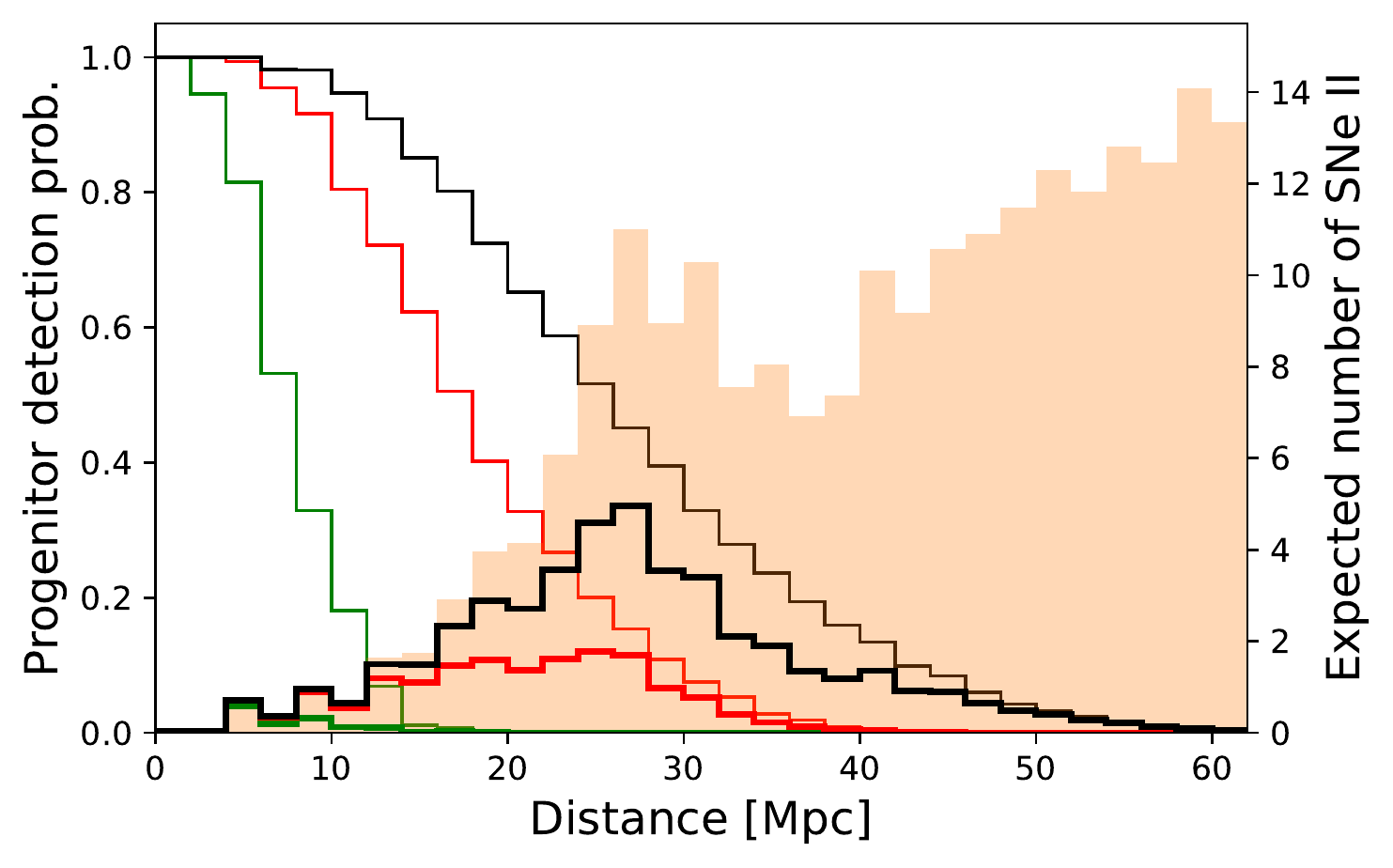}
\\
\caption{Probability that an RSG progenitor with an M4 spectrum is detected in the LSST $g$ (thin green line), $r$ (red), or $i$ (black) band and the expected number of bright, nearby SNe within the LSST footprint during the ten-year survey (orange distribution; right axis). The number of progenitor detections is given as the product of the two (thick lines; right axis) and we expect 46 RSG detections in the $i$ band, 17 in the $r$ band, and 1.4 in the $g$ band.}
\label{fig:lsst_det}
\end{figure}

\begin{deluxetable*}{lccccccc}
\tablewidth{\columnwidth}
\tablecaption{\label{tab:expected_lsst_det}Expected number of LSST detections for different progenitor types.}
\tablehead{\colhead{type} & \colhead{fraction} & \colhead{\#SNe} & \colhead{progenitor lumi.} & \colhead{spectral type} & \multicolumn{3}{c}{\# detections} \\
\colhead{} & \colhead{} & \colhead{} 
& \colhead{(mag$_{i}$)} & \colhead{} & \colhead{$g$} & \colhead{$r$} & \colhead{$i$}
}
\startdata
RSG & 0.56 & 273 & LF & M4 & $1.4$ & $17$ & $46$\\ 
 &  &  &  & M2 & $5.4$ & $44$ & $72$\\ 
YSG & 0.09 & 44 & $-5.4$ & A5 & $2.4$ & $5.2$ & $1.6$\\
BSG & 0.02 & 9.8 & $-5.6$ & B2 & $0.95$ & $2.3$ & $0.5$\\
\hline
SN Ib progenitor & 0.12 & 56 & $\leqslant-4.7$ & F2 & $\geqslant0.87$ & $\geqslant1.0$ & $\geqslant0.79$\\
SN Ic progenitor & 0.14 & 66 & $\leqslant-4.0$ & A5 & $\geqslant0.57$ & $\geqslant1.0$ & $\geqslant0.44$\\
\enddata
\tablecomments{The number of SNe within $70\,\text{Mpc}$ with both pre-SN and late-time observations during the ten-year survey for $18\,000\,\text{deg}^{2}$. The SN fractions in the second column are taken from \citet{li2011_loss_sn_fractions} and \citet{kleiser2011_sne_from_bsg}. For RSGs, the progenitor $i$-band magnitudes are drawn from the luminosity function by \citet{davies2020}. Due to the lack of a measured luminosity function, we assume that all YSGs are as bright as the progenitor of SN\,2008ax and BSGs as the one of SN\,1987A (forth column of the table; taken from Table~\ref{tab:other_progenitors}). The luminosities of SN Ib and SNe Ic progenitors are highly uncertain, but we expect at least one LSST detection if they are brighter than $-4.7\,\text{mag}$ and $-4.0\,\text{mag}$ in the $i$-band, respectively. In addition, we assume that the SN fading rates and host luminosities are similar to the ZTF sample (Sect.~\ref{sec:methods}).}
\end{deluxetable*}

Next, we convert the available number of LSST images, calculated in Sect.~\ref{sec:lsst_sim}, to limiting magnitudes while also considering the impact of the host background as described in Sect.~\ref{sec:lsst_sensitivity}. 
The most constraining limits are $\sim27.2\,\text{mag}$ in the $r$ band for SNe that have 70 LSST observations both before and after the SN. However, the median SN only has half as many observations in the shorter time window and the median limiting magnitude is, therefore, $26.2\,\text{mag}$ in the $r$ band and $25.9\,\text{mag}$ and $25.8\,\text{mag}$ in the $g$ and $i$ bands, respectively.

Figure~\ref{fig:lsst_det} shows the distance distribution of the closest SNe (orange distribution) and their detection probability in the LSST $g$, $r$, and $i$ bands. For the $i$ band the detection probability is close to one out to $\sim10\,\text{Mpc}$, i.e., even faint progenitor stars are detected within this distance. In the $g$ band, on the other hand, progenitors are so faint that they might remain undetected even for the closest SNe. The total expected number of RSG detections is given by the sum over the distribution: We expect about 46 RSG progenitor detections in the $i$ band, 17 in the $r$ band, and 1.4 in the $g$ band, if the stars have M4 spectra (see Table~\ref{tab:expected_lsst_det}). Neglecting the host background (see Sect.~\ref{sec:lsst_sensitivity}) would have yielded 56 $i$-band detections, e.g., $\sim30\%$ of the RSG progenitors remain undetected due to their bright host galaxies. For a less red M2 spectrum, we expect 72, 44, and 5 detections in the $i$, $r$, and $g$ bands, respectively. LSST is hence sensitive to the surface temperature and we expect more detections for slightly hotter progenitors.


Our search will also produce strong constraints for the high-luminosity end of the RSG luminosity function: The brightest detected RSG so far is SN\,2012ec with a bolometric magnitude of $-8.11\,\text{mag}$ (compare Table~\ref{tab:rsg_progenitors}), corresponding to an LSST $i$-band magnitude of $-7.28\,\text{mag}$. We expect that 100 LSST searches will be sensitive to as bright progenitor stars. These observations can help to establish whether a significant RSG problem is present.

LSST will also detect non-RSG progenitors. Table~\ref{tab:expected_lsst_det} lists the expected number of SNe within $70\,\text{Mpc}$ based on the volumetric SN fractions by \citet{li2011_loss_sn_fractions} and \citet{kleiser2011_sne_from_bsg}. The luminosity functions of these progenitor types are not well constrained, and we, therefore, assume that all YSGs are as bright as the progenitor of SN\,2008ax (see Table~\ref{tab:other_progenitors}), for BSGs we adopt the magnitude and spectrum of the progenitor of SN\,1987A. The last three columns of Table~\ref{tab:expected_lsst_det} show that we expect the detection of $\sim4$ YSG and $2$ BSG progenitors. In addition, we expect the detection of at least one SN Ib progenitor if they are brighter than $-4.7\,\text{mag}$ in the $i$ band, and SN Ic progenitors become detectable if they are at least as bright as $-4.0\,\text{mag}$.

\subsection{LSST detection rates for pre-SN eruptions}
\label{sec:lsst_precursors}

The impending core collapse can trigger pre-SN outbursts. These eruptions are well observed for SNe IIn \citep{ofek2014, strotjohann2021}, but also happen prior to other Types of SNe (see e.g. \citealt{ho2019, jacobson-galan2022}).
We here estimate rough LSST detection rates for outbursts that happen immediately before the SN explosion.

Bright outbursts can be detected even for distant SNe and we assume a homogeneous SN rate of $10^{-4}\,\text{Mpc}^{-3}\,\text{yr}^{-1}$ for distances larger than $70\,\text{Mpc}$ where the GLADE+ galaxy catalog is incomplete. As before, we only consider SNe with peak magnitudes brighter than $18.5\,\text{mag}$. Within the ten-year LSST survey, we expect the detection of $3\,300$ as bright core-collapse SNe and they are located at a median distance of $120\,\text{Mpc}$. 

\begin{figure}[t]
\centering
\includegraphics[width=0.45\textwidth]{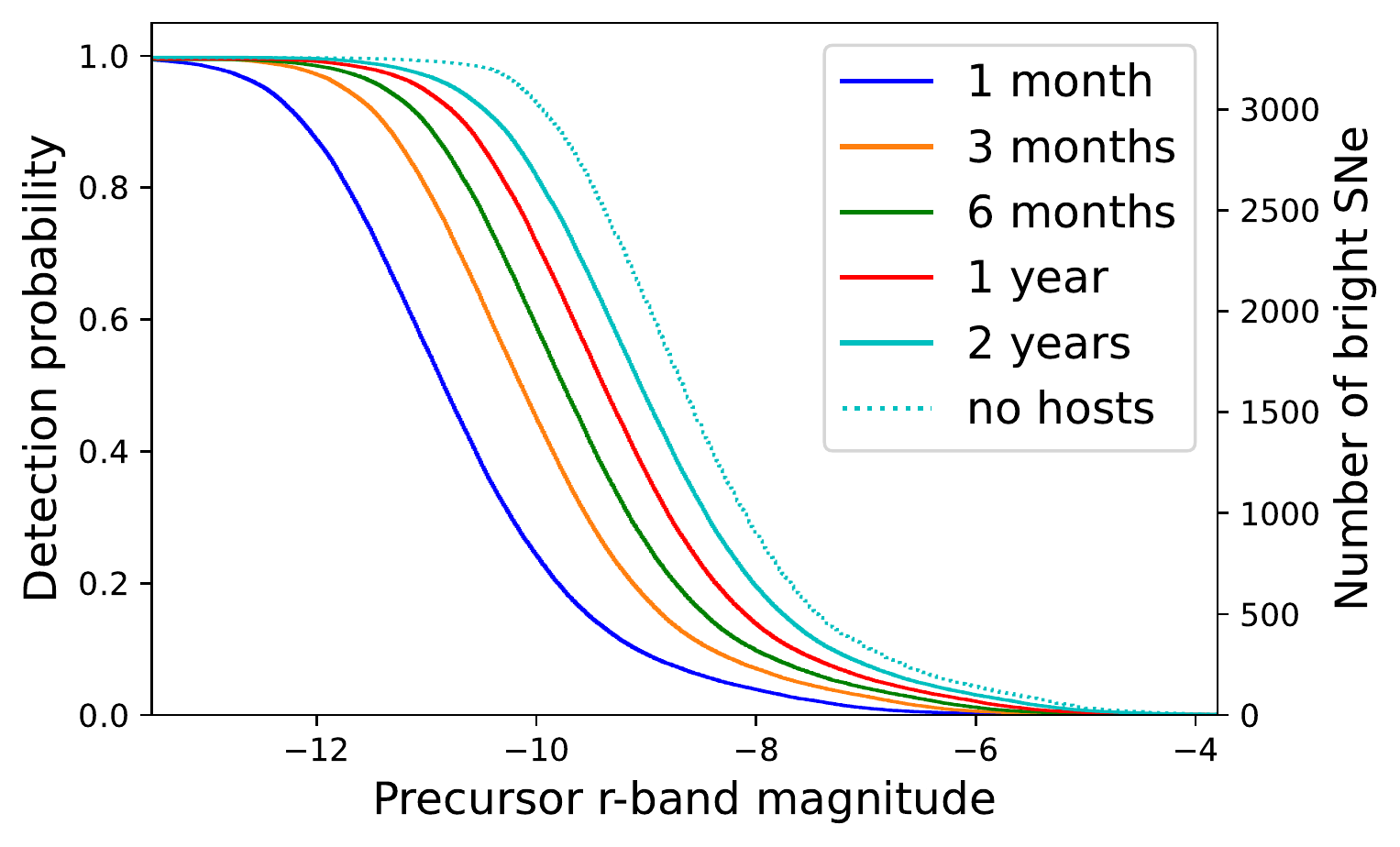}
\\
\caption{Detection probability for pre-SN outbursts depending on their luminosity and duration. The calculation was done for a population of SNe with peak luminosities brighter than $18.5\,\text{mag}$ and the right-hand axis shows the expected number of core-collapse SNe within the LSST footprint during the 10-year survey. The dotted, cyan line indicates that the average host galaxy reduces our sensitivity by $0.4\,\text{mag}$ (see also Fig.~\ref{fig:host_impact}).}
\label{fig:lsst_precursor_sensitivity}
\end{figure}

Figure~\ref{fig:lsst_precursor_sensitivity} shows our sensitivity to precursor eruptions depending on the outburst magnitude and duration. One-month-long outbursts are typically only present in one or two LSST $r$-band images and LSST can detect half of them if they are brighter than an absolute magnitude of $-10.3\,\text{mag}$. Longer-lasting outbursts are observed several times and combining several epochs increases our sensitivity. We search for outbursts as described in Sect.~\ref{sec:lsst_sensitivity}.  The only difference is that we compare observations obtained a few months or years before the SN to earlier pre-SN observations. and for For a two-year-long eruption, the median sensitivity improves to $-9.0\,\text{mag}$. The dotted, cyan line in Fig.~\ref{fig:lsst_precursor_sensitivity} visualizes that the average host reduces our sensitivity by $0.4\,\text{mag}$ (see also Fig.~\ref{fig:host_impact}).

SNe IIn make up about 9\% of all core-collapse SNe \citep{perley2020} and LSST will detect $\sim300$ SNe IIn with peak magnitudes brighter than $18.5\,\text{mag}$. \citet{strotjohann2021} found that $25\%$ of the progenitors have outbursts brighter than $-13\,\text{mag}$ and Fig.~\ref{fig:lsst_precursor_sensitivity} shows that as bright outbursts are detectable even in single LSST images. Therefore, we expect the detection of at least 75 such outbursts. A four-months-long outburst with a faint, absolute magnitude of $-11.3\,\text{mag}$ was detected prior to SN\,2020tlf, an SN II with long-lasting flash-spectroscopy features \citep{jacobson-galan2022}. We expect the detection of 1\,800 bright SNe II and similar outbursts would be detectable for 90\% of them. This large number of bright SNe will allow us to measure the rate, luminosity function, and timing of the outbursts which might contribute to revealing their triggering mechanism.

\section{Conclusion}
\label{sec:conclusion}

We propose searching for SN progenitor stars by combining a large number of images from ground-based, wide-field surveys to detect the star's disappearance. The advantages of this approach are that the data is readily available, it is collected systematically for a large fraction of the sky, usually in several bands, and late-time data ensures that the progenitor star has indeed vanished.

We test the proposed method by searching for the progenitor stars of 29 ZTF SNe with redshift $z<0.01$ that have already faded sufficiently (see Sect.~\ref{sec:methods}). We combine up to a few hundred observations before and after the SN and subtract the fluxes in these two time windows from each other. With the available amount of data, the background from host galaxies, and statistical and systematic errors in ZTF difference images, we reach limiting magnitudes down to $\sim23\,\text{mag}$ in the $g$ and $r$ bands. We do not detect any progenitor stars or long-lasting outbursts and detections reported in the literature are typically several magnitudes fainter than our limits (see Sect.~\ref{sec:compare_to_hst}). We estimate that ZTF is sensitive to the brightest YSGs and to the closest RSGs within $5\,\text{Mpc}$, which yields a detection rate of approximately one per decade. Additional data would likely improve the limiting magnitude for some fields, while others seem limited by various errors. Our search is, on average, $1\,\text{mag}$ less sensitive than expected for a ZTF-like survey, in part presumably because we combined relatively shallow difference images instead of producing dedicated difference images. Our method hence requires precise photometric calibration, long-term stability, and potentially improved image co-addition and source detection methods.

In Sect.~\ref{sec:lsst}, we quantify the expected number of progenitor detections with the upcoming LSST survey, assuming that the survey reaches its design specifications. We carefully consider the increased background due to bright host galaxies, but presume that they do not create residuals or other systematic errors during image subtraction. We can mitigate such errors empirically, as done for ZTF in Sect.~\ref{sec:results}. However, this would reduce the sensitivity of our search.

We estimate that LSST will yield 50 to 70 RSG progenitor detections. For half of the RSG progenitors, $r-i$ colors will allow us to constrain the surface temperature and distinguish between M2 and M4 type stars, reducing the statistical and systematic errors on the RSG luminosity function. The larger sample of RSG progenitors with measured bolometric corrections might help to establish whether the RSG problem is significant (impact of bolometric corrections shown by \citealt{davies2018}).

We also expect to detect several YSG and BSG progenitors (see Table~\ref{tab:expected_lsst_det}), but the number of detections is highly uncertain due to the unknown luminosity functions and surface temperatures and due to statistical fluctuations. The detection of SNe Ib and Ic progenitors is possible if their LSST $i$-band magnitudes are brighter than $-4.7\,\text{mag}$ or $-4.0\,\text{mag}$, respectively. The progenitors of stripped-envelope SNe could either be stripped by binary partners, or they could be very massive stars with strong winds, and progenitor detections in several bands might allow us to distinguish between these two scenarios. \citet{wu2022_massloss_sneIbc} predict that late nuclear burning stages trigger mass loss and brightening for stripped stars. Such events would boost the detection rate for these progenitors and offer direct information about processes in the stellar core.

LSST pre-SN light curves will be sensitive to precursor eruptions that could produce confined shells of material around the star (see, e.g., \citealt{tsuna2022_precursor_predictions}). Increased mass-loss is required to explain both flash-spectroscopy features (see e.g., \citealt{bruch2022}) and early SN light curves \citep{morazova2020, foerster2018_early_snlcs_csm}, but it is so far unclear how the star produces this environment (see e.g., \citealt{davies2022}). We expect the detection of more than $70$ bright outbursts prior to SNe IIn and many RSG outbursts if they are common and brighter than $-9\,\text{mag}$. In addition, low-amplitude variability or dimming events, for example observed by \citealt{szczgiel2012_variability_of_sn2011dh_progenitor} and \citealt{rui2019}, can be detected for the brightest progenitors.

In summary, we show that the planned LSST survey is well suited to detect both quiescent and flaring SN progenitor stars and might therefore provide answers to several open questions concerning the last evolutionary stages of massive stars.

\begin{acknowledgments}

We thank the anonymous referee, Schuyler Van Dyk, and Doron Kushnir for their comments on the manuscript. In addition, we are grateful to Morgan Fraser, Emma Beasor, Samantha Wu, Azalee Bostroem, Eva Laplace, and Dietrich Baade for helpful discussion. We would like to thank the organizers of the MIAPbP workshop on interacting supernovae.

N.L.S. is funded by the Deutsche Forschungsgemeinschaft (DFG, German Research Foundation) via the Walter Benjamin program – 461903330. This research was supported by the Munich Institute for Astro-, Particle and BioPhysics (MIAPbP) which is funded by the Deutsche Forschungsgemeinschaft under Germany's Excellence Strategy – EXC-2094 – 390783311.

E.O.O. is grateful for the support of grants from the Benozio center, Willner Family Leadership Institute, Ilan Gluzman (Secaucus NJ), Madame Olga Klein - Astrachan, Minerva foundation, Israel Science Foundation, BSF, Israel Ministry of Science, Yeda-Sela, and Weizmann-MIT.

This work is based on observations obtained with the Samuel Oschin Telescope 48-inch and the 60-inch Telescope at the Palomar Observatory as part of the Zwicky Transient Facility project. ZTF is supported by the National Science Foundation under Grants No. AST-1440341 and AST-2034437 and a collaboration including current partners Caltech, IPAC, the Weizmann Institute of Science, the Oskar Klein Center at Stockholm University, the University of Maryland, Deutsches Elektronen-Synchrotron and Humboldt University, the TANGO Consortium of Taiwan, the University of Wisconsin at Milwaukee, Trinity College Dublin, Lawrence Livermore National Laboratories, IN2P3, University of Warwick, Ruhr University Bochum, Northwestern University and former partners the University of Washington, Los Alamos National Laboratories, and Lawrence Berkeley National Laboratories. Operations are conducted by COO, IPAC, and UW.

This research has made use of the NASA/IPAC Extragalactic Database (NED), which is funded by the National Aeronautics and Space Administration and operated by the California Institute of Technology.

This research has used the Spanish Virtual Observatory (\url{https://svo.cab.inta-csic.es}) project funded by MCIN/AEI/10.13039/501100011033/ through grant PID2020-112949GB-I00.

\end{acknowledgments}

\appendix

\section{ZTF search}
\label{sec:ztf_details}

\subsection{Forced Photometry Pipeline and Quality Cuts} \label{sec:pipeline}

Instead of implementing our own image subtraction algorithm, we rely on difference images created by IPAC \citep{masci2019} based on the ZOGY algorithm \citep{zackay2016}. IPAC difference images are based on relatively shallow reference images that were produced by coadding 15 to 20 images. We download the IPAC difference images, their associated point spread functions (PSFs), and reference images using the \emph{ztfquery} package \citep{rigault2018} and calculate forced-photometry light curves for all SNe.

We first determine the exact SN position by calculating the median right ascension and declination of all ZTF alerts \citep{patterson2019} that were issued for the SN. Our forced-photometry pipeline is a modified version of the pipeline written by \citet{yao2019}, also used by \citet{strotjohann2021}. As demonstrated by \citet{strotjohann2021} 50 MCMC walkers yield a sufficiently precise flux measurement. The results are similar to light curves produced by the ZTF forced-photometry service \citep{masci2019}, but our pipeline returns additional parameters that we use for quality cuts.

After calculating the SN light curves, we reject less reliable data points according to the following criteria: (1) flagged images (i.e. the INFOBITS keyword in the header is not zero), (2) images with large residuals in the background region (a background standard deviation of $>25\,\text{counts}$), (3) images with bad pixels within the $7\times7$ pixels around the SN position, (4) images with a seeing of $>4\,\text{arcsec}$, (5) images with intermittent clouds that are flagged in the IPAC metadata tables\footnote{As described in \url{https://web.ipac.caltech.edu/staff/fmasci/ztf/extended_cautionary_notes.pdf}. The flag is only present in the meta tables because the image headers are written earlier.} (6) images based on deleted reference images. A few ZTF reference images were accidentally overwritten, such that later difference images are based on a new, different reference image. To avoid combining observations with different reference images we discard any difference images that were created earlier than the available reference image. All quality cuts and their impact on the amount of data are listed in Table~\ref{tab:ztf_cuts} and the fluxes of all observations that pass our quality cuts are given in Table~\ref{tab:fp_fluxes}.

\begin{deluxetable}{clcc}
\tablewidth{\columnwidth}
\tablecaption{\label{tab:ztf_cuts} Quality cuts for ZTF data.}
\tablehead{\colhead{} & \colhead{cut} & \colhead{discarded frac.} & \colhead{remaining obs.}}
\startdata
 & before cuts & -- & 35\,120 \\ 
1 & flagged images & 5.8\% & 33\,081 \\
2 & bkg. std $>25\,\text{counts}$ & 2.9\% & 32\,106 \\
3 & bad pixels & 0.1\% & 32\,082 \\
4 & seeing $\geqslant4\,\text{arcsec}$ & 2.5\% & 31\,285 \\
5 & cloudiness parameter & 1.4\% & 30\,848 \\
6 & reference overwritten & 2.8\% & 29\,981 \\
7 & $<15$ pre-SN images & 2.3\% & 29\,305 \\
\enddata
\end{deluxetable}

ZTF $i$-band observations are on average one magnitude less sensitive compared to the $g$ and $r$ bands and the images suffer from fringing. Therefore, we discard them here even though the $i$-band flux of RSGs are typically $\sim1.5\,\text{mag}$ brighter than their $r$-band flux (see Sec.~\ref{sec:compare_to_hst}).

Like \citet{yao2019}, we use the zeropoints from the image headers to calculate fluxes $f$ in units of maggies. These are converted to asinh magnitudes \citep{lupton1999} via:
\begin{eqnarray}
m_{\text{AB}} &=& -2.5 / \ln(10) (\text{arcsinh}(f/2 b) + \ln(b)) \\
\Delta m_{\text{AB}} &=& 2.5 / \ln(10) \times \Delta f/f \\
l_{\text{AB}} &=& -2.5 \log(5 \times \Delta f),
\end{eqnarray}
with a softening parameter $b=10^{-10}$. If the calculated magnitude is smaller than the limiting magnitude, the source is detected at the $5\,\sigma$ level. Otherwise, we use the $5\,\sigma$ upper limit as a non-detection, and the magnitude does not have a physical meaning \citep{lupton1999}.

In the following, we only consider objects with at least 20 pre-SN observations per field. For these sources, we do a baseline correction, to ensure that the weighted mean pre-SN flux is zero. This is done separately for each field or reference image, i.e., for observations with the same combination of ZTF field, CCD, quadrant, and filter.
We use the baseline correction only to calculate the SN fading time in Sect.~\ref{sec:sample}. When searching for progenitors in Sect.~\ref{sec:search} we subtract late from early fluxes, such that the baseline correction cancels out.

To check whether the errors of the flux can account for the scatter of the pre-explosion light curve, we calculate the reduced $\chi^2$ relative to the zero-flux level. If the result is larger than one, the flux errors are likely underestimated and we scale them up such that the reduced $\chi^2$ reaches one. If the reduced $\chi^2$ is smaller than one, the error bars are potentially too large and we do not modify them. The corrections are generally small with factors close to one. 

For the sample selection, described in Sect.~\ref{sec:sample}, we bin light curves in 7-day-long bins and also combine fluxes with different reference images. We use the median observation time as the time of the binned data point.
The flux per bin is the mean flux weighted by the flux errors and the error on the flux is calculated via error propagation because many bins do not contain enough data points to calculate bootstrap errors \citep{efron1982}. We make sure that observations before and after the estimated explosion date are never combined in the same bin.

\subsection{Additional error sources in ZTF images}
\label{sec:ztf_syserr}

One major error source for our analysis are bright host galaxies which can impact our search in several ways as image subtractions do not always work well for these environments: One effect is that the higher source noise increases the statistical errors in each pixel (as quantified in Sect.~\ref{sec:lsst_sensitivity}). This higher background results in larger errors when fitting the flux and is therefore considered automatically by the forced-photometry pipeline. SN\,2020oi and SN\,2018ivc have the brightest hosts (see Table~\ref{tab:fields}) and Fig.~\ref{fig:sn_lcs} shows that the typical flux upper limits are less constraining compared to the other SNe in the sample.

In addition to increased source noise, bright hosts can cause photometric errors or residuals in difference images, that are harder to quantify. Both registration errors, i.e., misalignments by the fraction of a pixel, and errors on the gain are more severe in such locations as the large fluxes in the science and reference images are subtracted from each other. We find that our quality cuts are to some degree sensitive to these problems: For positions on top of a bright galaxy such as SN\,2018ivc, most $r$-band observations are rejected, because they have poor PSF fits with a reduced $\chi^2>1.4$.

\begin{figure*}[t]
\centering
\includegraphics[width=\textwidth]{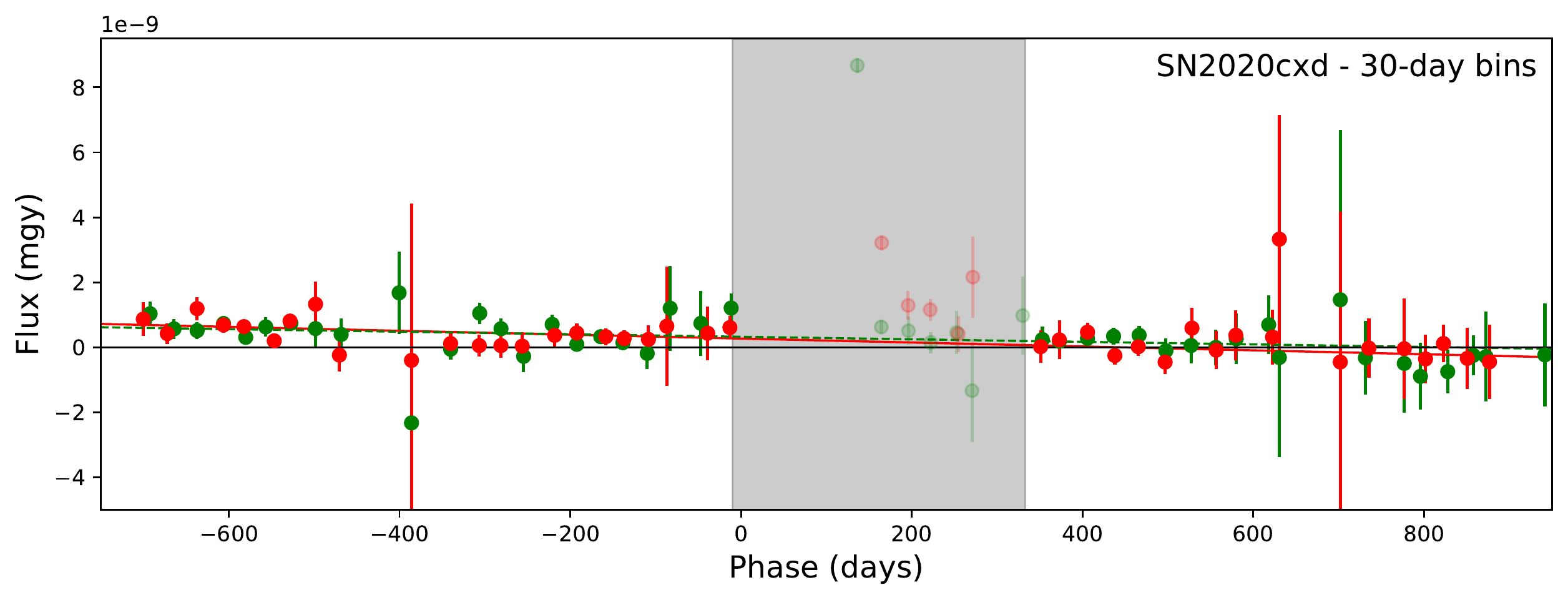}
\\
\caption{Variability at the position of SN\,2020cxd. We fit a straight line to the $g$- and $r$-band observations, excluding the SN itself (gray shaded area) and find that the fluxes in both bands decrease over the duration of the survey. The trend appears to continue after the SN and, therefore, we conclude that it is unrelated to the progenitor star. No similar variability was observed for 20 nearby positions in the same host galaxy.}
\label{fig:lc_sn2020cxd}
\end{figure*}

When searching for progenitor stars, we are comparing observations that were obtained several years apart and we observe for some sources that the residuals become larger due to long-term changes. This is also observed for SN\,2020cxd (see Fig.~\ref{fig:lc_sn2020cxd}) for which the background flux decreases both before and after the SN. We do not detect similar trends at other positions in the galaxy and it might be caused by a background source.
Another error source are the rather shallow reference images. In the progenitor search, we are trying to detect sources that are potentially fainter than the limiting magnitude of the reference image. In our search, we only compare difference images that were produced with the same reference image, such that the reference's impact cancels out to first order. However, when producing difference images the reference is convoluted with the PSF of the science image which could introduce residuals, e.g. if the average seeing is different in pre-explosion images compared to late-time images.
To mitigate the impact of these errors empirically, we repeat the progenitor search for 20 positions in the field where the host is similarly bright and if the resulting residuals are larger than expected, we scale up the error bars.

First, we measure the host flux at the SN position. Some reference images contain SN light and, therefore, we measure the host flux in late-time science images instead. For each field, we download 25 science images, that are not flagged, have a limiting magnitude of more than $19.5\,\text{mag}$, and a seeing of $<4\,\text{arcsec}$. Then, we measure the sky background in a $600 \times 600$ pixels large region around the SN position. We reject $2\,\sigma$ outliers to remove any sources in the image and calculate the median count rate for the remaining pixels. We then determine the host flux by calculating the median flux of the $7\times7$ pixels around the SN position, the same pixels that are used for the PSF fit, and we subtract the sky background from it. To obtain a result in physical units, we convert the fluxes to maggies using the zeropoint from the image header \citep{masci2019} and we normalize it to a flux per arcsec$^2$. In Table~\ref{tab:fields} we quote the median host flux of all 25 late-time images as the host flux and use the standard deviation as the uncertainty on the flux.

Next, we select positions within the same host galaxy that have similar fluxes. For this purpose, we take a cartesian grid of positions with separations of 7 pixels and 21 times 21 points. We estimate the host flux at each of these positions as described above and select the 20 positions with the flux that is closest to the flux at the SN position. Then, we obtain forced-photometry light curves for each position and run the progenitor search.

Each of the 20 background positions yields a flux residual, the early-time minus the late-time flux, and we calculate the standard deviation of these 20 residuals. If it is close to one the size of the error bars is appropriate, but if it is larger than one the residuals scatter more than expected and we scale up the error bars by the size of the standard deviation. All scaling factors are given in the seventh column of Table~\ref{tab:fields} and they are typically close to one. For SN\,2018ivc, the scaling factor is close to two due to its bright host and a few other fields also have relatively high scaling factors, but we cannot identify an obvious reason.

\subsection{Simulated sources}
\label{sec:simulated}

To verify our sensitivity we inject point sources into the ZTF images and rerun the search to check whether these sources are detected significantly. To save computational time, we do this test only for two fields that include SN\,2018hna, 8190141 and 8190142, which yield deep limits in the $g$ and $r$ bands (see Table~\ref{tab:fields}).

We simulate progenitors at the same 20 positions that we use to construct the background dataset in Sect.~\ref{sec:ztf_syserr}. When simulating a point source with a certain magnitude in a given image, we first calculate the expected number of photons based on the zeropoint and exposure time and simulate a Gaussian with the width of the measured seeing. We also consider atmospheric scintillations which shift the point source by the fraction of a pixel (the root mean squared displacement is given by Eq. 51 by \citealt{zackay2016}). We inject sources of various luminosities into the pre-explosion images, run the progenitor search, and check whether the simulated progenitors are detected with a significance of $5\,\sigma$.

\begin{figure}[t]
\centering
\includegraphics[width=0.45\textwidth]{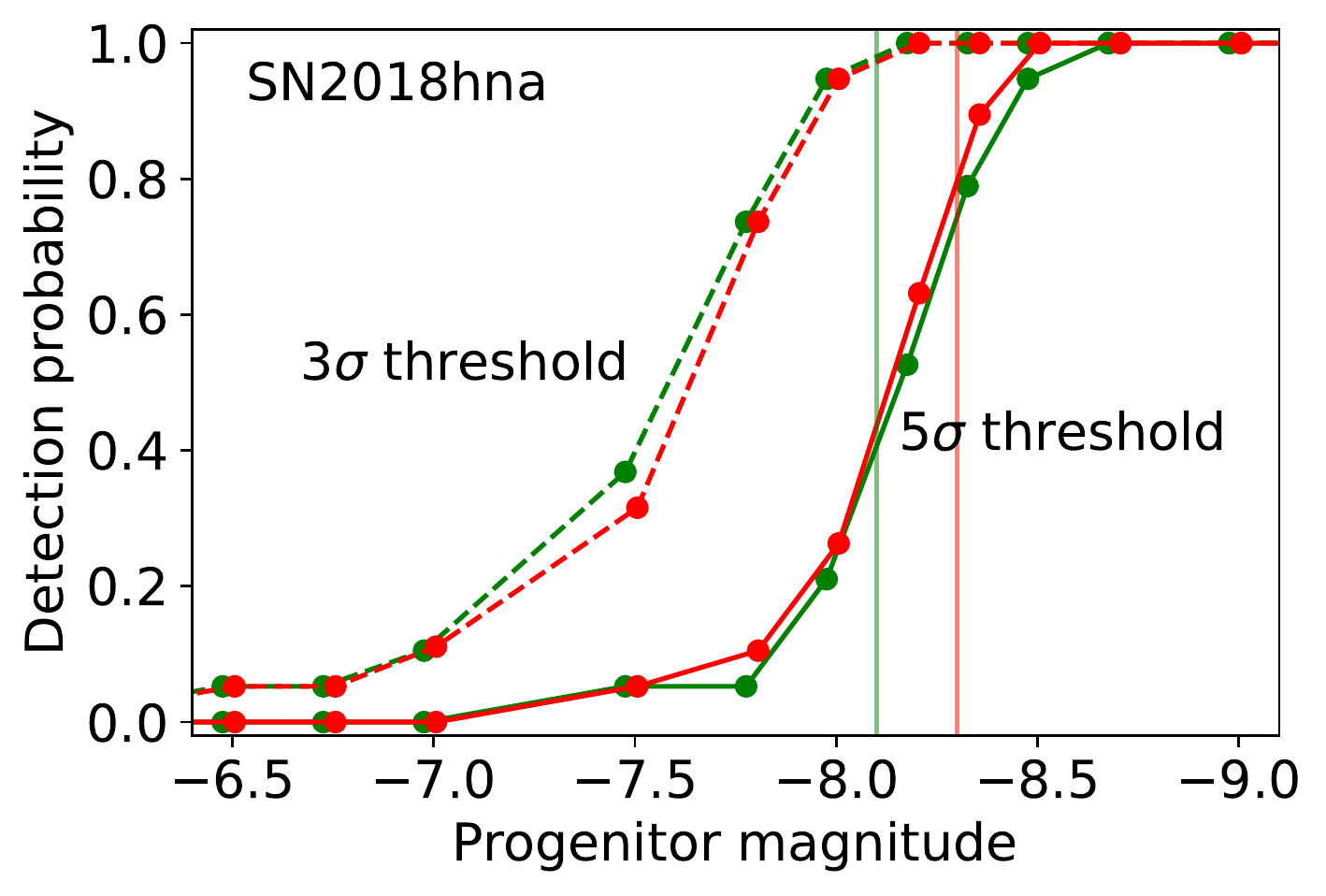}
\\
\caption{To confirm the sensitivity of our search, we simulate progenitor stars at 20 positions close to the location of SN\,2018hna that have similar host fluxes. The solid curves in the figure display the fraction of positions for which our search achieves $5\,\sigma$ detections. The two vertical lines show that the limits we obtained at the true SN position are comparable. A $3\,\sigma$ threshold (dashed lines) would increase the sensitivity by $\sim0.7\,\text{mag}$, but might yield false detections.}
\label{fig:lc_sn2018hna}
\end{figure}

Figure~\ref{fig:lc_sn2018hna} shows that we obtain $5\,\sigma$-detections if the progenitor is brighter than $-7.5\,\text{mag}$ to $-8.25\,\text{mag}$ in the $g$ or $r$ band. At the actual SN position, we measured $5\,\sigma$ limiting magnitude of $-8.1\,\text{mag}$ and $-8.3\,\text{mag}$ in the $g$ and $r$ band, respectively (see also Table~\ref{tab:fields} and Sect.~\ref{sec:results}). The limit that the progenitor search returns is, hence, comparable to the magnitudes at which we can detect simulated sources in the same field. We, therefore, conclude that the quoted limiting magnitudes are realistic. The dashed lines in the figure show that $3\,\sigma$ limiting magnitudes would be $\sim0.7\,\text{mag}$ more constraining. However, we here maintain the $5\,\sigma$ threshold to avoid false detections.

\section{Converting between different spectral bands}
\label{sec:bolo_corr}

\begin{figure*}[t]
\centering
\includegraphics[width=\textwidth]{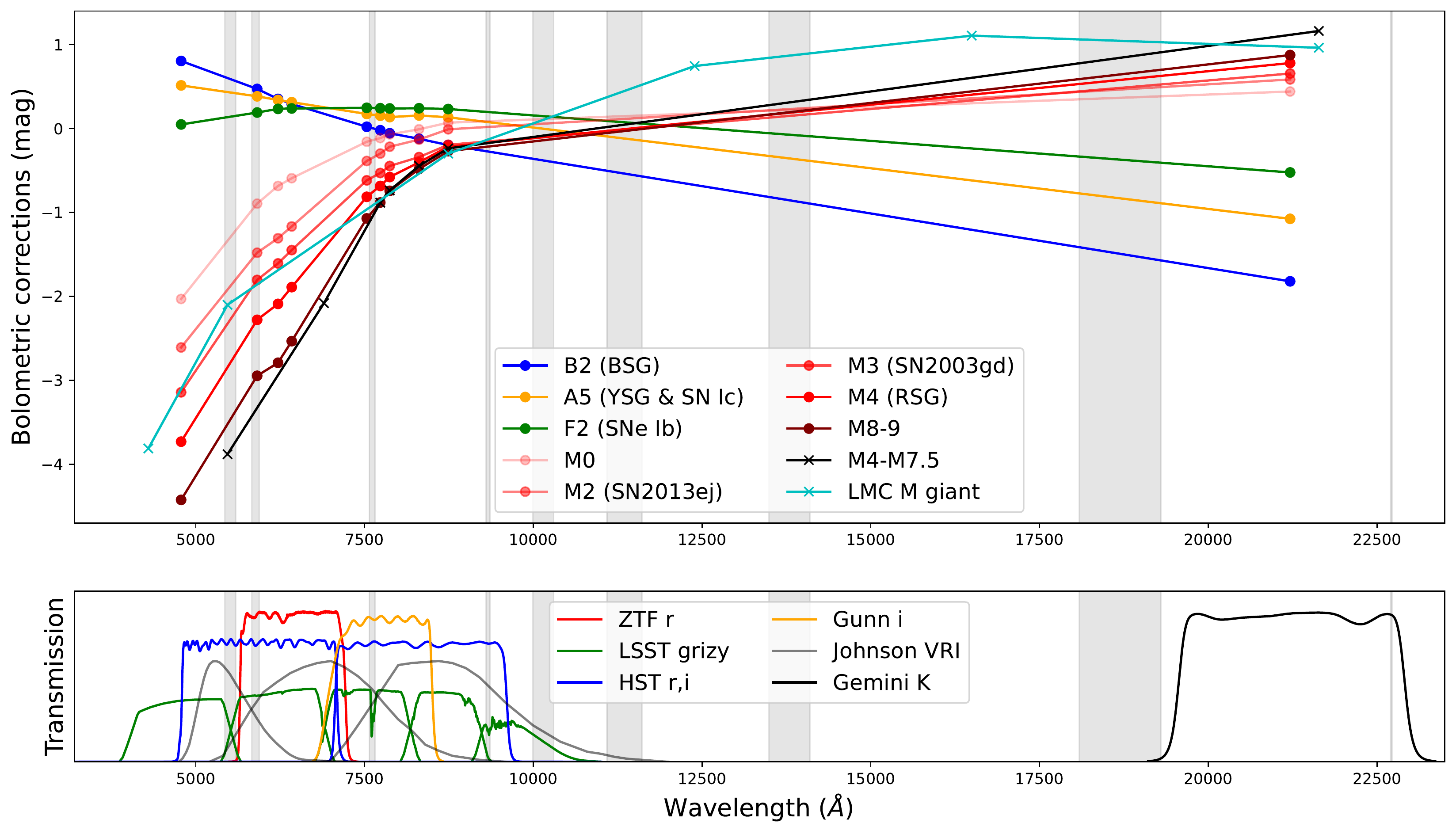}
\\
\caption{\label{fig:mag_conversions} Bolometric corrections based on XSL giant and supergiant spectra. Gray-shaded bands mark X-shooter chip gaps and telluric bands and we interpolate over them. In addition, we show the bolometric corrections by \citet{davies2018} as a black line and the average LMC M supergiant corrections in cyan. We use AB magnitudes throughout the paper. They can be converted to Vega magnitudes by subtracting 0.085 (F606W), 0.181 (ZTF $r$), 0.394 ($i'$), 0.418 ($I_C$), 0.445 (F814W), 0.487 ($I_J$), or $1.798\,\text{mag}$ ($K$), respectively.}
\label{fig:mgiant_fluxes}
\end{figure*}

Converting between different spectral bands requires assuming a spectral shape. We use observed stellar spectra from the third data release of the X-shooter spectral library (XSL\footnote{\url{http://xsl.u-strasbg.fr}}; \citealt{verro2022}). We only select dereddened spectra that are corrected for slit losses and have data over the entire wavelength range. We interpolate over telluric bands and chip gaps (wavelength ranges given in Fig.~A.3 of \citealt{lacon2021}) with a quadratic spline function. Supergiant spectra are not available for the coolest RSGs (more evolved than type M5) and we use giant spectra instead. We calculate bolometric magnitudes by integrating the XSL spectra between $3500\,\text{\AA}$ and $24800\,\text{\AA}$ and single-band magnitudes by integrating over the spectrum multiplied with the filter profile. Figure~\ref{fig:mag_conversions} shows the resulting bolometric corrections for supergiant or giant stars of different spectral types.

We assume that blue supergiant (BSG) progenitors have spectra similar to B2 stars \citep{walborn1989}. For YSG progenitors of SNe IIb we adopt an A5 star spectrum \citep{kilpatrick2022_sn2016gkg} and \citet{kilpatrick2021} found the only detected SN Ic progenitor has a similar $V-I$ color. The progenitors of SNe Ib are assumed to be similar to F2 stars \citep{xiang2019_sn2017ein_progenitor}. We caution that the true spectral shapes are largely unknown, such that the resulting bolometric corrections have large uncertainties.

\citet{davies2018} carefully consider the temperatures of RSG progenitors and conclude that they are most similar to the reddest RSGs. \citet{davies2018}, therefore, adopt an average M4-M7.5 type spectrum. However, none of the M-type supergiants in the XSL is as red (compare Fig.~\ref{fig:mag_conversions}). Instead of picking the reddest RSG in the catalog (shown as a dark red line in Fig.~\ref{fig:mag_conversions}), we adopt the spectrum of a slightly hotter M4 star for all RSGs, except the progenitors of SN\,2003gd and SN\,2013ej for which we assume M3 and M2 spectra following \citet{davies2018}. The spread of the red lines in Fig.~\ref{fig:mag_conversions} illustrate the size of the uncertainty for the entire range of M-type giants. 
Most RSG progenitors are detected in the F814W band where the bolometric corrections differ by up to $0.5\,\text{mag}$ between M supergiants of different types. However, the impact of the unknown progenitor temperatures is much larger in the $r$ band, where the cooler M9 stars are $2\,\text{mag}$ fainter than M0 supergiants. As a consequence, RSG observations in the $r$ band, including the ZTF data analyzed in Sect.~\ref{sec:methods}, have highly uncertain bolometric magnitudes.

\begin{deluxetable*}{lcccccll}
\tablecaption{\label{tab:rsg_progenitors}Limits and detections of RSG progenitor stars from the literature.}
\tablehead{\colhead{SN} & \colhead{dist. mod.} & \colhead{$E(B-V)$} & \colhead{band} & \colhead{mag} & \colhead{bol. mag} & \colhead{faded?} & \colhead{Ref.} \\
\colhead{}  & \colhead{(mag)} & \colhead{(mag)} & \colhead{} & \colhead{(mag)} & \colhead{(mag)} & \colhead{} & \colhead{}}
\startdata
SN\,1999an & $ 31.3 $ & $ 0.11 $ & $ F606W $ & $ >24.78 $ & $ >-9.11 $ & $-$ & 1, 2\\ 
SN\,1999br & $ 30.8 $ & $ 0.02 $ & $ F606W $ & $ >24.98 $ & $ >-8.08 $ & $-$ & 1, 2\\ 
SN\,1999em & $ 30.3 $ & $ 0.09 $ & $ I_c $ & $ >23.42 $ & $ >-7.66 $ & $-$ & 1, 2\\ 
SN\,1999gi & $ 30.0 $ & $ 0.17 $ & $ F606W $ & $ >24.98 $ & $ >-7.74 $ & $-$ & 1, 2\\ 
SN\,2001du & $ 31.3 $ & $ 0.16 $ & $ F814W $ & $ >24.69 $ & $ >-7.27 $ & $-$ & 1, 2\\ 
SN\,2002hh & $ 29.4 $ & $ 1.36 $ & $ i $ & $ >23.19 $ & $ >-9.51 $ & $-$ & 1, 2, 3, 4\\ 
SN\,2003gd (M3) & $ 29.8 $ & $ 0.14 $ & $ F814W $ & $ 24.44 \pm 0.04 $ & $ -5.96 \pm 0.04$ & yes & 1,5 \\ 
SN\,2004A & $ 31.5 $ & $ 0.21 $ & $ F814W $ & $ 24.8 \pm 0.12 $ & $ -7.47 \pm 0.12$ & yes & 1,5,6 \\ 
SN\,2004dg & $ 31.5 $ & $ 0.24 $ & $ F814W $ & $ >25.44 $ & $ >-6.86 $ & $-$ & 1 \\ 
SN\,2004et & $ 29.4 $ & $ 0.42 $ & $ I_J $ & $ 22.55 \pm 0.12 $ & $ -7.73 \pm 0.12 $ & yes & 1,2,4,7\\ 
SN\,2005cs & $ 29.6 $ & $ 0.16 $ & $ F814W $ & $ 24.06 \pm 0.07 $ & $ -6.22\pm 0.07$ & yes & 1, 5\\ 
SN\,2006bc & $ 30.8 $ & $ 0.21 $ & $ F814W $ & $ >24.89 $ & $ >-6.68 $ & $-$ & 1 \\ 
SN\,2006my & $ 31.7 $ & $ 0.49 $ & $ F814W $ & $ 25.3 \pm 0.13 $ & $ -7.64\pm 0.13$ & yes & 1, 5, 7\\ 
SN\,2006ov & $ 30.5 $ & $ 0.08 $ & $ F814W $ & $ >24.64 $ & $ >-6.38 $ & $-$ & 1, 7\\ 
SN\,2007aa & $ 31.6 $ & $ 0.03 $ & $ F814W $ & $ >24.88 $ & $ >-7.12 $ & $-$ & 1 \\ 
SN\,2008bk & $ 28.0 $ & $ 0.08 $ & $ K $ & $ 20.19 \pm 0.03 $ & $ -7.01 \pm 0.03 $ & yes & 1, 2, 5, 8 \\ 
SN\,2008cn & $ 32.6 $ & $ 0.33 $ & $ F814W $ & $ 25.57 \pm 0.09 $ & $ -7.97\pm 0.09$ & no & 1, 9 \\ 
SN\,2012A & $ 30.0 $ & $ 0.03 $ & $ K $ & $ 22.09 \pm 0.13 $ & $ -7.09 \pm 0.13$ & yes & 1, 10 \\ 
SN\,2012aw & $ 30.0 $ & $ 0.43 $ & $ K $ & $ 21.36 \pm 0.29 $ & $ -7.97 \pm 0.29$ & yes & 1, 11 \\ 
SN\,2012ec & $ 31.2 $ & $ 0.22 $ & $ F814W $ & $ 23.83 \pm 0.08 $  & $ -8.11\pm 0.08 $ & no & 1 \\ 
SN\,2013ej (M2) & $ 29.8 $ & $ 0.14 $ & $ F814W $ & $ 23.09 \pm 0.05 $ & $ -7.07\pm 0.05 $ & yes$^a$ & 1, 10 \\ 
SN\,2017eaw  & $ 29.4 $ & $ 0.3 $ & $ F606W $ & $ 26.48 \pm 0.05 $ & $ -6.03 \pm 0.05$ & no & 4, 12 \\ 
SN\,2018zd & $ 30.6 $ & $ 0.08 $ & $ F814W $ & $ 25.13 \pm 0.15 $ & $ -7.96 \pm 0.15 $ & yes & 10, 13 \\ 
SN\,2018aoq & $ 31.3 $ & $ 0.03 $ & $ F814W $ & $ 24.35 \pm 0.02 $ & $ -7.39\pm 0.02$ & no & 4 \\ 
SN\,2018ivc & $ 30.0 $ & $ 0.5 $ & $ F606W $ & $ >25.48 $ & $ >-8.13 $ & $-$ & 14 \\ 
SN\,2019mhm & $ 31.8 $ & $ 0.18 $ & $ F814W $ & $ >24.53 $ & $ >-8.01 $ & $-$ & 15 \\ 
SN\,2020fqv & $ 31.2 $ & $ 0.52 $ & $ F606W $ & $ >24.8 $ & $ >-10.04 $ & $-$ & 16 \\ 
SN\,2020jfo & $ 30.8 $ & $ 0.02 $ & $ F814W $ & $ 25.46 \pm 0.07 $ & $-5.77 \pm 0.07 $ & no & 17
\enddata
\tablecomments{All magnitudes are in the AB system. We adopt an M3 spectrum for SN\,2003gd, an M2 spectrum for SN\,201ej and an M4 spectrum for all other RSGs. $^a$\citet{vandyk2023_disapperance_of_progenitor_stars} find that the flux at the position of SN\,2013ej has faded below the pre-SN level in the $F814W$ band, but the SN itself and a background source are detected in late-time images. The progenitor luminosity is therefore uncertain.\\
(1) \citet{davies2018}; (2) \citet{smartt2009}; (3) \citet{kochanek2020}; (4) \citet{davies2020}; (5) \citet{maund2014}; (6) \citet{maund2017_progenitor_stellar_environment}; (7) \citet{crockett2011}; (8) \citet{vandyk2013_sn2008bk_progenitor_vanished}; (9) \citet{maund2015}; (10) \citet{vandyk2023_disapperance_of_progenitor_stars}; (11) \citet{fraser2016_disappearanceSN2012aw}; (12) \citet{vandyk2019_sn2017eaw}; (13) \citet{hiramatsu2021_sn2018zd}; (14) \citet{bostroem2020}; (15) \citet{vazquez2022_sn2019mjm_progenitor_nondet}; (16) \citet{tinyanont2022_sn2020fqv}; (17) \citet{sollerman2021}}
\end{deluxetable*}

\begin{deluxetable*}{llcccccccll}
\tablecaption{\label{tab:other_progenitors}Detections of partially stripped progenitor stars from the literature.}
\tablehead{\colhead{SN} & \colhead{type} & \colhead{dist. mod.} & \colhead{$E(B-V)$} & \colhead{band} & \colhead{mag} & \colhead{bol. mag} & \colhead{faded?} & \colhead{Ref.} \\
\colhead{} & \colhead{} & \colhead{(mag)} & \colhead{(mag)} & \colhead{} & \colhead{(mag)} & \colhead{(mag)} & \colhead{} & \colhead{}}
\startdata
SN\,1987A & BSG & $ 18.5 $ & $ 0.16 $ & $ V_j $ & $ 12.36 \pm 0.1 $ & $ -5.91 \pm 0.1 $ & yes & 1 \\ 
SN\,1993J & YSG & $ 27.8 $ & $ 0.05 $ & $ R_c $ & $ 20.05 \pm 0.11 $ & $ -7.75 \pm 0.11 $ & yes & 2, 3 \\ 
SN\,2008ax & YSG & $ 29.4 $ & $ 0.3 $ & $ F606W $ & $ 23.94 \pm 0.42 $ & $ -5.54 \pm 0.42 $ & yes & 4, 5 \\ 
SN\,2011dh & YSG & $ 29.5 $ & $ 0.07 $ & $ F814W $ & $ 21.64 \pm 0.03 $ & $ -7.91 \pm 0.03 $ & yes & 5, 6, 7\\ 
SN\,2013df & YSG & $ 31.1 $ & $ 0.09 $ & $ F555W $ & $ 24.51 \pm 0.07 $ & $ -6.51 \pm 0.07 $ & no & 5, 8\\ 
SN\,2016gkg & YSG & $ 32.1 $ & $ 0.09 $ & $ F606W $ & $ 24.4 \pm 0.18 $ & $ -7.67 \pm 0.18 $ & yes & 9, 10 \\ 
iPTF13bvn & SN Ib & 31.8 & 0.045 & $F814W$ & $25.88 \pm 0.24$ & $ -5.39 \pm 0.24 $ & yes & 11 \\
SN\,2019yvr & SN Ib & $ 30.8 $ & $ 0.51 $ & $ F635W $ & $ 24.90 \pm 0.02 $ & $ -6.09 \pm 0.02 $ & no & 12 \\ 
SN\,2017ein & SN Ic & $ 31.4 $ & $ 0.41 $ & $ F555W $ & $ 24.78 \pm 0.11 $ & $ -6.73 \pm 0.11 $ & no & 13 \\ 
\enddata
\tablecomments{Original progenitor detections and their bolometric magnitudes similar to Table~\ref{tab:rsg_progenitors}.\\
(1) \citet{walborn1989}; (2) \citet{aldering1994_sn1998j_progenitor}; (3) \citet{maund2004_sn1993j_binarydetection}; (4) \citet{folatelli2015}; (5) \citet{smartt2015}; (6) \citet{maund2015_sn2011dh_binarycompanion}; (7) \citet{maund2019}; (8) \citet{vandyk2014_sn2013df_progenitor}; (9) \citet{kilpatrick2022_sn2016gkg}; (10) \citet{vandyk2023_disapperance_of_progenitor_stars}; (11) \citet{eldridge2016_disappearance_Ib_progenitor_iPTF13bvn}; (12) \citet{kilpatrick2021}; (13) \citet{xiang2019_sn2017ein_progenitor}}
\end{deluxetable*}

\bibliography{references}{}

\begin{thebibliography}{}
\expandafter\ifx\csname natexlab\endcsname\relax\def\natexlab#1{#1}\fi
\providecommand{\url}[1]{\href{#1}{#1}}
\providecommand{\dodoi}[1]{doi:~\href{http://doi.org/#1}{\nolinkurl{#1}}}
\providecommand{\doeprint}[1]{\href{http://ascl.net/#1}{\nolinkurl{http://ascl.net/#1}}}
\providecommand{\doarXiv}[1]{\href{https://arxiv.org/abs/#1}{\nolinkurl{https://arxiv.org/abs/#1}}}

\bibitem[{{Abell} {et~al.}(2009){Abell}, {Allison}, {Anderson}, {Andrew},
  {Angel}, {Armus}, {Arnett}, {Asztalos}, {Axelrod}, {Bailey}, {Ballantyne},
  {Bankert}, {Barkhouse}, {Barr}, {Barrientos}, {Barth}, {Bartlett}, {Becker},
  {Becla}, {Beers}, {Bernstein}, {Biswas}, {Blanton}, {Bloom}, {Bochanski},
  {Boeshaar}, {Borne}, {Bradac}, {Brandt}, {Bridge}, {Brown}, {Brunner},
  {Bullock}, {Burgasser}, {Burge}, {Burke}, {Cargile}, {Chandrasekharan},
  {Chartas}, {Chesley}, {Chu}, {Cinabro}, {Claire}, {Claver}, {Clowe},
  {Connolly}, {Cook}, {Cooke}, {Cooray}, {Covey}, {Culliton}, {de Jong}, {de
  Vries}, {Debattista}, {Delgado}, {Dell'Antonio}, {Dhital}, {Di Stefano},
  {Dickinson}, {Dilday}, {Djorgovski}, {Dobler}, {Donalek}, {Dubois-Felsmann},
  {Durech}, {Eliasdottir}, {Eracleous}, {Eyer}, {Falco}, {Fan}, {Fassnacht},
  {Ferguson}, {Fernandez}, {Fields}, {Finkbeiner}, {Figueroa}, {Fox},
  {Francke}, {Frank}, {Frieman}, {Fromenteau}, {Furqan}, {Galaz}, {Gal-Yam},
  {Garnavich}, {Gawiser}, {Geary}, {Gee}, {Gibson}, {Gilmore}, {Grace},
  {Green}, {Gressler}, {Grillmair}, {Habib}, {Haggerty}, {Hamuy}, {Harris},
  {Hawley}, {Heavens}, {Hebb}, {Henry}, {Hileman}, {Hilton}, {Hoadley},
  {Holberg}, {Holman}, {Howell}, {Infante}, {Ivezic}, {Jacoby}, {Jain}, {R},
  {Jedicke}, {Jee}, {Garrett Jernigan}, {Jha}, {Johnston}, {Jones}, {Juric},
  {Kaasalainen}, {Styliani}, {Kafka}, {Kahn}, {Kaib}, {Kalirai}, {Kantor},
  {Kasliwal}, {Keeton}, {Kessler}, {Knezevic}, {Kowalski}, {Krabbendam},
  {Krughoff}, {Kulkarni}, {Kuhlman}, {Lacy}, {Lepine}, {Liang}, {Lien}, {Lira},
  {Long}, {Lorenz}, {Lotz}, {Lupton}, {Lutz}, {Macri}, {Mahabal}, {Mandelbaum},
  {Marshall}, {May}, {McGehee}, {Meadows}, {Meert}, {Milani}, {Miller},
  {Miller}, {Mills}, {Minniti}, {Monet}, {Mukadam}, {Nakar}, {Neill}, {Newman},
  {Nikolaev}, {Nordby}, {O'Connor}, {Oguri}, {Oliver}, {Olivier}, {Olsen},
  {Olsen}, {Olszewski}, {Oluseyi}, {Padilla}, {Parker}, {Pepper}, {Peterson},
  {Petry}, {Pinto}, {Pizagno}, {Popescu}, {Prsa}, {Radcka}, {Raddick},
  {Rasmussen}, {Rau}, {Rho}, {Rhoads}, {Richards}, {Ridgway}, {Robertson},
  {Roskar}, {Saha}, {Sarajedini}, {Scannapieco}, {Schalk}, {Schindler},
  {Schmidt}, {Schmidt}, {Schneider}, {Schumacher}, {Scranton}, {Sebag},
  {Seppala}, {Shemmer}, {Simon}, {Sivertz}, {Smith}, {Allyn Smith}, {Smith},
  {Spitz}, {Stanford}, {Stassun}, {Strader}, {Strauss}, {Stubbs}, {Sweeney},
  {Szalay}, {Szkody}, {Takada}, {Thorman}, {Trilling}, {Trimble}, {Tyson}, {Van
  Berg}, {Vanden Berk}, {VanderPlas}, {Verde}, {Vrsnak}, {Walkowicz},
  {Wandelt}, {Wang}, {Wang}, {Warner}, {Wechsler}, {West}, {Wiecha},
  {Williams}, {Willman}, {Wittman}, {Wolff}, {Wood-Vasey}, {Wozniak}, {Young},
  {Zentner}, \& {Zhan}}]{lsst2009_sciencebook}
{Abell}, P.~A., {Allison}, J., {Anderson}, S.~F., {et~al.} 2009, arXiv
  e-prints, arXiv:0912.0201.
\newblock \doarXiv{0912.0201}

\bibitem[{{Aldering} {et~al.}(1994){Aldering}, {Humphreys}, \&
  {Richmond}}]{aldering1994_sn1998j_progenitor}
{Aldering}, G., {Humphreys}, R.~M., \& {Richmond}, M. 1994, \aj, 107, 662,
  \dodoi{10.1086/116886}

\bibitem[{{Basinger} {et~al.}(2021){Basinger}, {Kochanek}, {Adams}, {Dai}, \&
  {Stanek}}]{basinger2021_lbt_disappeared_star}
{Basinger}, C.~M., {Kochanek}, C.~S., {Adams}, S.~M., {Dai}, X., \& {Stanek},
  K.~Z. 2021, \mnras, 508, 1156, \dodoi{10.1093/mnras/stab2620}

\bibitem[{{Bellm} {et~al.}(2019){Bellm}, {Kulkarni}, {Graham}, {Dekany},
  {Smith}, {Riddle}, {Masci}, {Helou}, {Prince}, {Adams}, {Barbarino},
  {Barlow}, {Bauer}, {Beck}, {Belicki}, {Biswas}, {Blagorodnova}, {Bodewits},
  {Bolin}, {Brinnel}, {Brooke}, {Bue}, {Bulla}, {Burruss}, {Cenko}, {Chang},
  {Connolly}, {Coughlin}, {Cromer}, {Cunningham}, {De}, {Delacroix}, {Desai},
  {Duev}, {Eadie}, {Farnham}, {Feeney}, {Feindt}, {Flynn}, {Franckowiak},
  {Frederick}, {Fremling}, {Gal-Yam}, {Gezari}, {Giomi}, {Goldstein},
  {Golkhou}, {Goobar}, {Groom}, {Hacopians}, {Hale}, {Henning}, {Ho}, {Hover},
  {Howell}, {Hung}, {Huppenkothen}, {Imel}, {Ip}, {Ivezi{\'c}}, {Jackson},
  {Jones}, {Juric}, {Kasliwal}, {Kaspi}, {Kaye}, {Kelley}, {Kowalski},
  {Kramer}, {Kupfer}, {Landry}, {Laher}, {Lee}, {Lin}, {Lin}, {Lunnan},
  {Giomi}, {Mahabal}, {Mao}, {Miller}, {Monkewitz}, {Murphy}, {Ngeow},
  {Nordin}, {Nugent}, {Ofek}, {Patterson}, {Penprase}, {Porter}, {Rauch},
  {Rebbapragada}, {Reiley}, {Rigault}, {Rodriguez}, {van Roestel}, {Rusholme},
  {van Santen}, {Schulze}, {Shupe}, {Singer}, {Soumagnac}, {Stein}, {Surace},
  {Sollerman}, {Szkody}, {Taddia}, {Terek}, {Van Sistine}, {van Velzen},
  {Vestrand}, {Walters}, {Ward}, {Ye}, {Yu}, {Yan}, \& {Zolkower}}]{bellm2019}
{Bellm}, E.~C., {Kulkarni}, S.~R., {Graham}, M.~J., {et~al.} 2019, \pasp, 131,
  018002, \dodoi{10.1088/1538-3873/aaecbe}

\bibitem[{{Blagorodnova} {et~al.}(2018){Blagorodnova}, {Neill}, {Walters},
  {Kulkarni}, {Fremling}, {Ben-Ami}, {Dekany}, {Fucik}, {Konidaris}, \&
  {Nash}}]{blagorodnova2018}
{Blagorodnova}, N., {Neill}, J.~D., {Walters}, R., {et~al.} 2018, \pasp, 130,
  035003, \dodoi{10.1088/1538-3873/aaa53f}

\bibitem[{{Bostroem} {et~al.}(2020){Bostroem}, {Valenti}, {Sand}, {Andrews},
  {Van Dyk}, {Galbany}, {Pooley}, {Amaro}, {Smith}, {Yang}, {Anupama},
  {Arcavi}, {Baron}, {Brown}, {Burke}, {Cartier}, {Hiramatsu}, {Dastidar},
  {DerKacy}, {Dong}, {Egami}, {Ertel}, {Filippenko}, {Fox}, {Haislip},
  {Hosseinzadeh}, {Howell}, {Gangopadhyay}, {Jha}, {Kouprianov}, {Kumar},
  {Lundquist}, {Milisavljevic}, {McCully}, {Milne}, {Misra}, {Reichart},
  {Sahu}, {Sai}, {Singh}, {Smith}, {Vinko}, {Wang}, {Wang}, {Wheeler},
  {Williams}, {Wyatt}, {Zhang}, \& {Zhang}}]{bostroem2020}
{Bostroem}, K.~A., {Valenti}, S., {Sand}, D.~J., {et~al.} 2020, \apj, 895, 31,
  \dodoi{10.3847/1538-4357/ab8945}

\bibitem[{{Bruch} {et~al.}(2021){Bruch}, {Gal-Yam}, {Schulze}, {Yaron}, {Yang},
  {Soumagnac}, {Rigault}, {Strotjohann}, {Ofek}, {Sollerman}, {Masci},
  {Barbarino}, {Ho}, {Fremling}, {Perley}, {Nordin}, {Cenko}, {Adams},
  {Adreoni}, {Bellm}, {Blagorodnova}, {Bulla}, {Burdge}, {De}, {Dhawan},
  {Drake}, {Duev}, {Dugas}, {Graham}, {Graham}, {Irani}, {Jencson},
  {Karamehmetoglu}, {Kasliwal}, {Kim}, {Kulkarni}, {Kupfer}, {Liang},
  {Mahabal}, {Miller}, {Prince}, {Riddle}, {Sharma}, {Smith}, {Taddia},
  {Taggart}, {Walters}, \& {Yan}}]{bruch2021}
{Bruch}, R.~J., {Gal-Yam}, A., {Schulze}, S., {et~al.} 2021, \apj, 912, 46,
  \dodoi{10.3847/1538-4357/abef05}

\bibitem[{{Bruch} {et~al.}(2022){Bruch}, {Gal-Yam}, {Yaron}, {Chen},
  {Strotjohann}, {Irani}, {Zimmerman}, {Schulze}, {Yang}, {Kim}, {Bulla},
  {Sollerman}, {Rigault}, {Ofek}, {Soumagnac}, {Masci}, {Fremling}, {Perley},
  {Nordin}, {Cenko}, {Ho}, {Adams}, {Adreoni}, {Bellm}, {Blagorodnova},
  {Burdge}, {De}, {Dekany}, {Dhawan}, {Drake}, {Duev}, {Graham}, {Graham},
  {Jencson}, {Karamehmetoglu}, {Kasliwal Shrinivas Kulkarni}, {Miller},
  {Neill}, {Prince}, {Riddle}, {Rusholme}, {Sharma}, {Smith}, {Sravan},
  {Taggart}, {Walters}, \& {Yan}}]{bruch2022}
{Bruch}, R.~J., {Gal-Yam}, A., {Yaron}, O., {et~al.} 2022, arXiv e-prints,
  arXiv:2212.03313.
\newblock \doarXiv{2212.03313}

\bibitem[{{Crockett} {et~al.}(2011){Crockett}, {Smartt}, {Pastorello},
  {Eldridge}, {Stephens}, {Maund}, \& {Mattila}}]{crockett2011}
{Crockett}, R.~M., {Smartt}, S.~J., {Pastorello}, A., {et~al.} 2011, \mnras,
  410, 2767, \dodoi{10.1111/j.1365-2966.2010.17652.x}

\bibitem[{{D{\'a}lya} {et~al.}(2022){D{\'a}lya}, {D{\'\i}az}, {Bouchet},
  {Frei}, {Jasche}, {Lavaux}, {Macas}, {Mukherjee}, {P{\'a}lfi}, {de Souza},
  {Wandelt}, {Bilicki}, \& {Raffai}}]{dalya2022}
{D{\'a}lya}, G., {D{\'\i}az}, R., {Bouchet}, F.~R., {et~al.} 2022, \mnras,
  \dodoi{10.1093/mnras/stac1443}

\bibitem[{{Davies} \& {Beasor}(2018)}]{davies2018}
{Davies}, B., \& {Beasor}, E.~R. 2018, \mnras, 474, 2116,
  \dodoi{10.1093/mnras/stx2734}

\bibitem[{{Davies} \& {Beasor}(2020)}]{davies2020}
---. 2020, \mnras, 493, 468, \dodoi{10.1093/mnras/staa174}

\bibitem[{{Davies} {et~al.}(2022){Davies}, {Plez}, \& {Petrault}}]{davies2022}
{Davies}, B., {Plez}, B., \& {Petrault}, M. 2022, \mnras, 517, 1483,
  \dodoi{10.1093/mnras/stac2427}

\bibitem[{{Dekany} {et~al.}(2020){Dekany}, {Smith}, {Riddle}, {Feeney},
  {Porter}, {Hale}, {Zolkower}, {Belicki}, {Kaye}, {Henning}, {Walters},
  {Cromer}, {Delacroix}, {Rodriguez}, {Reiley}, {Mao}, {Hover}, {Murphy},
  {Burruss}, {Baker}, {Kowalski}, {Reif}, {Mueller}, {Bellm}, {Graham}, \&
  {Kulkarni}}]{dekany2020_ztf_observing_system}
{Dekany}, R., {Smith}, R.~M., {Riddle}, R., {et~al.} 2020, \pasp, 132, 038001,
  \dodoi{10.1088/1538-3873/ab4ca2}

\bibitem[{{Dressel} \& {Marinelli}(2023)}]{dressel2023_wfc3handbook}
{Dressel}, L., \& {Marinelli}, M. 2023, {Wide Field Camera 3 Instrument
  Handbook, Version 15.0} (Baltimore: STScI)

\bibitem[{{Efron}(1982)}]{efron1982}
{Efron}, B. 1982, {The Jackknife, the Bootstrap and other resampling plans}

\bibitem[{{Eldridge} \&
  {Maund}(2016)}]{eldridge2016_disappearance_Ib_progenitor_iPTF13bvn}
{Eldridge}, J.~J., \& {Maund}, J.~R. 2016, \mnras, 461, L117,
  \dodoi{10.1093/mnrasl/slw099}

\bibitem[{{Elias-Rosa} {et~al.}(2011){Elias-Rosa}, {Van Dyk}, {Li},
  {Silverman}, {Foley}, {Ganeshalingam}, {Mauerhan}, {Kankare}, {Jha},
  {Filippenko}, {Beckman}, {Berger}, {Cuillandre}, \& {Smith}}]{elias-rosa2011}
{Elias-Rosa}, N., {Van Dyk}, S.~D., {Li}, W., {et~al.} 2011, \apj, 742, 6,
  \dodoi{10.1088/0004-637X/742/1/6}

\bibitem[{{Ergon} {et~al.}(2015){Ergon}, {Jerkstrand}, {Sollerman},
  {Elias-Rosa}, {Fransson}, {Fraser}, {Pastorello}, {Kotak}, {Taubenberger},
  {Tomasella}, {Valenti}, {Benetti}, {Helou}, {Kasliwal}, {Maund}, {Smartt}, \&
  {Spyromilio}}]{ergon2015_sn2011dh}
{Ergon}, M., {Jerkstrand}, A., {Sollerman}, J., {et~al.} 2015, \aap, 580, A142,
  \dodoi{10.1051/0004-6361/201424592}

\bibitem[{{Folatelli} {et~al.}(2015){Folatelli}, {Bersten}, {Kuncarayakti},
  {Benvenuto}, {Maeda}, \& {Nomoto}}]{folatelli2015}
{Folatelli}, G., {Bersten}, M.~C., {Kuncarayakti}, H., {et~al.} 2015, \apj,
  811, 147, \dodoi{10.1088/0004-637X/811/2/147}

\bibitem[{{F{\"o}rster} {et~al.}(2018){F{\"o}rster}, {Moriya}, {Maureira},
  {Anderson}, {Blinnikov}, {Bufano}, {Cabrera-Vives}, {Clocchiatti}, {de
  Jaeger}, {Est{\'e}vez}, {Galbany}, {Gonz{\'a}lez-Gait{\'a}n}, {Gr{\"a}fener},
  {Hamuy}, {Hsiao}, {Huentelemu}, {Huijse}, {Kuncarayakti}, {Mart{\'\i}nez},
  {Medina}, {Olivares E.}, {Pignata}, {Razza}, {Reyes}, {San Mart{\'\i}n},
  {Smith}, {Vera}, {Vivas}, {de Ugarte Postigo}, {Yoon}, {Ashall}, {Fraser},
  {Gal-Yam}, {Kankare}, {Le Guillou}, {Mazzali}, {Walton}, \&
  {Young}}]{foerster2018_early_snlcs_csm}
{F{\"o}rster}, F., {Moriya}, T.~J., {Maureira}, J.~C., {et~al.} 2018, Nature
  Astronomy, 2, 808, \dodoi{10.1038/s41550-018-0563-4}

\bibitem[{{Fraser}(2016)}]{fraser2016_disappearanceSN2012aw}
{Fraser}, M. 2016, \mnras, 456, L16, \dodoi{10.1093/mnrasl/slv168}

\bibitem[{{Fremling} {et~al.}(2020){Fremling}, {Miller}, {Sharma}, {Dugas},
  {Perley}, {Taggart}, {Sollerman}, {Goobar}, {Graham}, {Neill}, {Nordin},
  {Rigault}, {Walters}, {Andreoni}, {Bagdasaryan}, {Belicki}, {Cannella},
  {Bellm}, {Cenko}, {De}, {Dekany}, {Frederick}, {Golkhou}, {Graham}, {Helou},
  {Ho}, {Kasliwal}, {Kupfer}, {Laher}, {Mahabal}, {Masci}, {Riddle},
  {Rusholme}, {Schulze}, {Shupe}, {Smith}, {van Velzen}, {Yan}, {Yao},
  {Zhuang}, \& {Kulkarni}}]{fremling2020}
{Fremling}, C., {Miller}, A.~A., {Sharma}, Y., {et~al.} 2020, \apj, 895, 32,
  \dodoi{10.3847/1538-4357/ab8943}

\bibitem[{{Gagliano} {et~al.}(2022){Gagliano}, {Izzo}, {Kilpatrick}, {Mockler},
  {Jacobson-Gal{\'a}n}, {Terreran}, {Dimitriadis}, {Zenati}, {Auchettl},
  {Drout}, {Narayan}, {Foley}, {Margutti}, {Rest}, {Jones}, {Aganze}, {Aleo},
  {Burgasser}, {Coulter}, {Gerasimov}, {Gall}, {Hjorth}, {Hsu}, {Magnier},
  {Mandel}, {Piro}, {Rojas-Bravo}, {Siebert}, {Stacey}, {Stroh}, {Swift},
  {Taggart}, {Tinyanont}, \& {Tinyanont}}]{gagliano2022}
{Gagliano}, A., {Izzo}, L., {Kilpatrick}, C.~D., {et~al.} 2022, \apj, 924, 55,
  \dodoi{10.3847/1538-4357/ac35ec}

\bibitem[{{Gal-Yam} {et~al.}(2007){Gal-Yam}, {Leonard}, {Fox}, {Cenko},
  {Soderberg}, {Moon}, {Sand}, {Caltech Core Collapse Program}, {Li},
  {Filippenko}, {Aldering}, \& {Copin}}]{gal-yam2007_2005gl_lbv_progenitor}
{Gal-Yam}, A., {Leonard}, D.~C., {Fox}, D.~B., {et~al.} 2007, \apj, 656, 372,
  \dodoi{10.1086/510523}

\bibitem[{{Graham} {et~al.}(2019){Graham}, {Kulkarni}, {Bellm}, {Adams},
  {Barbarino}, {Blagorodnova}, {Bodewits}, {Bolin}, {Brady}, {Cenko}, {Chang},
  {Coughlin}, {De}, {Eadie}, {Farnham}, {Feindt}, {Franckowiak}, {Fremling},
  {Gezari}, {Ghosh}, {Goldstein}, {Golkhou}, {Goobar}, {Ho}, {Huppenkothen},
  {Ivezi{\'c}}, {Jones}, {Juric}, {Kaplan}, {Kasliwal}, {Kelley}, {Kupfer},
  {Lee}, {Lin}, {Lunnan}, {Mahabal}, {Miller}, {Ngeow}, {Nugent}, {Ofek},
  {Prince}, {Rauch}, {van Roestel}, {Schulze}, {Singer}, {Sollerman}, {Taddia},
  {Yan}, {Ye}, {Yu}, {Barlow}, {Bauer}, {Beck}, {Belicki}, {Biswas}, {Brinnel},
  {Brooke}, {Bue}, {Bulla}, {Burruss}, {Connolly}, {Cromer}, {Cunningham},
  {Dekany}, {Delacroix}, {Desai}, {Duev}, {Feeney}, {Flynn}, {Frederick},
  {Gal-Yam}, {Giomi}, {Groom}, {Hacopians}, {Hale}, {Helou}, {Henning},
  {Hover}, {Hillenbrand}, {Howell}, {Hung}, {Imel}, {Ip}, {Jackson}, {Kaspi},
  {Kaye}, {Kowalski}, {Kramer}, {Kuhn}, {Landry}, {Laher}, {Mao}, {Masci},
  {Monkewitz}, {Murphy}, {Nordin}, {Patterson}, {Penprase}, {Porter},
  {Rebbapragada}, {Reiley}, {Riddle}, {Rigault}, {Rodriguez}, {Rusholme}, {van
  Santen}, {Shupe}, {Smith}, {Soumagnac}, {Stein}, {Surace}, {Szkody}, {Terek},
  {Van Sistine}, {van Velzen}, {Vestrand}, {Walters}, {Ward}, {Zhang}, \&
  {Zolkower}}]{graham2019}
{Graham}, M.~J., {Kulkarni}, S.~R., {Bellm}, E.~C., {et~al.} 2019, \pasp, 131,
  078001, \dodoi{10.1088/1538-3873/ab006c}

\bibitem[{{Hiramatsu} {et~al.}(2021){Hiramatsu}, {Howell}, {Van Dyk},
  {Goldberg}, {Maeda}, {Moriya}, {Tominaga}, {Nomoto}, {Hosseinzadeh},
  {Arcavi}, {McCully}, {Burke}, {Bostroem}, {Valenti}, {Dong}, {Brown},
  {Andrews}, {Bilinski}, {Williams}, {Smith}, {Smith}, {Sand}, {Anand}, {Xu},
  {Filippenko}, {Bersten}, {Folatelli}, {Kelly}, {Noguchi}, \&
  {Itagaki}}]{hiramatsu2021_sn2018zd}
{Hiramatsu}, D., {Howell}, D.~A., {Van Dyk}, S.~D., {et~al.} 2021, Nature
  Astronomy, 5, 903, \dodoi{10.1038/s41550-021-01384-2}

\bibitem[{{Ho} {et~al.}(2019){Ho}, {Goldstein}, {Schulze}, {Khatami}, {Perley},
  {Ergon}, {Gal-Yam}, {Corsi}, {Andreoni}, {Barbarino}, {Bellm},
  {Blagorodnova}, {Bright}, {Burns}, {Cenko}, {Cunningham}, {De}, {Dekany},
  {Dugas}, {Fender}, {Fransson}, {Fremling}, {Goldstein}, {Graham}, {Hale},
  {Horesh}, {Hung}, {Kasliwal}, {Kuin}, {Kulkarni}, {Kupfer}, {Lunnan},
  {Masci}, {Ngeow}, {Nugent}, {Ofek}, {Patterson}, {Petitpas}, {Rusholme},
  {Sai}, {Sfaradi}, {Shupe}, {Sollerman}, {Soumagnac}, {Tachibana}, {Taddia},
  {Walters}, {Wang}, {Yao}, \& {Zhang}}]{ho2019}
{Ho}, A. Y.~Q., {Goldstein}, D.~A., {Schulze}, S., {et~al.} 2019, \apj, 887,
  169, \dodoi{10.3847/1538-4357/ab55ec}

\bibitem[{{IRSA}(2022)}]{irsa}
{IRSA}. 2022, Zwicky Transient Facility Image Service,  IPAC,
  \dodoi{10.26131/IRSA539}.
\newblock
  \url{https://catcopy.ipac.caltech.edu/dois/doi.php?id=10.26131/IRSA539}

\bibitem[{{Ivezi{\'c}} {et~al.}(2019){Ivezi{\'c}}, {Kahn}, {Tyson}, {Abel},
  {Acosta}, {Allsman}, {Alonso}, {AlSayyad}, {Anderson}, {Andrew}, {Angel},
  {Angeli}, {Ansari}, {Antilogus}, {Araujo}, {Armstrong}, {Arndt}, {Astier},
  {Aubourg}, {Auza}, {Axelrod}, {Bard}, {Barr}, {Barrau}, {Bartlett}, {Bauer},
  {Bauman}, {Baumont}, {Bechtol}, {Bechtol}, {Becker}, {Becla}, {Beldica},
  {Bellavia}, {Bianco}, {Biswas}, {Blanc}, {Blazek}, {Blandford}, {Bloom},
  {Bogart}, {Bond}, {Booth}, {Borgland}, {Borne}, {Bosch}, {Boutigny},
  {Brackett}, {Bradshaw}, {Brandt}, {Brown}, {Bullock}, {Burchat}, {Burke},
  {Cagnoli}, {Calabrese}, {Callahan}, {Callen}, {Carlin}, {Carlson},
  {Chandrasekharan}, {Charles-Emerson}, {Chesley}, {Cheu}, {Chiang}, {Chiang},
  {Chirino}, {Chow}, {Ciardi}, {Claver}, {Cohen-Tanugi}, {Cockrum}, {Coles},
  {Connolly}, {Cook}, {Cooray}, {Covey}, {Cribbs}, {Cui}, {Cutri}, {Daly},
  {Daniel}, {Daruich}, {Daubard}, {Daues}, {Dawson}, {Delgado}, {Dellapenna},
  {de Peyster}, {de Val-Borro}, {Digel}, {Doherty}, {Dubois},
  {Dubois-Felsmann}, {Durech}, {Economou}, {Eifler}, {Eracleous}, {Emmons},
  {Fausti Neto}, {Ferguson}, {Figueroa}, {Fisher-Levine}, {Focke}, {Foss},
  {Frank}, {Freemon}, {Gangler}, {Gawiser}, {Geary}, {Gee}, {Geha}, {Gessner},
  {Gibson}, {Gilmore}, {Glanzman}, {Glick}, {Goldina}, {Goldstein}, {Goodenow},
  {Graham}, {Gressler}, {Gris}, {Guy}, {Guyonnet}, {Haller}, {Harris},
  {Hascall}, {Haupt}, {Hernandez}, {Herrmann}, {Hileman}, {Hoblitt}, {Hodgson},
  {Hogan}, {Howard}, {Huang}, {Huffer}, {Ingraham}, {Innes}, {Jacoby}, {Jain},
  {Jammes}, {Jee}, {Jenness}, {Jernigan}, {Jevremovi{\'c}}, {Johns}, {Johnson},
  {Johnson}, {Jones}, {Juramy-Gilles}, {Juri{\'c}}, {Kalirai}, {Kallivayalil},
  {Kalmbach}, {Kantor}, {Karst}, {Kasliwal}, {Kelly}, {Kessler}, {Kinnison},
  {Kirkby}, {Knox}, {Kotov}, {Krabbendam}, {Krughoff}, {Kub{\'a}nek},
  {Kuczewski}, {Kulkarni}, {Ku}, {Kurita}, {Lage}, {Lambert}, {Lange},
  {Langton}, {Le Guillou}, {Levine}, {Liang}, {Lim}, {Lintott}, {Long},
  {Lopez}, {Lotz}, {Lupton}, {Lust}, {MacArthur}, {Mahabal}, {Mandelbaum},
  {Markiewicz}, {Marsh}, {Marshall}, {Marshall}, {May}, {McKercher}, {McQueen},
  {Meyers}, {Migliore}, {Miller}, {Mills}, {Miraval}, {Moeyens}, {Moolekamp},
  {Monet}, {Moniez}, {Monkewitz}, {Montgomery}, {Morrison}, {Mueller},
  {Muller}, {Mu{\~n}oz Arancibia}, {Neill}, {Newbry}, {Nief}, {Nomerotski},
  {Nordby}, {O'Connor}, {Oliver}, {Olivier}, {Olsen}, {O'Mullane}, {Ortiz},
  {Osier}, {Owen}, {Pain}, {Palecek}, {Parejko}, {Parsons}, {Pease},
  {Peterson}, {Peterson}, {Petravick}, {Libby Petrick}, {Petry},
  {Pierfederici}, {Pietrowicz}, {Pike}, {Pinto}, {Plante}, {Plate}, {Plutchak},
  {Price}, {Prouza}, {Radeka}, {Rajagopal}, {Rasmussen}, {Regnault}, {Reil},
  {Reiss}, {Reuter}, {Ridgway}, {Riot}, {Ritz}, {Robinson}, {Roby}, {Roodman},
  {Rosing}, {Roucelle}, {Rumore}, {Russo}, {Saha}, {Sassolas}, {Schalk},
  {Schellart}, {Schindler}, {Schmidt}, {Schneider}, {Schneider}, {Schoening},
  {Schumacher}, {Schwamb}, {Sebag}, {Selvy}, {Sembroski}, {Seppala}, {Serio},
  {Serrano}, {Shaw}, {Shipsey}, {Sick}, {Silvestri}, {Slater}, {Smith},
  {Smith}, {Sobhani}, {Soldahl}, {Storrie-Lombardi}, {Stover}, {Strauss},
  {Street}, {Stubbs}, {Sullivan}, {Sweeney}, {Swinbank}, {Szalay}, {Takacs},
  {Tether}, {Thaler}, {Thayer}, {Thomas}, {Thornton}, {Thukral}, {Tice},
  {Trilling}, {Turri}, {Van Berg}, {Vanden Berk}, {Vetter}, {Virieux},
  {Vucina}, {Wahl}, {Walkowicz}, {Walsh}, {Walter}, {Wang}, {Wang}, {Warner},
  {Wiecha}, {Willman}, {Winters}, {Wittman}, {Wolff}, {Wood-Vasey}, {Wu},
  {Xin}, {Yoachim}, \& {Zhan}}]{ivezic2019}
{Ivezi{\'c}}, {\v{Z}}., {Kahn}, S.~M., {Tyson}, J.~A., {et~al.} 2019, \apj,
  873, 111, \dodoi{10.3847/1538-4357/ab042c}

\bibitem[{{Jacobson-Gal{\'a}n} {et~al.}(2020){Jacobson-Gal{\'a}n}, {Margutti},
  {Kilpatrick}, {Hiramatsu}, {Perets}, {Khatami}, {Foley}, {Raymond}, {Yoon},
  {Bobrick}, {Zenati}, {Galbany}, {Andrews}, {Brown}, {Cartier}, {Coppejans},
  {Dimitriadis}, {Dobson}, {Hajela}, {Howell}, {Kuncarayakti}, {Milisavljevic},
  {Rahman}, {Rojas-Bravo}, {Sand}, {Shepherd}, {Smartt}, {Stacey}, {Stroh},
  {Swift}, {Terreran}, {Vinko}, {Wang}, {Anderson}, {Baron}, {Berger},
  {Blanchard}, {Burke}, {Coulter}, {DeMarchi}, {DerKacy}, {Fremling}, {Gomez},
  {Gromadzki}, {Hosseinzadeh}, {Kasen}, {Kriskovics}, {McCully},
  {M{\"u}ller-Bravo}, {Nicholl}, {Ordasi}, {Pellegrino}, {Piro}, {P{\'a}l},
  {Ren}, {Rest}, {Rich}, {Sai}, {S{\'a}rneczky}, {Shen}, {Short}, {Siebert},
  {Stauffer}, {Szak{\'a}ts}, {Zhang}, {Zhang}, \& {Zhang}}]{jacobson_galan2020}
{Jacobson-Gal{\'a}n}, W.~V., {Margutti}, R., {Kilpatrick}, C.~D., {et~al.}
  2020, \apj, 898, 166, \dodoi{10.3847/1538-4357/ab9e66}

\bibitem[{{Jacobson-Gal{\'a}n}
  {et~al.}(2022{\natexlab{a}}){Jacobson-Gal{\'a}n}, {Dessart}, {Jones},
  {Margutti}, {Coppejans}, {Dimitriadis}, {Foley}, {Kilpatrick}, {Matthews},
  {Rest}, {Terreran}, {Aleo}, {Auchettl}, {Blanchard}, {Coulter}, {Davis}, {de
  Boer}, {DeMarchi}, {Drout}, {Earl}, {Gagliano}, {Gall}, {Hjorth}, {Huber},
  {Ibik}, {Milisavljevic}, {Pan}, {Rest}, {Ridden-Harper}, {Rojas-Bravo},
  {Siebert}, {Smith}, {Taggart}, {Tinyanont}, {Wang}, \&
  {Zenati}}]{jacobson-galan2022}
{Jacobson-Gal{\'a}n}, W.~V., {Dessart}, L., {Jones}, D.~O., {et~al.}
  2022{\natexlab{a}}, \apj, 924, 15, \dodoi{10.3847/1538-4357/ac3f3a}

\bibitem[{{Jacobson-Gal{\'a}n}
  {et~al.}(2022{\natexlab{b}}){Jacobson-Gal{\'a}n}, {Venkatraman}, {Margutti},
  {Khatami}, {Terreran}, {Foley}, {Angulo}, {Angus}, {Auchettl}, {Blanchard},
  {Bobrick}, {Bright}, {Brout}, {Chambers}, {Couch}, {Coulter}, {Clever},
  {Davis}, {de Boer}, {DeMarchi}, {Dodd}, {Jones}, {Johnson}, {Kilpatrick},
  {Khetan}, {Lai}, {Langeroodi}, {Lin}, {Magnier}, {Milisavljevic}, {Perets},
  {Pierel}, {Raymond}, {Rest}, {Rest}, {Ridden-Harper}, {Shen}, {Siebert},
  {Smith}, {Taggart}, {Tinyanont}, {Valdes}, {Villar}, {Wang}, {Yadavalli},
  {Zenati}, \& {Zenteno}}]{jacobson-galan2022_sn2021gno}
{Jacobson-Gal{\'a}n}, W.~V., {Venkatraman}, P., {Margutti}, R., {et~al.}
  2022{\natexlab{b}}, \apj, 932, 58, \dodoi{10.3847/1538-4357/ac67dc}

\bibitem[{{Khazov} {et~al.}(2016){Khazov}, {Yaron}, {Gal-Yam}, {Manulis},
  {Rubin}, {Kulkarni}, {Arcavi}, {Kasliwal}, {Ofek}, {Cao}, {Perley},
  {Sollerman}, {Horesh}, {Sullivan}, {Filippenko}, {Nugent}, {Howell}, {Cenko},
  {Silverman}, {Ebeling}, {Taddia}, {Johansson}, {Laher}, {Surace},
  {Rebbapragada}, {Wozniak}, \& {Matheson}}]{khazov2016}
{Khazov}, D., {Yaron}, O., {Gal-Yam}, A., {et~al.} 2016, \apj, 818, 3,
  \dodoi{10.3847/0004-637X/818/1/3}

\bibitem[{{Kilpatrick} {et~al.}(2022){Kilpatrick}, {Coulter}, {Foley}, {Piro},
  {Rest}, {Rojas-Bravo}, \& {Siebert}}]{kilpatrick2022_sn2016gkg}
{Kilpatrick}, C.~D., {Coulter}, D.~A., {Foley}, R.~J., {et~al.} 2022, \apj,
  936, 111, \dodoi{10.3847/1538-4357/ac8a4c}

\bibitem[{{Kilpatrick} {et~al.}(2021){Kilpatrick}, {Drout}, {Auchettl},
  {Dimitriadis}, {Foley}, {Jones}, {DeMarchi}, {French}, {Gall}, {Hjorth},
  {Jacobson-Gal{\'a}n}, {Margutti}, {Piro}, {Ramirez-Ruiz}, {Rest}, \&
  {Rojas-Bravo}}]{kilpatrick2021}
{Kilpatrick}, C.~D., {Drout}, M.~R., {Auchettl}, K., {et~al.} 2021, \mnras,
  504, 2073, \dodoi{10.1093/mnras/stab838}

\bibitem[{{Kim} {et~al.}(2022){Kim}, {Rigault}, {Neill}, {Briday}, {Copin},
  {Lezmy}, {Nicolas}, {Riddle}, {Sharma}, {Smith}, {Sollerman}, \&
  {Walters}}]{kim2022}
{Kim}, Y.~L., {Rigault}, M., {Neill}, J.~D., {et~al.} 2022, \pasp, 134, 024505,
  \dodoi{10.1088/1538-3873/ac50a0}

\bibitem[{{Kleiser} {et~al.}(2011){Kleiser}, {Poznanski}, {Kasen}, {Young},
  {Chornock}, {Filippenko}, {Challis}, {Ganeshalingam}, {Kirshner}, {Li},
  {Matheson}, {Nugent}, \& {Silverman}}]{kleiser2011_sne_from_bsg}
{Kleiser}, I. K.~W., {Poznanski}, D., {Kasen}, D., {et~al.} 2011, \mnras, 415,
  372, \dodoi{10.1111/j.1365-2966.2011.18708.x}

\bibitem[{{Kochanek}(2020)}]{kochanek2020}
{Kochanek}, C.~S. 2020, \mnras, 493, 4945, \dodoi{10.1093/mnras/staa605}

\bibitem[{{Kochanek} {et~al.}(2008){Kochanek}, {Beacom}, {Kistler}, {Prieto},
  {Stanek}, {Thompson}, \& {Y{\"u}ksel}}]{kochanek2008_failedsne}
{Kochanek}, C.~S., {Beacom}, J.~F., {Kistler}, M.~D., {et~al.} 2008, \apj, 684,
  1336, \dodoi{10.1086/590053}

\bibitem[{{Kochanek} {et~al.}(2012){Kochanek}, {Khan}, \& {Dai}}]{kochanek2012}
{Kochanek}, C.~S., {Khan}, R., \& {Dai}, X. 2012, \apj, 759, 20,
  \dodoi{10.1088/0004-637X/759/1/20}

\bibitem[{{Kozyreva} {et~al.}(2022){Kozyreva}, {Janka}, {Kresse},
  {Taubenberger}, \& {Baklanov}}]{kozyreva2022}
{Kozyreva}, A., {Janka}, H.-T., {Kresse}, D., {Taubenberger}, S., \&
  {Baklanov}, P. 2022, \mnras, \dodoi{10.1093/mnras/stac1518}

\bibitem[{{Lan{\c{c}}on} {et~al.}(2021){Lan{\c{c}}on}, {Gonneau}, {Verro},
  {Prugniel}, {Arentsen}, {Trager}, {Peletier}, {Chen}, {Coelho},
  {Falc{\'o}n-Barroso}, {Hauschildt}, {Husser}, {Jain}, {Lyubenova}, {Martins},
  {S{\'a}nchez Bl{\'a}zquez}, \& {Vazdekis}}]{lacon2021}
{Lan{\c{c}}on}, A., {Gonneau}, A., {Verro}, K., {et~al.} 2021, \aap, 649, A97,
  \dodoi{10.1051/0004-6361/202039371}

\bibitem[{{Li} {et~al.}(2006){Li}, {Van Dyk}, {Filippenko}, {Cuillandre},
  {Jha}, {Bloom}, {Riess}, \& {Livio}}]{li2006_rsgproblem}
{Li}, W., {Van Dyk}, S.~D., {Filippenko}, A.~V., {et~al.} 2006, \apj, 641,
  1060, \dodoi{10.1086/499916}

\bibitem[{{Li} {et~al.}(2011){Li}, {Leaman}, {Chornock}, {Filippenko},
  {Poznanski}, {Ganeshalingam}, {Wang}, {Modjaz}, {Jha}, {Foley}, \&
  {Smith}}]{li2011_loss_sn_fractions}
{Li}, W., {Leaman}, J., {Chornock}, R., {et~al.} 2011, \mnras, 412, 1441,
  \dodoi{10.1111/j.1365-2966.2011.18160.x}

\bibitem[{{Lupton} {et~al.}(1999){Lupton}, {Gunn}, \& {Szalay}}]{lupton1999}
{Lupton}, R.~H., {Gunn}, J.~E., \& {Szalay}, A.~S. 1999, \aj, 118, 1406,
  \dodoi{10.1086/301004}

\bibitem[{{Margutti} {et~al.}(2017){Margutti}, {Kamble}, {Milisavljevic},
  {Zapartas}, {de Mink}, {Drout}, {Chornock}, {Risaliti}, {Zauderer},
  {Bietenholz}, {Cantiello}, {Chakraborti}, {Chomiuk}, {Fong}, {Grefenstette},
  {Guidorzi}, {Kirshner}, {Parrent}, {Patnaude}, {Soderberg}, {Gehrels}, \&
  {Harrison}}]{margutti2017}
{Margutti}, R., {Kamble}, A., {Milisavljevic}, D., {et~al.} 2017, \apj, 835,
  140, \dodoi{10.3847/1538-4357/835/2/140}

\bibitem[{{Masci} {et~al.}(2019){Masci}, {Laher}, {Rusholme}, {Shupe}, {Groom},
  {Surace}, {Jackson}, {Monkewitz}, {Beck}, {Flynn}, {Terek}, {Landry},
  {Hacopians}, {Desai}, {Howell}, {Brooke}, {Imel}, {Wachter}, {Ye}, {Lin},
  {Cenko}, {Cunningham}, {Rebbapragada}, {Bue}, {Miller}, {Mahabal}, {Bellm},
  {Patterson}, {Juri{\'c}}, {Golkhou}, {Ofek}, {Walters}, {Graham}, {Kasliwal},
  {Dekany}, {Kupfer}, {Burdge}, {Cannella}, {Barlow}, {Van Sistine}, {Giomi},
  {Fremling}, {Blagorodnova}, {Levitan}, {Riddle}, {Smith}, {Helou}, {Prince},
  \& {Kulkarni}}]{masci2019}
{Masci}, F.~J., {Laher}, R.~R., {Rusholme}, B., {et~al.} 2019, \pasp, 131,
  018003, \dodoi{10.1088/1538-3873/aae8ac}

\bibitem[{{Maund}(2017)}]{maund2017_progenitor_stellar_environment}
{Maund}, J.~R. 2017, \mnras, 469, 2202, \dodoi{10.1093/mnras/stx879}

\bibitem[{{Maund}(2019)}]{maund2019}
---. 2019, \apj, 883, 86, \dodoi{10.3847/1538-4357/ab2386}

\bibitem[{{Maund} {et~al.}(2015{\natexlab{a}}){Maund}, {Fraser}, {Reilly},
  {Ergon}, \& {Mattila}}]{maund2015}
{Maund}, J.~R., {Fraser}, M., {Reilly}, E., {Ergon}, M., \& {Mattila}, S.
  2015{\natexlab{a}}, \mnras, 447, 3207, \dodoi{10.1093/mnras/stu2658}

\bibitem[{{Maund} {et~al.}(2014){Maund}, {Reilly}, \& {Mattila}}]{maund2014}
{Maund}, J.~R., {Reilly}, E., \& {Mattila}, S. 2014, \mnras, 438, 938,
  \dodoi{10.1093/mnras/stt2131}

\bibitem[{{Maund} \& {Smartt}(2009)}]{maund2009_progenitor_disappearance}
{Maund}, J.~R., \& {Smartt}, S.~J. 2009, Science, 324, 486,
  \dodoi{10.1126/science.1170198}

\bibitem[{{Maund} {et~al.}(2004){Maund}, {Smartt}, {Kudritzki},
  {Podsiadlowski}, \& {Gilmore}}]{maund2004_sn1993j_binarydetection}
{Maund}, J.~R., {Smartt}, S.~J., {Kudritzki}, R.~P., {Podsiadlowski}, P., \&
  {Gilmore}, G.~F. 2004, \nat, 427, 129, \dodoi{10.1038/nature02161}

\bibitem[{{Maund} {et~al.}(2015{\natexlab{b}}){Maund}, {Arcavi}, {Ergon},
  {Eldridge}, {Georgy}, {Cenko}, {Horesh}, {Izzard}, \&
  {Stancliffe}}]{maund2015_sn2011dh_binarycompanion}
{Maund}, J.~R., {Arcavi}, I., {Ergon}, M., {et~al.} 2015{\natexlab{b}}, \mnras,
  454, 2580, \dodoi{10.1093/mnras/stv2098}

\bibitem[{{Morozova} {et~al.}(2020){Morozova}, {Piro}, {Fuller}, \& {Van
  Dyk}}]{morazova2020}
{Morozova}, V., {Piro}, A.~L., {Fuller}, J., \& {Van Dyk}, S.~D. 2020, \apjl,
  891, L32, \dodoi{10.3847/2041-8213/ab77c8}

\bibitem[{{Neustadt} {et~al.}(2021){Neustadt}, {Kochanek}, {Stanek},
  {Basinger}, {Jayasinghe}, {Garling}, {Adams}, \&
  {Gerke}}]{neustadt2021_LBT_missingBSG}
{Neustadt}, J.~M.~M., {Kochanek}, C.~S., {Stanek}, K.~Z., {et~al.} 2021,
  \mnras, 508, 516, \dodoi{10.1093/mnras/stab2605}

\bibitem[{{Ofek} \& {Ben-Ami}(2020)}]{ofek2020_seeing_limited_surveys}
{Ofek}, E.~O., \& {Ben-Ami}, S. 2020, \pasp, 132, 125004,
  \dodoi{10.1088/1538-3873/abc14c}

\bibitem[{{Ofek} {et~al.}(2014){Ofek}, {Sullivan}, {Shaviv}, {Steinbok},
  {Arcavi}, {Gal-Yam}, {Tal}, {Kulkarni}, {Nugent}, {Ben-Ami}, {Kasliwal},
  {Cenko}, {Laher}, {Surace}, {Bloom}, {Filippenko}, {Silverman}, \&
  {Yaron}}]{ofek2014}
{Ofek}, E.~O., {Sullivan}, M., {Shaviv}, N.~J., {et~al.} 2014, \apj, 789, 104,
  \dodoi{10.1088/0004-637X/789/2/104}

\bibitem[{{Pastorello} {et~al.}(2007){Pastorello}, {Smartt}, {Mattila},
  {Eldridge}, {Young}, {Itagaki}, {Yamaoka}, {Navasardyan}, {Valenti}, {Patat},
  {Agnoletto}, {Augusteijn}, {Benetti}, {Cappellaro}, {Boles}, {Bonnet-Bidaud},
  {Botticella}, {Bufano}, {Cao}, {Deng}, {Dennefeld}, {Elias-Rosa},
  {Harutyunyan}, {Keenan}, {Iijima}, {Lorenzi}, {Mazzali}, {Meng}, {Nakano},
  {Nielsen}, {Smoker}, {Stanishev}, {Turatto}, {Xu}, \&
  {Zampieri}}]{pastorello2007}
{Pastorello}, A., {Smartt}, S.~J., {Mattila}, S., {et~al.} 2007, \nat, 447,
  829, \dodoi{10.1038/nature05825}

\bibitem[{{Patterson} {et~al.}(2019){Patterson}, {Bellm}, {Rusholme}, {Masci},
  {Juric}, {Krughoff}, {Golkhou}, {Graham}, {Kulkarni}, {Helou}, \& {Zwicky
  Transient Facility Collaboration}}]{patterson2019}
{Patterson}, M.~T., {Bellm}, E.~C., {Rusholme}, B., {et~al.} 2019, \pasp, 131,
  018001, \dodoi{10.1088/1538-3873/aae904}

\bibitem[{{Patton} \& {Sukhbold}(2020)}]{patton2020}
{Patton}, R.~A., \& {Sukhbold}, T. 2020, \mnras, 499, 2803,
  \dodoi{10.1093/mnras/staa3029}

\bibitem[{{Perley} {et~al.}(2020){Perley}, {Fremling}, {Sollerman}, {Miller},
  {Dahiwale}, {Sharma}, {Bellm}, {Biswas}, {Brink}, {Bruch}, {De}, {Dekany},
  {Drake}, {Duev}, {Filippenko}, {Gal-Yam}, {Goobar}, {Graham}, {Graham}, {Ho},
  {Irani}, {Kasliwal}, {Kim}, {Kulkarni}, {Mahabal}, {Masci}, {Modak}, {Neill},
  {Nordin}, {Riddle}, {Soumagnac}, {Strotjohann}, {Schulze}, {Taggart},
  {Tzanidakis}, {Walters}, \& {Yan}}]{perley2020}
{Perley}, D.~A., {Fremling}, C., {Sollerman}, J., {et~al.} 2020, \apj, 904, 35,
  \dodoi{10.3847/1538-4357/abbd98}

\bibitem[{{Reynolds} {et~al.}(2015){Reynolds}, {Fraser}, \&
  {Gilmore}}]{reynolds2015_HST_disappeared_star}
{Reynolds}, T.~M., {Fraser}, M., \& {Gilmore}, G. 2015, \mnras, 453, 2885,
  \dodoi{10.1093/mnras/stv1809}

\bibitem[{{Rigault}(2018)}]{rigault2018}
{Rigault}, M. 2018, {ztfquery, a python tool to access ZTF data}, doi,  Zenodo,
  \dodoi{10.5281/zenodo.1345222}

\bibitem[{{Rigault} {et~al.}(2019){Rigault}, {Neill}, {Blagorodnova}, {Dugas},
  {Feeney}, {Walters}, {Brinnel}, {Copin}, {Fremling}, {Nordin}, \&
  {Sollerman}}]{rigault2019}
{Rigault}, M., {Neill}, J.~D., {Blagorodnova}, N., {et~al.} 2019, \aap, 627,
  A115, \dodoi{10.1051/0004-6361/201935344}

\bibitem[{{Rizzo Smith} {et~al.}(2023){Rizzo Smith}, {Kochanek}, \&
  {Neustadt}}]{smith2023_latetime_nearbysne}
{Rizzo Smith}, M., {Kochanek}, C.~S., \& {Neustadt}, J.~M.~M. 2023, \mnras,
  523, 1474, \dodoi{10.1093/mnras/stad1483}

\bibitem[{{Rodr{\'\i}guez}(2022)}]{rodriguez2022}
{Rodr{\'\i}guez}, {\'O}. 2022, \mnras, 515, 897, \dodoi{10.1093/mnras/stac1831}

\bibitem[{{Rui} {et~al.}(2019){Rui}, {Wang}, {Mo}, {Xiang}, {Zhang}, {Maund},
  {Gal-Yam}, {Wang}, \& {Zhang}}]{rui2019}
{Rui}, L., {Wang}, X., {Mo}, J., {et~al.} 2019, \mnras, 485, 1990,
  \dodoi{10.1093/mnras/stz503}

\bibitem[{{Seitenzahl} {et~al.}(2014){Seitenzahl}, {Timmes}, \&
  {Magkotsios}}]{seitenzahl2014}
{Seitenzahl}, I.~R., {Timmes}, F.~X., \& {Magkotsios}, G. 2014, \apj, 792, 10,
  \dodoi{10.1088/0004-637X/792/1/10}

\bibitem[{{Singh} {et~al.}(2019){Singh}, {Sahu}, {Anupama}, {Kumar}, {Kumar},
  {Yamanaka}, {Baklanov}, {Tominaga}, {Blinnikov}, {Maeda}, {Dutta},
  {Bhalerao}, {Anche}, {Barway}, {Akitaya}, {Nakaoka}, {Kawabata}, {Kawabata},
  {Sasada}, {Takagi}, {Maehara}, {Isogai}, {Kino}, {Taguchi}, \&
  {Nagao}}]{singh2019}
{Singh}, A., {Sahu}, D.~K., {Anupama}, G.~C., {et~al.} 2019, \apjl, 882, L15,
  \dodoi{10.3847/2041-8213/ab3d44}

\bibitem[{{Smartt}(2015)}]{smartt2015}
{Smartt}, S.~J. 2015, \pasa, 32, e016, \dodoi{10.1017/pasa.2015.17}

\bibitem[{{Smartt} {et~al.}(2009){Smartt}, {Eldridge}, {Crockett}, \&
  {Maund}}]{smartt2009}
{Smartt}, S.~J., {Eldridge}, J.~J., {Crockett}, R.~M., \& {Maund}, J.~R. 2009,
  \mnras, 395, 1409, \dodoi{10.1111/j.1365-2966.2009.14506.x}

\bibitem[{{Smith}(2017)}]{smith2017_lbv_fate}
{Smith}, N. 2017, Philosophical Transactions of the Royal Society of London
  Series A, 375, 20160268, \dodoi{10.1098/rsta.2016.0268}

\bibitem[{{Sollerman} {et~al.}(2020){Sollerman}, {Fransson}, {Barbarino},
  {Fremling}, {Horesh}, {Kool}, {Schulze}, {Sfaradi}, {Yang}, {Bellm},
  {Burruss}, {Cunningham}, {De}, {Drake}, {Golkhou}, {Green}, {Kasliwal},
  {Kulkarni}, {Kupfer}, {Laher}, {Masci}, {Rodriguez}, {Rusholme}, {Williams},
  {Yan}, \& {Zolkower}}]{sollerman2020}
{Sollerman}, J., {Fransson}, C., {Barbarino}, C., {et~al.} 2020, \aap, 643,
  A79, \dodoi{10.1051/0004-6361/202038960}

\bibitem[{{Sollerman} {et~al.}(2021){Sollerman}, {Yang}, {Schulze},
  {Strotjohann}, {Jerkstrand}, {Van Dyk}, {Kool}, {Barbarino}, {Brink},
  {Bruch}, {De}, {Filippenko}, {Fremling}, {Patra}, {Perley}, {Yan}, {Yang},
  {Andreoni}, {Campbell}, {Coughlin}, {Kasliwal}, {Kim}, {Rigault}, {Shin},
  {Tzanidakis}, {Ashley}, {Moore}, \& {Travouillon}}]{sollerman2021}
{Sollerman}, J., {Yang}, S., {Schulze}, S., {et~al.} 2021, \aap, 655, A105,
  \dodoi{10.1051/0004-6361/202141374}

\bibitem[{{Strotjohann} {et~al.}(2015){Strotjohann}, {Ofek}, {Gal-Yam},
  {Sullivan}, {Kulkarni}, {Shaviv}, {Fremling}, {Kasliwal}, {Nugent}, {Cao},
  {Arcavi}, {Sollerman}, {Filippenko}, {Yaron}, {Laher}, \&
  {Surace}}]{strotjohann2015_precursors_sneIIb}
{Strotjohann}, N.~L., {Ofek}, E.~O., {Gal-Yam}, A., {et~al.} 2015, \apj, 811,
  117, \dodoi{10.1088/0004-637X/811/2/117}

\bibitem[{{Strotjohann} {et~al.}(2021){Strotjohann}, {Ofek}, {Gal-Yam},
  {Bruch}, {Schulze}, {Shaviv}, {Sollerman}, {Filippenko}, {Yaron}, {Fremling},
  {Nordin}, {Kool}, {Perley}, {Ho}, {Yang}, {Yao}, {Soumagnac}, {Graham},
  {Barbarino}, {Tartaglia}, {De}, {Goldstein}, {Cook}, {Brink}, {Taggart},
  {Yan}, {Lunnan}, {Kasliwal}, {Kulkarni}, {Nugent}, {Masci}, {Rosnet},
  {Adams}, {Andreoni}, {Bagdasaryan}, {Bellm}, {Burdge}, {Duev}, {Dugas},
  {Frederick}, {Goldwasser}, {Hankins}, {Irani}, {Karambelkar}, {Kupfer},
  {Liang}, {Neill}, {Porter}, {Riddle}, {Sharma}, {Short}, {Taddia},
  {Tzanidakis}, {van Roestel}, {Walters}, \& {Zhuang}}]{strotjohann2021}
---. 2021, \apj, 907, 99, \dodoi{10.3847/1538-4357/abd032}

\bibitem[{{Sun} {et~al.}(2022){Sun}, {Maund}, {Crowther}, {Hirai}, {Kashapov},
  {Liu}, {Liu}, \& {Zapartas}}]{sun2022_sn2019yvr_binaryprogenitor}
{Sun}, N.-C., {Maund}, J.~R., {Crowther}, P.~A., {et~al.} 2022, \mnras, 510,
  3701, \dodoi{10.1093/mnras/stab3768}

\bibitem[{{Szczygie{\l}} {et~al.}(2012){Szczygie{\l}}, {Gerke}, {Kochanek}, \&
  {Stanek}}]{szczgiel2012_variability_of_sn2011dh_progenitor}
{Szczygie{\l}}, D.~M., {Gerke}, J.~R., {Kochanek}, C.~S., \& {Stanek}, K.~Z.
  2012, \apj, 747, 23, \dodoi{10.1088/0004-637X/747/1/23}

\bibitem[{{Tinyanont} {et~al.}(2022){Tinyanont}, {Ridden-Harper}, {Foley},
  {Morozova}, {Kilpatrick}, {Dimitriadis}, {DeMarchi}, {Gagliano},
  {Jacobson-Gal{\'a}n}, {Messick}, {Pierel}, {Piro}, {Ramirez-Ruiz}, {Siebert},
  {Chambers}, {Clever}, {Coulter}, {De}, {Hankins}, {Hung}, {Jha}, {Jimenez
  Angel}, {Jones}, {Kasliwal}, {Lin}, {Marques-Chaves}, {Margutti}, {Moore},
  {P{\'e}rez-Fournon}, {Poidevin}, {Rest}, {Shirley}, {Smith}, {Strasburger},
  {Swift}, {Wainscoat}, {Wang}, \& {Zenati}}]{tinyanont2022_sn2020fqv}
{Tinyanont}, S., {Ridden-Harper}, R., {Foley}, R.~J., {et~al.} 2022, \mnras,
  512, 2777, \dodoi{10.1093/mnras/stab2887}

\bibitem[{{Tsuna} {et~al.}(2022){Tsuna}, {Takei}, \&
  {Shigeyama}}]{tsuna2022_precursor_predictions}
{Tsuna}, D., {Takei}, Y., \& {Shigeyama}, T. 2022, arXiv e-prints,
  arXiv:2208.08256, \dodoi{10.48550/arXiv.2208.08256}

\bibitem[{{Valerin} {et~al.}(2022){Valerin}, {Pumo}, {Pastorello}, {Reguitti},
  {Elias-Rosa}, {G{\'u}tierrez}, {Kankare}, {Fraser}, {Mazzali}, {Howell},
  {Kotak}, {Galbany}, {Williams}, {Cai}, {Salmaso}, {Pinter},
  {M{\"u}ller-Bravo}, {Burke}, {Padilla Gonzalez}, {Hiramatsu}, {McCully},
  {Newsome}, \& {Pellegrino}}]{valerin2022}
{Valerin}, G., {Pumo}, M.~L., {Pastorello}, A., {et~al.} 2022, \mnras, 513,
  4983, \dodoi{10.1093/mnras/stac1182}

\bibitem[{{Van Dyk}(2013)}]{vandyk2013_sn2008bk_progenitor_vanished}
{Van Dyk}, S.~D. 2013, \aj, 146, 24, \dodoi{10.1088/0004-6256/146/2/24}

\bibitem[{{Van Dyk}(2017)}]{vandyk2017}
---. 2017, Philosophical Transactions of the Royal Society of London Series A,
  375, 20160277, \dodoi{10.1098/rsta.2016.0277}

\bibitem[{{Van Dyk} {et~al.}(2014){Van Dyk}, {Zheng}, {Fox}, {Cenko}, {Clubb},
  {Filippenko}, {Foley}, {Miller}, {Smith}, {Kelly}, {Lee}, {Ben-Ami}, \&
  {Gal-Yam}}]{vandyk2014_sn2013df_progenitor}
{Van Dyk}, S.~D., {Zheng}, W., {Fox}, O.~D., {et~al.} 2014, \aj, 147, 37,
  \dodoi{10.1088/0004-6256/147/2/37}

\bibitem[{{Van Dyk} {et~al.}(2019){Van Dyk}, {Zheng}, {Maund}, {Brink},
  {Srinivasan}, {Andrews}, {Smith}, {Leonard}, {Morozova}, {Filippenko},
  {Conner}, {Milisavljevic}, {de Jaeger}, {Long}, {Isaacson}, {Crossfield},
  {Kosiarek}, {Howard}, {Fox}, {Kelly}, {Piro}, {Littlefair}, {Dhillon},
  {Wilson}, {Butterley}, {Yunus}, {Channa}, {Jeffers}, {Falcon}, {Ross},
  {Hestenes}, {Stegman}, {Zhang}, \& {Kumar}}]{vandyk2019_sn2017eaw}
{Van Dyk}, S.~D., {Zheng}, W., {Maund}, J.~R., {et~al.} 2019, \apj, 875, 136,
  \dodoi{10.3847/1538-4357/ab1136}

\bibitem[{{Van Dyk} {et~al.}(2023){Van Dyk}, {de Graw}, {Baer-Way}, {Zheng},
  {Filippenko}, {Fox}, {Smith}, {Brink}, {de Jaeger}, {Kelly}, \&
  {Vasylyev}}]{vandyk2023_disapperance_of_progenitor_stars}
{Van Dyk}, S.~D., {de Graw}, A., {Baer-Way}, R., {et~al.} 2023, \mnras, 519,
  471, \dodoi{10.1093/mnras/stac3549}

\bibitem[{{Vazquez} {et~al.}(2022){Vazquez}, {Kilpatrick}, {Dimitriadis},
  {Foley}, {Piro}, {Rest}, \&
  {Rojas-Bravo}}]{vazquez2022_sn2019mjm_progenitor_nondet}
{Vazquez}, J., {Kilpatrick}, C.~D., {Dimitriadis}, G., {et~al.} 2022, arXiv
  e-prints, arXiv:2210.05131.
\newblock \doarXiv{2210.05131}

\bibitem[{{Verro} {et~al.}(2022){Verro}, {Trager}, {Peletier}, {Lan{\c{c}}on},
  {Gonneau}, {Vazdekis}, {Prugniel}, {Chen}, {Coelho},
  {S{\'a}nchez-Bl{\'a}zquez}, {Martins}, {Arentsen}, {Lyubenova},
  {Falc{\'o}n-Barroso}, \& {Dries}}]{verro2022}
{Verro}, K., {Trager}, S.~C., {Peletier}, R.~F., {et~al.} 2022, \aap, 660, A34,
  \dodoi{10.1051/0004-6361/202142388}

\bibitem[{{Walborn} {et~al.}(1989){Walborn}, {Prevot}, {Prevot}, {Wamsteker},
  {Gonzalez}, {Gilmozzi}, \& {Fitzpatrick}}]{walborn1989}
{Walborn}, N.~R., {Prevot}, M.~L., {Prevot}, L., {et~al.} 1989, \aap, 219, 229

\bibitem[{{Weil} {et~al.}(2020){Weil}, {Fesen}, {Patnaude}, \&
  {Milisavljevic}}]{weil2020_sn2017eaw_latetime}
{Weil}, K.~E., {Fesen}, R.~A., {Patnaude}, D.~J., \& {Milisavljevic}, D. 2020,
  \apj, 900, 11, \dodoi{10.3847/1538-4357/aba4b1}

\bibitem[{{Wu} \& {Fuller}(2022)}]{wu2022_massloss_sneIbc}
{Wu}, S.~C., \& {Fuller}, J. 2022, \apjl, 940, L27,
  \dodoi{10.3847/2041-8213/ac9b3d}

\bibitem[{{Xiang} {et~al.}(2019){Xiang}, {Wang}, {Mo}, {Wang}, {Smartt},
  {Fraser}, {Ehgamberdiev}, {Mirzaqulov}, {Zhang}, {Zhang}, {Vinko}, {Wheeler},
  {Hosseinzadeh}, {Howell}, {McCully}, {DerKacy}, {Baron}, {Brown}, {Zhang},
  {Bi}, {Song}, {Zhang}, {Rest}, {Nomoto}, {Tolstov}, \&
  {Blinnikov}}]{xiang2019_sn2017ein_progenitor}
{Xiang}, D., {Wang}, X., {Mo}, J., {et~al.} 2019, \apj, 871, 176,
  \dodoi{10.3847/1538-4357/aaf8b0}

\bibitem[{{Yang} {et~al.}(2021){Yang}, {Sollerman}, {Strotjohann}, {Schulze},
  {Lunnan}, {Kool}, {Fremling}, {Perley}, {Ofek}, {Schweyer}, {Bellm},
  {Kasliwal}, {Masci}, {Rigault}, \& {Yang}}]{yang2021}
{Yang}, S., {Sollerman}, J., {Strotjohann}, N.~L., {et~al.} 2021, \aap, 655,
  A90, \dodoi{10.1051/0004-6361/202141244}

\bibitem[{{Yao} {et~al.}(2019){Yao}, {Miller}, {Kulkarni}, {Bulla}, {Masci},
  {Goldstein}, {Goobar}, {Nugent}, {Dugas}, {Blagorodnova}, {Neill}, {Rigault},
  {Sollerman}, {Nordin}, {Bellm}, {Cenko}, {De}, {Dhawan}, {Feindt},
  {Fremling}, {Gatkine}, {Graham}, {Graham}, {Ho}, {Hung}, {Kasliwal},
  {Kupfer}, {Laher}, {Perley}, {Rusholme}, {Shupe}, {Soumagnac}, {Taggart},
  {Walters}, \& {Yan}}]{yao2019}
{Yao}, Y., {Miller}, A.~A., {Kulkarni}, S.~R., {et~al.} 2019, \apj, 886, 152,
  \dodoi{10.3847/1538-4357/ab4cf5}

\bibitem[{{Zackay} {et~al.}(2016){Zackay}, {Ofek}, \& {Gal-Yam}}]{zackay2016}
{Zackay}, B., {Ofek}, E.~O., \& {Gal-Yam}, A. 2016, \apj, 830, 27,
  \dodoi{10.3847/0004-637X/830/1/27}

\end{thebibliography}
\bibliographystyle{aasjournal}

\end{document}